\documentclass[acmsmall]{acmart}

\startPage{1}

\setcopyright{none}
\setcopyright{rightsretained}
\acmDOI{10.1145/3649841}
\acmYear{2024}
\copyrightyear{2024}
\acmSubmissionID{oopslaa24main-p119-p}
\acmJournal{PACMPL}
\acmVolume{8}
\acmNumber{OOPSLA1}
\acmArticle{124}
\acmMonth{4}
\received{21-OCT-2023}
\received[accepted]{2024-02-24}
\usepackage{booktabs}   %
\usepackage{subcaption} %
\usepackage[utf8]{inputenc} %
\usepackage[T1]{fontenc}    %
\usepackage{hyperref}       %
\usepackage{url}            %
\usepackage{booktabs}       %
\usepackage{amsfonts}       %
\usepackage{nicefrac}       %
\usepackage{microtype}      %
\usepackage{xcolor}         %
\usepackage{caption}
\usepackage{mdframed}
\usepackage{graphicx}
\usepackage{enumitem}
\usepackage{algorithmic}
\usepackage{algorithm}
\usepackage{pifont}%
\newcommand{\cmark}{\ding{51}}%
\newcommand{\xmark}{\ding{55}}%
\usepackage{makecell}
\usepackage[frozencache]{minted}
\usepackage{amsmath}
\usepackage{wrapfig}
\usepackage{multirow}
\usepackage{listings}
\usepackage{tabularx} %
\usepackage{lipsum}   %
\usepackage{svg}

\renewrobustcmd{\bfseries}{\fontseries{b}\selectfont}
\renewrobustcmd{\boldmath}{}
\newrobustcmd{\B}{\bfseries}

\usepackage{xcolor}
\usepackage{tikz}
\usepackage{listings}
\usepackage{xspace}
\usepackage{minted}
\usepackage{proof}
\usepackage{amsmath}

\newcommand{\tool}{\framework}
\newcommand{\framework}{\textsc{TorchQL}}
\newcommand{\physionet}{Physionet-2012 Challenge}
\newcommand{\saits}{SAITS}
\newcommand{\cityscapes}{\textsc{Cityscapes}}

\newcommand{\join}{\texttt{join}\xspace}
\newcommand{\filter}{\texttt{filter}\xspace}
\newcommand{\cols}{\texttt{project}\xspace}
\newcommand{\project}{\texttt{project}\xspace}
\newcommand{\groupby}{\texttt{group\_by}\xspace}
\newcommand{\orderby}{\texttt{order\_by}\xspace}
\newcommand{\flatten}{\texttt{flatten}\xspace}
\newcommand{\unique}{\texttt{unique}\xspace}
\newcommand{\reduce}{\texttt{reduce}\xspace}

\newcommand{\wilds}{iWildCam}
\newcommand{\Alpaca}{Alpaca}

\newcommand{\production}{\mathrel{::=}}

\newcommand{\ctx}[2]{#2 \vdash #1}
\newcommand{\step}[2]{#1 ~\triangleright~ #2}
\newcommand{\infr}[3][]{\infer[\textsc{[#1]}]{#3}{#2}}
\newcommand{\iand}{\qquad}
\newcommand{\gor}{\ \ \ $|$\ \ \ }

\definecolor{mdbgreen}{rgb}{0,0.46,0}
\definecolor{mdbblue}{rgb}{0.02,0.42,0.74}
\definecolor{mdblightgrey}{rgb}{0.94,0.94,0.94}%
\definecolor{mdbgreyblue}{rgb}{0.3,0.4,0.6}%
\definecolor{mdbcyan}{rgb}{0.1,0.4,0.6}%
\definecolor{mdbpurple}{rgb}{0.71,0,0.85}%
\definecolor{mdbyellow}{rgb}{0.9,0.6,0.05}%
\definecolor{mdborange}{rgb}{1,0.36,0.03}%
\definecolor{mdbred}{rgb}{0.6,0.2,0.0}%

\newcommand{\changed}[1]{#1}

\lstdefinelanguage{mypython}{
    keywords={class,def,return,if,elif,else,lambda,and,for,in,while,int},keywordstyle=\color{blue},
    commentstyle=\color{mdbpurple},
    stringstyle=\color{mdborange},
    morekeywords=[2]{Database,Query},
    keywordstyle=[2]\color{mdbgreyblue},
    morekeywords=[3]{register,join,filter,cols,groupby,orderby,flatten},
    keywordstyle=[3]\color{mdbred},
    morecomment=[l]{\#},
    morestring=[b]",
    morestring=[b]'
}
\begin{document}

\title{TorchQL: A Programming Framework for Integrity Constraints in Machine Learning}

\author{Aaditya Naik}
\orcid{0000-0002-3100-0455}
\affiliation{%
  \institution{University of Pennsylvania}
  \city{Philadelphia}
  \country{USA}
}
\email{asnaik@seas.upenn.edu}

\author{Adam Stein}
\orcid{0000-0003-1887-100X}
\affiliation{%
  \institution{University of Pennsylvania}
  \city{Philadelphia}
  \country{USA}
}
\email{steinad@seas.upenn.edu}

\author{Yinjun Wu}
\orcid{0000-0002-9770-5765}
\affiliation{%
  \institution{University of Pennsylvania}
  \city{Philadelphia}
  \country{USA}
}
\email{wuyinjun@seas.upenn.edu}

\author{Mayur Naik}
\orcid{0000-0003-1348-8618}
\affiliation{%
  \institution{University of Pennsylvania}
  \city{Philadelphia}
  \country{USA}
}
\email{mhnaik@seas.upenn.edu}

\author{Eric Wong}
\orcid{0000-0002-8568-6659}
\affiliation{%
  \institution{University of Pennsylvania}
  \city{Philadelphia}
  \country{USA}
}
\email{exwong@cis.upenn.edu}

\begin{abstract}
 Finding errors in machine learning applications requires a thorough exploration of their behavior over data.
  Existing approaches used by practitioners are often ad-hoc and lack the abstractions needed to scale this process.
  We present \framework{}, a programming framework to evaluate and improve the correctness of machine learning applications.
  \framework{} allows users to write queries to specify and check integrity constraints over machine learning models and datasets.
  It seamlessly integrates relational algebra with functional programming to allow for highly expressive queries using only eight intuitive operators.
  We evaluate \tool{} on diverse use-cases including finding critical temporal inconsistencies in objects detected across video frames in autonomous driving, finding data imputation errors in time-series medical records, finding data labeling errors in real-world images, and evaluating biases and constraining outputs of language models.
  Our experiments show that \framework{} enables up to 13x faster query executions than baselines like Pandas and MongoDB, and up to 40\% shorter queries than native Python.
  We also conduct a user study and find that \framework{} is natural enough for developers familiar with Python to specify complex integrity constraints.
\end{abstract}



\maketitle

\section{Introduction}
\label{sec:intro}

Machine learning models can fail in unexpected and harmful ways.
Examples include fatalities caused by self-driving vehicles \cite{wakabayashi2018}, vision models performing worse on people with darker skin \cite{wilson2019predictive}, language models producing text containing offensive stereotypes \cite{abid2021persistent}, and medical diagnosis models degrading in performance when used in new hospitals \cite{zech2018variable}.
Identifying and avoiding such behaviors is crucial to ensuring performance, reliability, and trustworthiness of machine learning applications.

\begin{figure}
    \centering
    \begin{subfigure}[b]{0.58\textwidth}
        \includegraphics[width=3in]{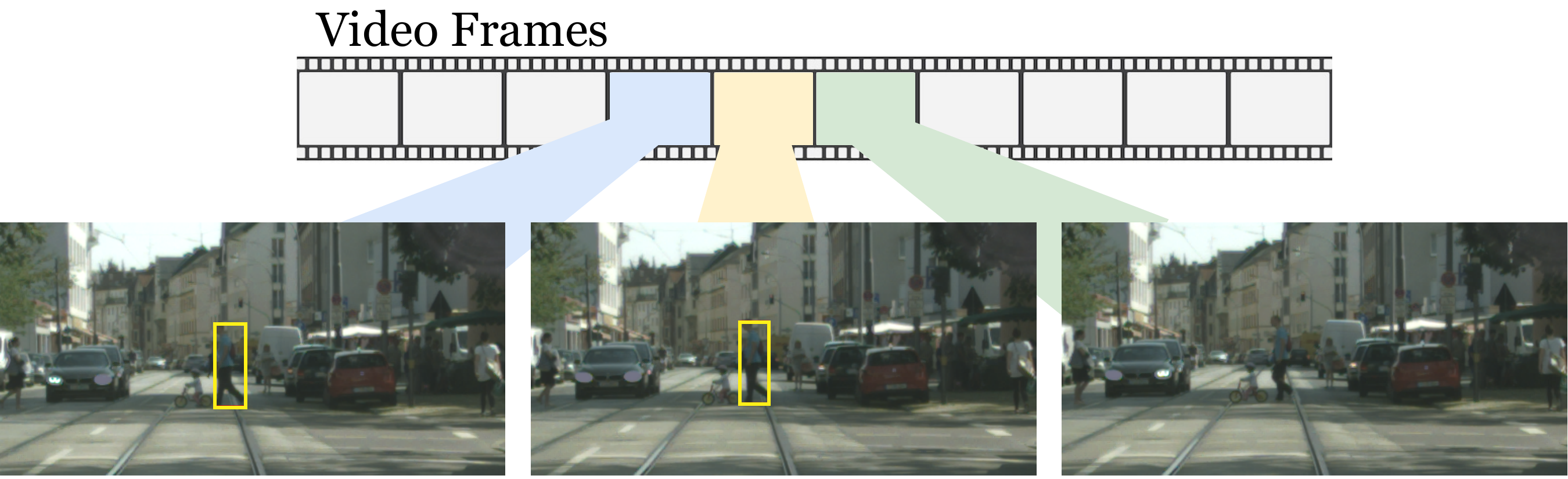}
        \caption{Model fault visualization.}
        \label{fig:f1}
    \end{subfigure}
    \hfill
    \begin{subfigure}[b]{0.4\textwidth}
    \begin{minted}[fontsize=\scriptsize,breaklines]{Python}
Query('temporal_consistency', base='preds')
  .join('preds', key=fid_plus1, fkey=fid)
  .join('preds', key=fid_plus2, fkey=fid)
  .project(match_three_bboxes)
  .project(zip_adj_preds).flatten()
  .filter(lambda sid, frames, lb, bx: (
     lb[0] == lb[1] and lb[2] == "No Match"
     and is_center(bx[0])))
    \end{minted}
        \caption{Integrity constraint as a query.}
        \label{fig:f2}
    \end{subfigure}
    \caption{(a) shows three consecutive frames in a video from the \cityscapes{} self-driving object-detection dataset.
    OneFormer, a state-of-the-art model, detects pedestrians in the center of the first two frames but not in the third. We formalize this fault as a violation of an integrity constraint.
    (b) shows a \tool{} query to find all such violations over a dataset named {\footnotesize \texttt{`preds'}} consisting of video frames along with the model's predictions.}
    \label{fig:self-driving-bug}
    \vspace{-0.1in}
\end{figure}

Finding and characterizing these behaviors is difficult.
As a running example, consider Figure~\ref{fig:f1}, which depicts three consecutive video frames from the \cityscapes{} self-driving dataset \cite{cordts2016cityscapes}. A state-of-the-art vision model, OneFormer \cite{jain2022oneformer}, fails to predict a pedestrian in the third frame. This is a safety-critical issue since predicted objects should not suddenly disappear, especially not pedestrians. This model failure, however, is difficult to find since the error occurs sparsely within the 15K validation samples and it has minimal impact on most numeric evaluation metrics, so even the highest-performing models exhibit such faults.

Similar to the above example, errors often exhibit sparsely in model predictions over vast datasets.
As such, in order to identify and characterize these errors, one must sift through large numbers of correct predictions.
Moreover, simply identifying the errors may not be enough; oftentimes we must identify the underlying cause of these errors.
In general, erroneous model predictions can be cast as violations of certain properties.
For instance, the error in Figure~\ref{fig:f1} violates the property of object permanence across three adjacent frames in a video sequence.

We desire a general mechanism for specifying and checking such properties over data.
\textit{Integrity Constraints}, originally studied in the context of databases \cite{godfrey1998integrity}, constitute such a mechanism.
As such, we can cast the problem of detecting and avoiding these errors as the problem of specifying appropriate integrity constraints and evaluating them over model predictions.
For example, the error in Figure~\ref{fig:f1} can be cast as the violation of the constraint: \textit{for every sequence of three consecutive frames, a person detected in the first two frames should be detected in the third}.

Existing machine learning frameworks lack adequate abstractions to support integrity constraints.
Writing such constraints from scratch in Python typically also requires writing the infrastructure code needed to execute them.
This places a burden on the user who may not know the constraint a priori and may want to test out multiple candidate constraints.
Furthermore, once these constraints are specified, executing them over large scale data may require further optimizing the code, which is another burden on the user.

While frameworks such as Pandas and MongoDB can potentially abstract away some of the required infrastructure, they either have ad-hoc support for Python or come with complex interfaces and strict schema requirements.
Furthermore, they may not naturally support the vast number of possible modalities that machine learning models operate over, from bounding boxes over video frames to unstructured text or audio data.
This is also a challenge for systems such as LMQL \cite{beurer2023prompting} that seek to enforce integrity constraints over model outputs at runtime but are only applicable to the domain of language modeling.

We therefore observe that any effective framework for programming integrity constraints must address the following challenges:
\begin{enumerate}[leftmargin=*,itemsep=2pt]
    \item \textit{Scalability:} it must be able to check integrity constraints in large-scale machine learning settings,
    \item \textit{Interactivity:} it must support interactive testing and inspection of constraint violations, and
    \item \textit{Expressivity:} it must be able to support integrity constraints  specified over a diverse landscape of models, datasets, and use-cases.
\end{enumerate}

In this paper, we present \tool{}, a framework that satisfies the above criteria.
\tool{} enables users to specify integrity constraints as {\em queries}.
For this purpose, it introduces a query language that seamlessly integrates relational algebra with functional programming.
Figure~\ref{fig:f2} shows an example query that specifies the constraint that detects the error in Figure~\ref{fig:f1}.
\tool{} executes the queries over a database representing datasets and model predictions to find violations of the corresponding integrity constraints.
This allows developers to quickly and scalably test potential bugs on large datasets and models. 
Furthermore, the compositionality of \tool{} queries enables the rapid refinement of previous iterations to interactively prototype new potential issues and reduce false alarms among discovered faults.

\tool{} queries are easy to write yet highly expressive, capable of representing complex faults with only eight operators and simple user-defined functions.
For example, the query in Figure \ref{fig:f2} uses only four table operators---\join, \project, \filter and \flatten---along with three user-defined functions to search for violations of the object permanence constraint.
This query finds only 627 frames containing such errors out of 15,000, allowing it to effectively detect errors in large-scale machine learning tasks, while systems like Python require considerable optimizations to do so without timing out.
We discuss this example in more detail in Section \ref{sec:case_study_self_driving}.

We demonstrate how \tool{} enables the use of integrity constraints for detecting and fixing correctness problems in a diverse set of machine learning tasks---object detection, data imputation, image classification, text generation, and natural language reasoning---across domains involving self-driving videos, time-series medical records, and large language models (LLMs). We show the effectiveness of \tool{} on these tasks and domains through five extensive case studies. Moreover, our evaluation shows that \tool{} enables faster query executions (up to 13x best-case speed-ups) than baseline systems such as Pandas and MongoDB, and more concise queries (up to 40\% shorter) than native Python. We also conduct a user study with 10 users to evaluate the usability of \tool{}.
We find that they were able to quickly learn \tool{} to program integrity constraints and find model mispredictions.

We summarize the contributions of the paper:
\begin{enumerate}[leftmargin=*,itemsep=2pt]
\item We propose integrity constraints as a means for discovering and characterizing faults in general machine learning tasks.
\item We develop \framework{}\footnote{The \tool{} system is currently available at \url{https://github.com/TorchQL/torchql/}.}, a framework 
for programming and checking integrity constraints over models and datasets in a manner that is \textit{scalable}, \textit{interactive}, and \textit{expressive}. At its core, \framework{} 
provides abstractions to specify and evaluate integrity constraints as database queries.
\item We perform extensive experiments that demonstrate the efficiency and conciseness of \tool{} for integrity constraint queries in a multitude of use-cases across a variety of domains. 
\item We also complement our experiments with a user study and multiple case studies to validate the usability and expressiveness of \framework{}.
\end{enumerate}

\section{Illustrative Overview}
\label{sec:overview}

In this section, we illustrate how \tool{} is used to
iteratively discover and check integrity constraints to detect model faults, and contrast with programming the same constraints in native Python.
This process is done offline, once the model is trained, but before it is deployed.
In Section~\ref{sec:case_studies}, we discuss other use-cases, such as using the constraints to detect model faults at runtime.

Consider the error in Figure~\ref{fig:self-driving-bug} where a self-driving object detector fails to detect a pedestrian in the third frame of a video sequence.
Since this is a critical bug, a machine learning practitioner would want to find other instances of this error, both within the training data as well as when the model is deployed.
We therefore seek to \textit{characterize} this error as an integrity constraint violation.
Before we can accomplish this, however, we must first discover the integrity constraint itself.

\begin{wrapfigure}{r}{0.4\textwidth}
\begin{mdframed}
\begin{minted}[fontsize=\scriptsize, breaklines]{Python}
preds = ...  # frames and predictions
db = Database()
db.register(preds, 'preds')
q = Query(...)  # creating a query
result = q(db)  # running the query
\end{minted}
\end{mdframed}
\caption{Initializing a \tool{} database and running queries over it.}
\label{fig:db_init}
\vspace{-0.1in}
\end{wrapfigure}

We first initialize a \tool{} database \texttt{db} as shown in Figure~\ref{fig:db_init} and populate it with a table named \texttt{`preds'} containing predictions of OneFormer over 15,000 frames.
Any query \texttt{q} we write can then be evaluated over this table by invoking it over \texttt{db}.

Since we do not know the constraint beforehand, we start with a simple initial constraint: frames in which at least one pedestrian is predicted.
This can be written as a simple \framework{} query as shown in Figure~\ref{fig:r1-tool}.
Here, we specify the query name (\texttt{people\_count}) and the table over which the query operates (\texttt{preds}).
Each prediction consists of the bounding boxes and their corresponding labels for each frame.
We supply a \textit{user-defined function} to the \texttt{filter} operator, which executes it over each frame when the query is run, and keep the predictions that satisfy the condition.

We also show the same constraint implemented in native Python in Figure~\ref{fig:r1-python}.
The lack of query abstractions in Python means that we need to write a \texttt{for} loop, an \texttt{if} condition, and manually add satisfying predictions to the data structure (\texttt{people\_count}).
As such, the lack of query abstractions often necessitates one to implement both \textit{what} the constraint is, as well as \textit{how} to evaluate it.

\begin{figure}[t]
    \centering
    \begin{subfigure}[t]{0.49\textwidth}
        \begin{mdframed}
        \begin{minted}[fontsize=\scriptsize,breaklines]{Python}
Query('people_count', base='preds')
  .filter(lambda p:
    count_people(p['labels']) != 0)

        \end{minted}
        \end{mdframed}
        \caption{\tool{}}
        \label{fig:r1-tool}
    \end{subfigure}
    \hfill
    \begin{subfigure}[t]{0.49\textwidth}
        \begin{mdframed}
        \begin{minted}[fontsize=\scriptsize,breaklines]{Python}
people_count = []
for p in preds:
  if count_people(p['labels']) != 0:
    people_count.append(p)
        \end{minted}
        \end{mdframed}
        \caption{Native Python}
        \label{fig:r1-python}
    \end{subfigure}
\vspace{-0.1in}
    \caption{Queries for finding individual frames in which at least one pedestrian is predicted.}
    \label{fig:r1}
\vspace{-0.05in}
\end{figure}
\begin{figure}[t]
    \centering
    \begin{subfigure}[t]{0.49\textwidth}
        \begin{mdframed}
        \begin{minted}[fontsize=\scriptsize,breaklines]{Python}
Query('changing_people', base='preds')  
  .join('preds', key=fid_plus_1, fkey=fid)
  .filter(lambda f1, f2:
    count_people(f1['labels']) !=
    count_people(f2['labels']))


        \end{minted}
        \end{mdframed}
        \caption{\tool{}}
        \label{fig:r2-tool}
    \end{subfigure}
    \hfill
    \begin{subfigure}[t]{0.49\textwidth}
        \begin{mdframed}
        \begin{minted}[fontsize=\scriptsize,breaklines]{Python}
changing_people = []
for p in preds:
  for q in preds:
    if fid_plus_1(p) == fid(q):
      if count_people(p['labels']) !=
         count_people(q['labels']):
        changing_people.append([p, q])
        \end{minted}
        \end{mdframed}
        \caption{Native Python}
        \label{fig:r2-python}
    \end{subfigure}
    \vspace{-0.1in}
    \caption{Queries for finding pairs of consecutive frames with different predicted pedestrian counts.}
    \label{fig:r2}
    \vspace{-0.05in}
\end{figure}
\begin{figure}[t]
    \centering
    \begin{subfigure}[t]{0.49\textwidth}
        \begin{mdframed}
        \begin{minted}[fontsize=\scriptsize,breaklines]{Python}
Query('temporal_consistency', base='preds')
  .join('preds', key=fid_plus_1, fkey=fid)
  .join('preds', key=fid_plus_2, fkey=fid) 
  .project(match_three_bboxes)
  .project(zip_adj_preds)
  .flatten()
  .filter(lambda sid, frames, lb, bx:
    lb[0] == lb[1] and
    lb[2] == "No Match" and
    is_center(bx[0]))




        \end{minted}
        \end{mdframed}
    \caption{\tool{}}
    \label{fig:r3-tool}
    \end{subfigure}
    \hfill
    \begin{subfigure}[t]{0.49\textwidth}
        \begin{mdframed}
        \begin{minted}[fontsize=\scriptsize,breaklines]{Python}
temporal_consistency = []
for p in preds:
  for q in preds:
    for r in preds:
      if fid_plus_1(p) == fid(q) and
         fid_plus_2(p) == fid(r):
        bboxes = match_three_bboxes(p, q, r)
        bboxes = zip_adj_preds(*bboxes)
        for box in bboxes:
          for (sid, frames, lb, bx) in box:
            if lb[0] == lb[1] and
               lb[2] == "No Match" and
               is_center(bx[0]):
              temporal_consistency.append(box)
        \end{minted}
        \end{mdframed}
        \caption{Native Python}
        \label{fig:r3-python}
    \end{subfigure}
    \vspace{-0.1in}
    \caption{Queries for finding sequences of three consecutive frames where a pedestrian is detected in the center of the first two frames but not in the third.}
    \label{fig:enter-label}
    \vspace{-0.1in}
\end{figure}

Evaluating this constraint gives us 4,591 frames out of the total 15,000 frames, many of them false positives.
This means that the constraint is too broad and must be refined further.
A subsequent hypothesis may look for two consecutive predictions which differ in the number of people detected.
In such a case, it is possible that some people predicted in the first frame are missed in the second.

We can construct the query in Figure~\ref{fig:r2-tool} to represent this hypothesis by chaining more table operators to the original query.
First, we use a \join operator to join the table of pedestrians with the same table shifted forward in time by one frame to get a new table containing pairs of consecutive predictions.
Here too, we use user-defined functions to specify the key and foreign key of the join.
This gives us a table of consecutive predictions.
Observe that \tool{} allows its queries to directly run on the objects being queried themselves.
This allows us to seamlessly integrate relational algebra with lambda functions, as queries can contain and execute arbitrary Python functions (e.g., \texttt{fid} and \texttt{fid\_plus\_1}) over these objects without needing to define an explicit schema.

We can then apply another \filter operator to select only pairs of predictions with a differing number of pedestrians.
This now reduces the number of filtered frames to 3,638 from the 4,591 that were filtered out in the initial query.
As was the case with the previous iteration, however, the Python implementation (Figure~\ref{fig:r2-python}) requires writing multiple looping and conditional statements to mimic the \join and \filter operators.

Our latest query produced a table of consecutive predictions with a differing number of people.
However, such predictions also include people that did not go missing, such as those exiting or entering the second frame.
We therefore need to identify the exact persons that have eluded the object detector.
This requires converting our latest table of consecutive predictions into a table of consecutive persons.
We also wish to do so over three frames instead of two to avoid even more false positives, and restrict our constraint to predictions in the center of the frame to avoid issues with people leaving or entering the frame.

The \tool{} query identifying all such sequences of three consecutive frames is shown in Figure~\ref{fig:r3-tool}.
Here, the sequences are such that the pedestrian is detected in the center of the first frame, detected again in the second frame, but not in the third.

Our first step is to join an additional `preds' table to get a set of three consecutive predictions instead of two.
We again use the \join{} operator, similar to how we used it in the previous query.
Although we have a set of three consecutive predictions, each prediction contains an unordered list of detected objects.
As a result, it is not immediately obvious which bounding boxes refer to the same person.
We therefore use a well-developed tool from object detection for aligning the bounding boxes between frames \citep{bolya2020tide}.
We can apply this complex function, denoted as \texttt{match\_three\_bboxes} to all of our prediction sequences using the \cols table operator, which applies a function to every row in a table to create a new table.

We further use a combination of other user-defined functions, table operators, and a final filter to extract consecutive objects where the object is detected as a person in the first two frames but not in the third.
This final constraint refines our search of critical errors from 3638 to 627 frames out of 15,000.
In contrast, the Python implementation shown in Figure~\ref{fig:r3-python} requires the additional join to be explicitly set up as a nested loop, and more loops to mimic the \flatten operator.
Moreover, the Python implementation now fails to scale to the 15,000 frames, timing out after 10 minutes, while the \tool{} query executes in around 43 seconds.
While the Python implementation can be optimized, doing so adds 24 more lines of code, imposing additional user burden.

We thus show that throughout this process, \tool{} allowed us to quickly and easily execute and iterate over prototype queries on not only the original dataset but also on intermediate tables produced while trying to program the constraint to detect temporal inconsistencies. On the other hand, iterating over constraints in Python is accompanied by setting up the infrastructure to scalably execute the constraint, \changed{an issue prevalent in machine learning codebases such as the ones mentioned in Appendix~\ref{sec:prevalence}}. Our final query involves the composition of 4 unique table operators, each of which is parameterized by a custom lambda function to handle various data types. This process is representative of how the scalability, interactivity, and expressivity of \tool{} allows us to program integrity constraints to discover errors in machine learning applications.

\section{The \framework{} Framework}

This section presents the \tool{} framework.
We first describe the underlying data model that queries operate over and then present the syntax and semantics of queries.

\subsection{Data Model}

While integrity constraints can be programmed using existing systems like Pandas or native Python, as we illustrated in Section~\ref{sec:overview}, they are ill-equipped for our needs.
If arbitrary Python objects are needed for a debugging task, Python is preferable to Pandas, since the latter does not support querying over arbitrary objects.
Conversely, if the objects adhere to a rigid structure or can be easily flattened into a table of primitive datatypes, it is more efficient to take advantage of Pandas querying abstractions.
Other candidates include NoSQL database systems like MongoDB and its in-memory counterpart, ArangoDB.
However, they are designed to store and query over key-value pairs and do not support user-defined functions (UDFs) that allow users to leverage external modules and express arbitrarily complex queries.
We summarize these tradeoffs in Table~\ref{tab:comparison}.

\begin{table}
    \centering
    \small
    \caption{Comparison of querying systems.}
    \begin{tabular}{llccc}\toprule
    Querying Systems & Data Representation & \makecell{Flexible \\ Schema} & \makecell{User-Defined \\ Functions} & \makecell{Querying \\ Abstractions}\\
    \midrule
    Pandas & {\footnotesize {\tt numpy.array}} tables & \xmark & \cmark & \cmark \\
    MongoDB \cite{banker2016mongodb} & key-value stores & \cmark  & \xmark &  \cmark \\
    \textbf{\framework{}} & object collections & \cmark  & \cmark & \cmark \\
    \makecell{Python, OMG \cite{kang2018model}} & arbitrary objects & \cmark  & \cmark & \xmark \\
    \bottomrule
    \end{tabular}
    \label{tab:comparison}
\end{table}

In order to be usable, \tool{} must support queries over arbitrary Python objects without requiring much data wrangling.
\tool{} therefore uses an object-relational representation to represent the data being queried.
It inherits Python's data model and considers an \textit{object} ($o$) to be a fundamental abstraction of data.
As such, each object may either be a Python primitive or instantiation of a Python class, a list or set of other objects, or a collection of key-value pairs.

These objects may be further organized into lists with other similar objects, referred to as \textit{tables}~($T$).
Objects in the same table are oftentimes related with each other, but need not conform to any schema, like frames from a video, or prompts for a large language model.
Tables can be further assigned names and organized into a collection called a \textit{database} ($D$).
We show the semantic domains for objects, tables, and databases used by \tool{} in Figure~\ref{fig:domains}.

An object-relational representation has several advantages over other common representations.
First, while it requires objects being queried to be contained within a collection, it has no restrictions on the structure of the objects being queried.
This allows for the flexibility to query over models and datasets without the associated overhead of converting the data into a compatible representation.

Moreover, storing these objects within collections allows for enough structure within each table to provide succinct yet powerful querying abstractions over them akin to their relational algebra counterparts.
Operations such as joining tables or grouping rows turn into single-line specifications without the need for setting up low-level infrastructure that would otherwise be required.

An object-relational representation also allows for the support of executing UDFs over the objects within each table.
\tool{} supports this by allowing users to supplement each relational algebra abstraction with one or more UDFs if needed.
This also allows the user to leverage external libraries, machine learning models, and other tools to write their queries within the \tool{} framework.

\subsection{The Query Language}\label{sec: ql}

We now describe \tool{}'s query language.
We first present its syntax, shown in Figure~\ref{fig:syntax}, followed by the operational semantics, shown in Figure~\ref{fig:semantics}.

\subsubsection{Syntax}
\label{sec:syntax}
A \tool{} \textit{program} comprises a sequence of statements each of which defines a named base table or a named query.
Each query is a chain of table operators that define a sequence of transformations to apply to one or more tables in the database.
\tool{} provides eight unique table operators.
These operators, with the exception of \flatten and \unique, are empowered by user-defined functions (UDFs).
\changed{UDFs are arbitrary Python functions that enable powerful transformations over each object of a given table, or in the case of \reduce{}, over the entire table. Note that since UDFs are arbitrary functions, they may result in potentially non-terminating queries depending on their definitions.}
We discuss these operators in more detail in Section~\ref{sec:operator_semantics}, after describing the program semantics.

\begin{figure}
  \centering
  \footnotesize
    \begin{subfigure}[t]{0.4\textwidth}
\begin{tabular}{r@{\ \ \ }r@{\ \ }c@{\ \ }l}
\text{(object)} & $o$ & $\production$ & $p$         \\
                &     &\gor& $[o_1, \ldots, o_n]$   \\
                &     &\gor& $\{o_1, \ldots, o_n\}$ \\
                &     &\gor& $\{p_1: o_1, \ldots, p_n: o_n\}$ \\
\text{(table)}      & $T$ & $\production$ & $[o_1, \ldots, o_n ]$ \\
    \text{(database)}   & $D$ & $\production$ &  $[n_1 \rightarrow T_1, \ldots, n_k \rightarrow T_k]$
\end{tabular}
\caption{Semantic Domains.}
\label{fig:domains}
\end{subfigure}
\begin{subfigure}[t]{0.59\textwidth}
\begin{minipage}{0.25\textwidth}
\begin{tabular}{r@{\ \ \ }r@{\ \ }c@{\ \ }l}
    \text{(name)}       & $n$ \\
    \text{(function)}   & $f$ \\
    \text{(query)} & $Q$ & $\production$ &  $n$ \ \ \ $|$\ \ \ $Q.a$ \\
    \text{(statement)}      & $S$ & $\production$ &
    \texttt{register}($n$, $T$)     \\
    & &\gor&  $n$ $\leftarrow$ $Q$  \\
    \text{(program)}      & $P$ & $\production$ & $\epsilon$\ \ \ $|$\ \ \ $S$; $P$
\end{tabular}
\end{minipage}
\hspace{0.45in}
\begin{minipage}{0.32\textwidth}
\begin{tabular}{r@{\ \ \ }r@{\ \ }c@{\ \ }l}
\text{(operator)}   & $a$ & $\production$ &  \join($n$, $f_1$, $f_2$) \\
                    &     &  $|$ & \filter($f$) $|$ \flatten() \\
                    &     &  $|$ & \cols($f$) \\
                    &     &  $|$          & \orderby($f$) \\
                    &     &  $|$          & \groupby($f$)  \\
                    &     &  $|$          & \unique() \ $|$\ \reduce($f$)
\end{tabular}
\end{minipage}
\caption{Abstract syntax.}
\label{fig:syntax}
\end{subfigure}

    \vspace{-0.05in}
    \caption{Core language of \framework{}.}
    \label{fig:grammar}
    \vspace{-0.1in}
\end{figure}

\begin{figure}
\footnotesize
\textbf{Program semantics}
\hfill
\begingroup\setlength{\jot}{5pt}
\begin{gather*}
\infr[Program\_E]
    {}
    {\step{\ctx{\epsilon}{D}}{D}}
\qquad
\infr[Query\_Op]
    {\step{\ctx{Q}{D}}{T} \iand \step{\ctx{a}{D, T}}{T'}}
    {\step{\ctx{Q.a}{D}}{T'}}
\qquad
\infr[Query\_n]
    {}
    {\step{\ctx{n}{D}}{D[n]}}
\\
\infr[Program\_Reg]
    {\step{\ctx{P}{D[n\mapsto T]}}{D'}}
    {\step{\ctx{\texttt{register}(n, T); P}{D}}{D'}}
\qquad
\infr[Program\_Query]
    {\step{\ctx{Q}{D}}{T} \iand \step{\ctx{P}{D[n\mapsto T]}}{D'}}
    {\step{\ctx{n \leftarrow Q; P}{D}}{D'}}
\end{gather*}
\endgroup

\vspace{5pt}
\textbf{Operator semantics}
\hfill
\footnotesize
\boxed{
\begin{array}{rclrclrcl}
    f &:& \mathbb{O} \rightarrow \mathbb{O}, & s &:& \mathbb{O} \rightarrow \textsc{Bool},  & g &:& \mathbb{T} \rightarrow \mathbb{T},\\
    \sigma_s &:& \mathbb{T} \rightarrow \mathbb{T}, & \cdot\bowtie_{f_\textsc{k}, f_\textsc{fk}}\cdot &:& \mathbb{T} \rightarrow \mathbb{T}, & \cdot\ \texttt{++}\ \cdot &:&  \mathbb{T} \rightarrow \mathbb{T}
\end{array}
}
\vspace{5pt}
\begingroup\setlength{\jot}{5pt}
\begin{gather*}
\infr[Join]
    {}
    {\step{\ctx{\join(n, f_\textsc{k}, f_{\textsc{fk}})}{D, T}}{T\bowtie_{f_\textsc{k}, f_\textsc{fk}} D[n]}}
\\
\infr[Project1]
    {\step{\ctx{\project(f)}{D, T}}{T'}}
    {\step{\ctx{\project(f)}{D, t:T}}{f(t) : T'}}
\qquad
\infr[Project2]
    {}
    {\step{\ctx{\project(f)}{D, []}}{[]}}
\\
\infr[Reduce]
    {}
    {\step{\ctx{\reduce(g)}{D, T}}{g(T)}}
\qquad
\infr[Filter]
    {}
    {\step{\ctx{\filter(s)}{D, T}}{\sigma_s T}}
\\
\infr[Order1]
    {\begin{array}{c}
    \step{\ctx{\filter(\lambda x. f(x)\leq f(t))}{D, T}}{T_{\leq x}}\\
    \step{\ctx{\filter(\lambda x. f(x) > f(t))}{D, T}}{T_{> x}}\\
    \step{\ctx{\orderby(f)}{D, T_{\leq x}}}{L}\\
    \step{\ctx{\orderby(f)}{D, T_{> x}}}{R}
    \end{array}
    }
    {\step{\ctx{\orderby(f)}{D, t:T}}{L \texttt{++} [t] \texttt{++} R}}
\qquad
\infr[Group1]
    {
    \begin{array}{c}
    \step{\ctx{\filter(\lambda x. f(x) = f(t))}{D, T}}{T_{=}}\\
    \step{\ctx{\filter(\lambda x. f(x)\neq f(t))}{D, T}}{T_{\neq}}\\
    \step{\ctx{\groupby(f)}{{D, T_{\neq}}}}{T'}
    \end{array}
    }
    {\step{\ctx{\groupby(f)}{D, t:T}}{(f(t), t:T_{=}):T'}}
\\
\infr[Order2    ]
    {}
    {\step{\ctx{\orderby(f)}{D, []}}{[]}}
\qquad
\infr[Group2]
    {}
    {\step{\ctx{\groupby(f)}{D, []}}{[]}}
\\
\infr[Unique1]
    {\step{\ctx{\filter(\lambda x. x \neq t)}{D, T}}{T_{\neq}} & \step{\ctx{\unique()}{T_{\neq}, D}}{T'}}
    {\step{\ctx{\unique()}{D, t:T}}{t:T'}}
\qquad
\infr[Unique2]
    {}
    {\step{\ctx{\unique()}{D, []}}{[]}}
\\
\infr[Flatten1]
    {\step{\ctx{\flatten()}{D, T}}{T'}}
    {\step{\ctx{\flatten()}{D, t:T}}{[t] \texttt{++} T'}}
\qquad
\infr[Flatten2]
    {}
    {\step{\ctx{\flatten()}{D, []}}{[]}}
\end{gather*}
\endgroup
\vspace{-0.15in}
\caption{Operational semantics of \framework{}.}
\vspace{-0.1in}
\label{fig:semantics}
\end{figure}

\subsubsection{Semantics}
\label{sec:program_semantics}
A \tool{} program is interpreted as a sequence of database transformations induced by its constituent statements.
Statement \texttt{register}$(n, T)$ adds base table $T$ with the name $n$ to the database whereas statement $n\leftarrow Q$ runs query $Q$ over tables in the database to result in a new table which is added to the database with the name $n$. In particular, $Q$ is comprised of the name $n$ of an existing table and a chain of zero or more table operators $a_1.a_2\ldots{}a_k$.
When executing~$Q$ over database $D$, table $T_0 = D[n]$ is first retrieved and then
the table operations specified in $Q$ sequentially transform the table. Starting from table $T_0$, operation $a_i$ transforms table $T_{i-1}$ into $T_i$ to eventually produce the output table $T_k$.

These semantics are vital for \tool{} to be effective at interactively programming integrity constraints.
Iterating over previous hypotheses requires the results of executed queries to be stored and usable across programming sessions.
Moreover, since the \tool{} language allows a single query to be arbitrarily complex, these semantics allow for simpler queries by decomposing complex operations into multiple queries.
Queries can also act as preprocessors, allowing for features to be extracted from unstructured data before using those features to program integrity constraints.
Sequentially executing and storing results of previous statements allows for these use cases.

\subsubsection{Operators}
\label{sec:operator_semantics}
We now describe the operator semantics of \tool{}.
The eight table operators are largely drawn from relational algebra and natively support complex operations over tables.
In general, each operator takes as input a table, and produces a table as output.
With the exception of \flatten{} and \unique{}, users can supplement these operators with UDFs.
However, the operators differ in how these supplied UDFs are executed over the objects within each table.

\vspace{0.05in}
(1) {\em Join.}
This operator composes objects from two tables into a single table.
The composition is achieved through the supplied UDFs $f_\textsc{K}: \mathbb{O} \rightarrow \mathbb{O}$ and $f_\textsc{FK}: \mathbb{O} \rightarrow \mathbb{O}$, as shown in \textsc{Join} in Figure~\ref{fig:semantics}:
\vspace{-0.04in}
$$
T_i \bowtie_{f_\textsc{k}, f_\textsc{fk}} T_j = [ [o_i, o_j] | o_i \in T_i, o_j \in T_j, f_\textsc{k}(o_i) = f_\textsc{fk}(o_j) ]
$$
Here, $\mathbb{O}$ denotes the domain of all objects.
Since the key-foreign key pairs for the tables are defined by the results of UDFs, one can join tables based on values that can be results of arbitrarily complex operations not existing within the tables.
An example of this is in the \join of the query in Figure~\ref{fig:self-driving-bug}, where the function \texttt{fid\_plus\_1} returns the frame ID of the next frame, while \texttt{fid} returns the frame ID of the current frame.
Using these functions allows one to join the table containing all the frames with itself to produce a table containing pairs of consecutive frames.

This join is a form of equijoin, and can thus be implemented as a \textit{hash join}, where we use a hash table to join the objects from table $T_i$ to those of $T_j$.
Using a hash join algorithm allows us to achieve this with a complexity of $O(m+n)$, where $m = |T_i|$ and $n = |T_j|$, as opposed to typical join algorithms that perform with complexity $O(mn)$.

\vspace{0.05in}
(2) {\em Project.}
This operator transforms each object in a table according to the supplied UDF $f: \mathbb{O} \rightarrow \mathbb{O}$ as described in \textsc{Project1} and \textsc{Project2} in Figure~\ref{fig:semantics}.
This allows us to perform powerful and flexible transformations on the rows of the table, such as including image transformations using image processing libraries, or analyzing the inferences made by external models as shown in the query in Figure~\ref{fig:cllm-impl} where the \cols{} function is used to generate prompts for the LLMs.
    
\vspace{0.05in}
(3) {\em Filter.}
This operator uses the supplied UDF $s: \mathbb{O} \rightarrow \textsc{Bool}$ to filter out rows from a table.
This is depicted in \textsc{Filter} in Figure~\ref{fig:semantics} using the selection operator $\sigma$ from relational algebra.

\vspace{0.05in}
(4) {\em Order By.}
This operator reorders the objects of the table $T$ according to the supplied UDF $f: \mathbb{O} \rightarrow \mathbb{O}$ as shown in \textsc{Order1} and \textsc{Order2} in Figure~\ref{fig:semantics}.
Again, the UDF allows the reordering of objects without the need to explicitly populate the table with the values to order the objects by.

\vspace{0.05in}
(5) {\em Group By.}
This operator allows grouping objects of a table into subtables, where each subtable contains the objects that produce the same value when passed to the UDF $f: \mathbb{O} \rightarrow \mathbb{O}$.
This operator is unique in the sense that its result contains nested tables that can be further manipulated using table operators.
This allows us to perform powerful group-by operations on the table, such as grouping by the hour of the day or the sequence ID as shown in the query in Figure~\ref{fig:wilds_q_1}.

\vspace{0.05in}
(6) {\em Flatten.}
This operator flattens each object $o_j \in T$ where $o_j$ is a collection  $[ o_{j1}, o_{j2}, o_{j3}, \ldots ]$ to produce table $T'$ such that each element $o_{jm} \in T'$.
If $o_j$ is not a collection, then it is left unchanged.

\vspace{0.05in}
(7) {\em Unique.}
This operator returns a table $T'$ where duplicate rows from $T$ are removed.

\vspace{0.05in}
(8) {\em Reduce.}
This operator is different from the previously discussed operators in that it runs the supplied UDF $g: \mathbb{T} \rightarrow \mathbb{T}$ over the \textit{entire table} rather than over individual objects within that table.
Here, $\mathbb{T}$ is the domain of tables.
This not only allows for general aggregation functions like count, length, min, max, and others, but also for recursive functions over tables to be run as a part of a \tool{} query without having to switch to native Python.

The use of UDFs within these operators enables a versatile querying system.
First, since the UDFs can be specified directly over objects, users do not need to wrangle the data to fit a schema in order to efficiently query a model's predictions.
Second, unlike in other querying systems, extracting features necessary for building these queries now becomes integrated into the query writing process, rather than being a separate component that practitioners independently refine.
Finally, since UDFs can be arbitrary Python code, \tool{} is Turing complete. %

\section{Implementation}
\label{sec:implementation}

\changed{\tool{} has been implemented in Python which is the primary language used for machine learning pipelines.}
The underlying database system is designed to be in-memory.
This design choice eliminates the overhead of storing and retrieving objects from disk while executing queries, thereby reducing the turnaround time for each query.
As such, all tables being queried over are permanently loaded in the main memory for as long as the programming session lasts.
While the data for the tables may be stored to disk and retrieved, they need to be loaded into memory before writing queries over them.

When writing queries, the tables that they access must exist in the database before they can be executed.
Therefore, when a database is newly initialized, as is the case when starting a programming session for a new task, it must be populated with these tables using the {\tt register} function.
This is typically used for loading in the training, validation, or test datasets, after which model predictions can be obtained over them via \tool{} queries.

\changed{The \tool{} database engine compiles each TorchQL query, comprising a pipeline of operators, into an executable sequence of table operations. The engine applies relevant optimizations while compiling the query. \tool{} relies on hash data structures to perform most of these optimizations, though batch-driven optimizations are also discussed later. One example where hashing allows for \tool{} to optimize queries is in the \join{} operation. Since \tool{} joins are analogous to SQL equijoins, they are implemented as hash joins, where the results of the key and foreign key functions are hashed. This results in a complexity of $O(M+N)$, where $M$ and $N$ are the number of rows in the tables being joined, as opposed to $O(MN)$ in the naive implementation. Hashing similarly allows TorchQL to optimize the grouping and unique operations.}

In order to seamlessly integrate the data-loading process, as well as the interaction between \tool{} queries and machine learning models, we design \tool{} to support conventional machine learning frameworks, specifically PyTorch.
To do so, \tool{} tables inherit the base PyTorch \texttt{Dataset} class.
In other words, \tool{} tables are instantiations of PyTorch \texttt{Datasets}.
This has a few advantages while working with PyTorch machine learning pipelines.

\begin{wrapfigure}{r}{0.36\textwidth}
\begin{mdframed}
\begin{minted}[fontsize=\scriptsize,breaklines]{Python}
train_data = datasets.MNIST(
  root = 'data', train = True,
  transform = ToTensor(),
  download = True,
)
db = Database("mnist")
db.register(train_data, "train")
\end{minted}
\end{mdframed}
\vspace{-0.05in}
\caption{Loading a PyTorch dataset.}
\label{fig:register}
\vspace{-0.05in}
\end{wrapfigure}
For one, users can directly register PyTorch datasets into \tool{} databases without needing to cast them as \tool{} tables explicitly, allowing them to directly query their data.
For instance, we can download the training data for the MNIST dataset using PyTorch's API and directly load it into a \tool{} database as shown in Figure~\ref{fig:register}.

This also allows \tool{}'s database engine \changed{to perform batch-driven optimizations by leveraging PyTorch's Dataloader to efficiently iterate over batches of objects, as well as PyTorch's support for batch processing and vectorized tensor operations.
This results in optimizing the execution of queries whose UDFs work with batched tensors.}

\section{Evaluation Setup}
\label{sec:benchmarks}

We now discuss the setup for evaluating the effectiveness of \tool{} as an integrity constraint programming framework.
As discussed in Section~\ref{sec:intro}, we require \tool{} to be scalable, interactive, and expressive.
We therefore evaluate it by answering the following research questions:
\begin{enumerate}[itemsep=2pt]
    \item[\textbf{RQ1.}] \textbf{Expressivity:} Can \tool{} be used to program integrity constraints in diverse settings?
    \item[\textbf{RQ2.}] \textbf{Performance:} Are \tool{} queries concise and efficient on large-scale data?
    \item[\textbf{RQ3.}] \textbf{Usability:} Is \tool{} intuitive and easy to use for users unfamiliar with the system?
\end{enumerate}

For \textbf{RQ1}, we write queries for five machine learning tasks over domains such as self-driving videos, time-series healthcare data, trap-camera images, and natural language text.
Out of 182 queries that were written, we present and analyze 11 queries in-depth across five case studies in Section~\ref{sec:case_studies}.
The tasks are summarized in Table~\ref{tab: summary} and the chosen queries are described in Table~\ref{tab:tasks}.

For \textbf{RQ2}, we implement seven of these queries in three baseline systems Python, Pandas, and ArangoDB to compare their conciseness and efficiency over their corresponding datasets and models.
We present our findings in Section~\ref{sec: quantitative}.
For \textbf{RQ3}, we conduct a user study with 10 participants writing three queries of increasing complexity in \tool{}.
We present the results in Section~\ref{sec:user_study}.

In the rest of this section, we describe each machine learning task from Table~\ref{tab:tasks} including the chosen datasets and models, the correctness problems of interest, and the integrity constraints and queries to address them.
We then proceed to the sections that answer each research question.

\begin{table*}
    \centering
    \footnotesize
    \caption{Summary of the datasets and models used in the experiments.}
    \begin{tabular}{
    cccccc
    }
    \toprule
  {\bf Task} &  Dataset & Dataset Size & Model & Application Domain
  \\ \midrule
  {\bf \makecell{Object \\ Detection}} & \makecell{\cityscapes{} validation set \\ \citep{cordts2016cityscapes}} & 15K & \makecell{OneFormer \\ \citep{jain2022oneformer}} & Self-Driving Videos
  \\ \midrule
  {\bf \makecell{Data \\ Imputation}} & \makecell{\physionet\ \\ \cite{silva2012predicting}} & 4000 & \makecell{\saits\ \\ \cite{du2023saits}} & Time-Series Healthcare Data
  \\ \midrule
  {\bf \makecell{Image \\ Classification}} &\makecell{\wilds{} training set \\ \cite{beery2020iwildcam} } & 121K & \makecell{ResNet50 \\ \cite{he2016deep}} & Trap-Camera Images
  \\ \midrule
  {\bf \makecell{Text \\ Generation}} & \makecell{Alpaca \\ \cite{alpaca}} &52K &\makecell{T5 \\ \cite{raffel2020exploring}} & Natural Language \\ \midrule
  {\bf \makecell{Natural Language \\ Reasoning}} & \makecell{GSM8K\\ \citep{cobbe2021training}\\Date Understanding\\ \citep{bigbench}} & \makecell{1319\\ \\369\\ \phantom{L}} & \makecell{Mistral 7B\\ \citep{mistral}} & Natural Language
  \\ \bottomrule
    \end{tabular}
    \label{tab: summary}
\end{table*}

\newcommand{\descwidth}{6.22cm}
\newcommand{\descvspace}{1.2pt}
\begin{table*}
    \centering
    \footnotesize
    \caption{Queries written for evaluating \tool{}. Queries with an asterisk are only used in case studies.
    }
    \begin{tabular}{
    @{}c@{\ \ \ }c c c@{\ \ \ }c@{\ \ \ }c@{\ \ \ }c@{\ \ \ }c@{\ \ \ }c@{\ \ \ }c@{}
    }
    \toprule
  \textbf{Task} & {\bf Query} & {\bf Query Description} & \multicolumn{7}{c}{\bf Query Operators} \\[5pt]
  & & & \rotatebox[origin=c]{80}{\textit{join}} & \rotatebox[origin=c]{80}{\textit{project}} & \rotatebox[origin=c]{80}{\textit{filter}} & \rotatebox[origin=c]{80}{\textit{group}} & \rotatebox[origin=c]{80}{\textit{order}} & \rotatebox[origin=c]{80}{\textit{flatten}} & \rotatebox[origin=c]{80}{\textit{UDFs}}
  \\ \midrule
  \multirow{2}{*}{\parbox{0.18\linewidth}{\vspace{0.15in}\bf \makecell{Object \\ Detection}}} &
  S-Q1 &\parbox{\descwidth}{\vspace{\descvspace} Retrieve all sequences of three continuous frames from the Cityscapes dataset in which an object appearing in the first two frames, and the center of the first, is not detected in the third frame by the OneFormer model. \vspace{\descvspace}} & 2 & 2 & 1 & 0 & 0 & 1 & 3
  \\ \cmidrule{2-10}
    & S-Q2 &\parbox{\descwidth}{\vspace{\descvspace} Retrieve all vehicles predicted in three consecutive frames of the Cityscapes dataset by the OneFormer model and compute their speed across the frames. \vspace{\descvspace}} & 1 & 1 & 1 & 0 & 0 & 0 & 2
  \\ \midrule
  \multirow{3}{*}{\parbox{0.18\linewidth}{\vspace{0.54in} \bf \makecell{Data \\ Imputation}}} & T-Q1 &\parbox{\descwidth}{\vspace{\descvspace} Compute the difference between all non-missing pairs of temporally consecutive feature values and compute the 99th percentile difference by following \cite{naik2023machine}, for each feature. \vspace{\descvspace}} & 1 & 2 & 0 & 0 & 0 & 0 & 2
  \\ \cmidrule{2-10}
  & T-Q2 &\parbox{\descwidth}{\vspace{\descvspace} Compute the interquartile range of non-missing ground-truth values across time for each feature and use it to detect outliers in the \saits{} model's imputations. \vspace{\descvspace}} & 1 & 2 & 0 & 0 & 0 & 1 & 1
  \\ \cmidrule{2-10}
  & T-Q3 &\parbox{\descwidth}{\vspace{\descvspace} Compute the interquartile range of all values (including imputed ones) across time for each feature and use it to detect outliers in the \saits{} model's imputations. \vspace{\descvspace}} & 0 & 1 & 0 & 0 & 0 & 0 & 1
  \\ \midrule
  \multirow{3}{*}{\parbox{0.18\linewidth}{\bf \makecell{Image \\ Classification}}} & I-Q1 &\parbox{\descwidth}{\vspace{\descvspace} Retrieve trap camera video sequences containing frames with more than one unique ground-truth animal. \vspace{\descvspace}} & 0 & 1 & 1 & 1 & 1 & 0 & 0
  \\ \cmidrule{2-10}
  & I-Q2* & \parbox{\descwidth}{\vspace{\descvspace} Find empty trap camera frames where a ResNet model predicts an animal. \vspace{\descvspace}} & 0 & 1 & 0 & 1 & 0 & 1 & 1 \\ \midrule
  \multirow{2}{*}{\parbox{0.18\linewidth}{\vspace{0.2in}\bf \makecell{Text \\ Generation}}} & B-Q1 &\parbox{\descwidth}{\vspace{\descvspace} Retrieve adjectives used to describe the nouns `farmer' and `engineer' in the T5 \Alpaca{} dataset. \vspace{\descvspace}} & 0 & 2 & 0 & 1 & 0 & 0 & 2 \\
  \cmidrule{2-10}
  & B-Q2* & \parbox{\descwidth}{\vspace{\descvspace} From sets of adjectives biased towards farmers and engineers, find the adjectives which GPT-3.5 consistently chooses to describe farmers or engineers.} & 1 & 5 & 0 & 0 & 0 & 2 & 3 \\ \midrule
  \multirow{2}{*}{\parbox{0.18\linewidth}{\vspace{0.2in} \bf \makecell{Natural Language \\ Reasoning}}} & L-Q1* &\parbox{\descwidth}{\vspace{\descvspace} Use the Mistral-7B LLM to answer and provide chain-of-thought reasoning for the GSM8K arithmetic reasoning questions, guaranteeing that the answer is an integer.} & 0 & 6 & 1 & 1 & 0 & 1 & 2 \\
  \cmidrule{2-10}
  & L-Q2* &\parbox{\descwidth}{\vspace{\descvspace} Use the Mistral-7B LLM to answer and provide chain-of-thought reasoning for the Date understanding dataset, guaranteeing that the answer is a valid date format. \vspace{\descvspace}} & 0 & 6 & 1 & 1 & 0 & 1 & 2 \\
  \bottomrule
    \end{tabular}
    \label{tab:tasks}
    \vspace{-0.15in}
\end{table*}

\subsection{Object Detection}

\paragraph{Overview and Setup.}
In this task, the goal is to predict bounding boxes and their labels for individual frames in self-driving videos.
We consider the validation set of the \cityscapes{} dataset, which contains 15,000 video frames,
and use the OneFormer~\citep{jain2022oneformer} model for predicting bounding boxes and their labels for each video frame.

\vspace{-0.05in}
\paragraph{Correctness Problems.}
Object detection models, in general, are evaluated using metrics like mean average precision (mAP).
This metric considers a prediction correct if it has an Intersection over Union (IoU) ratio of at least 0.5.
It then aggregates the precision of all predictions over all the frames.
However, in cases where critical errors are rare and sparsely distributed over the predictions, this may cause the accurate predictions to overshadow the severe errors.

\vspace{-0.05in}
\paragraph{Integrity Constraints.}
Given this issue with the mAP metric, we seek to discover critical model faults by programming integrity constraints like the one shown in Figure~\ref{fig:f2}.
We wrote a total of 57 queries to investigate various potential integrity constraints and analyze two such constraints in Section~\ref{sec:case_study_self_driving}.
The first, denoted S-Q1 in Table~\ref{tab: summary}, discovers all objects from all the 15K frames that occur in the center of one frame, somewhere in the next consecutive frame, and are not detected in the third consecutive frame.
A similar query was also investigated by \citet{kang2018model}.
The second, S-Q2, aims to find vehicles with extremely high speeds across three frames. These objects with outlier speeds are likely to indicate incorrect bounding box predictions since most objects (including moving cars) have limited movement between frames.

\subsection{Data Imputation}

\begin{figure}
    \centering
    \includegraphics[width=0.8\textwidth]{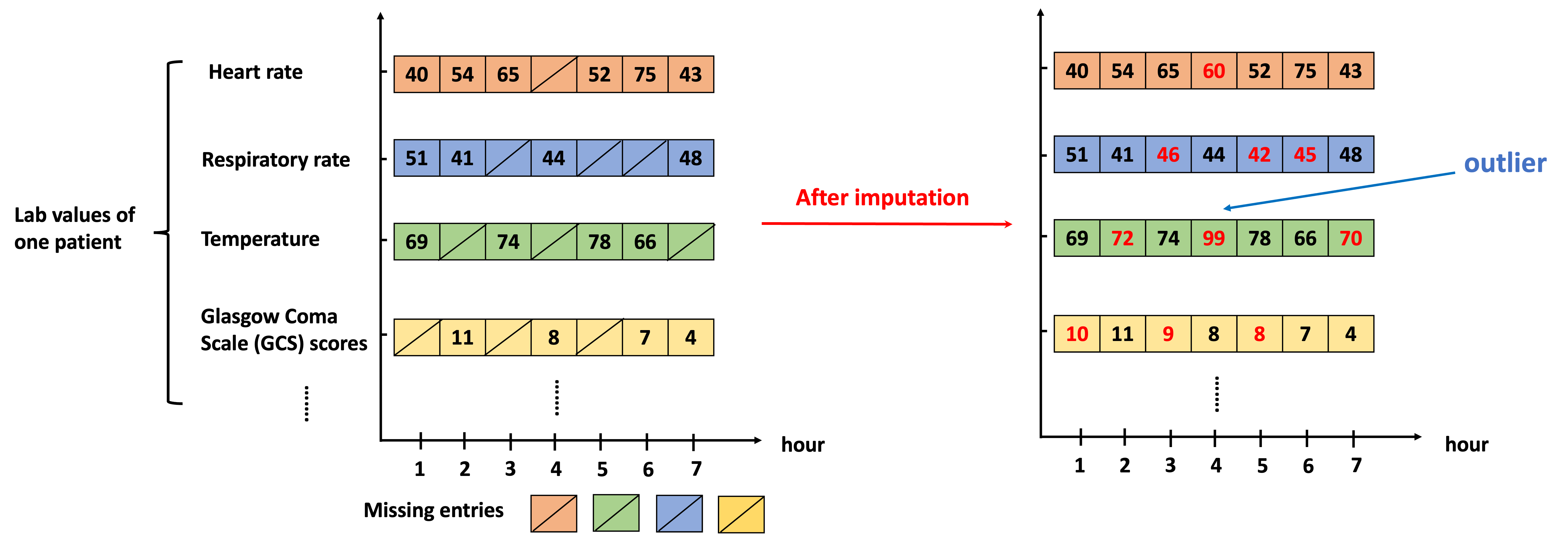}
    \vspace{-0.05in}
    \caption{A medical time series sample consisting of lab values of one patient over time. 
    We show the original data with missing values on the left and the values imputed by \saits{} on the right in red.
    The imputed value ``99'' for ``Temperature'' is an outlier since it deviates too far from other non-missing data for this feature.
    }
    \label{fig: time_series_example}
    \vspace{-0.1in}
\end{figure}

\paragraph{Overview and Setup.}
In this task, the goal is to predict missing values within time-series healthcare data.
We consider the \physionet{}'s~\cite{silva2012predicting} medical time-series dataset.
Each sample in this dataset is composed of multiple univariate time series, each of which includes records of one feature across multiple time stamps, as illustrated in Figure \ref{fig: time_series_example}.
The dataset contains 4000 time-series samples and each sample consists of 35 lab values (features), collected hourly, within a 48-hour time window.
Overall, up to 80\% of the values in the samples are missing.
We use the state-of-the-art model \saits~\cite{du2023saits} to impute these missing values.

\vspace{-0.05in}
\paragraph{Correctness Problems.}
Missing data in healthcare settings is prevalent due to irregular patient visits or clinical errors \cite{lee2017big}.
Imputing this data is necessary to train models for downstream use-cases.
As a result, it is important that the imputed data is as accurate as possible.
As discussed in \cite{you2018time}, medical data such as the sample in Figure \ref{fig: time_series_example} should respect domain knowledge, such as the Glasgow Coma Scale's range of 1 to 15, as well as common sense, such as the assumption of smooth variations in univariate time series data (henceforth called the {\em smoothness assumption}).
We therefore aim to find imputed values that do not respect these assumptions.
For instance, Figure~\ref{fig: time_series_example} shows an imputed temperature value (99) at the 4th hour, which significantly exceeds neighboring non-missing temperature values and causes an erratic spike in the temperature values.

\vspace{-0.05in}
\paragraph{Integrity Constraints.}
We wrote a total of around 30 queries to find different violations of this smoothness assumption.
We investigate three of them T-Q1, T-Q2, and T-Q3 in Table \ref{tab:tasks}, which describe two variations of the smoothness assumption.

The first such variant is that the difference between two continuous entries along the temporal dimension must be smooth and insignificant.
One can use domain knowledge to determine a threshold to filter out those consecutive values with large differences, such as the outlier highlighted in Figure \ref{fig: time_series_example}.
In the absence of such domain knowledge, however, we formulate T-Q1 to collect all pairs of non-missing consecutive values from the entire dataset, and compute their differences.
We then use the 99th percentile of all these gaps as the estimated threshold, in accordance with the method outlined in \cite{naik2023machine}. 

The other variant of the smoothness constraint that we specify captures the closeness between each imputed value and other entries within a time window. Specifically, we craft T-Q2 and T-Q3 to determine whether one imputed entry is an outlier or not 
with respect to other entries within a univariate time series.
We discuss T-Q2 and T-Q3 in more detail in Section \ref{sec:case_study_time_series}.

\subsection{Image Classification}

\paragraph{Overview and Setup.}
In this task, the goal is to classify the animal present in individual frames of videos captured by trap cameras.
We evaluate the \wilds{} dataset \citep{beery2020iwildcam} and the predictions of a ResNet model \citep{he2016deep} trained on the \wilds{} training dataset.

\vspace{-0.05in}
\paragraph{Correctness Problems.}
The \wilds{} dataset poses several challenges to machine learning models.
We present two such problems that we investigate in more detail.
First, models must have the ability to classify a frame as empty when no animal is present.
As such, it is necessary for empty frames to be correctly labeled.
However, there are several instances in \wilds{} where empty frames are mislabeled as containing an animal.
Second, the model tends to be inconsistent in its predictions.
Over a sequence of frames, there is typically a single animal that moves around.
However, the model sometimes predicts different animals in different frames within the same sequence.

\vspace{-0.05in}
\paragraph{Integrity Constraints.}
We wrote around 25 queries to capture various integrity constraints over the model predictions including whether nocturnal animals were predicted during the daytime, whether multiple animals were predicted across the same sequence of frames, and finding correlations between mispredicted classes and the ground truths.
We analyze two of these queries.

The first, query I-Q1, aims to find sequences where the predictions by the model vary across frames, but the ground truth remains the same.
The second, I-Q2, aims to find empty frames that are mislabeled to contain animals.
Since the ground truth cannot be relied on, we compute the pixel differences between pairs of consecutive frames to determine frames that have no movement and thus are likely empty.
We investigate I-Q2 in more detail in Section \ref{sec:case_study_wilds}.

\subsection{Text Generation} 

\paragraph{Overview and Setup.}
In this task, the goal is to generate text (as a sequence of tokens) given a prompt.
In particular, we consider the Alpaca instruction fine-tuning dataset \cite{alpaca} (\Alpaca\ for short),
which consists of 52K prompts and their responses from LLMs.
In this experiment, we focus on the model responses from the GPT-3.5 (\verb|gpt-3.5-turbo|) and T5\footnote{\url{https://huggingface.co/lmsys/fastchat-t5-3b-v1.0}} models.

\vspace{-0.05in}
\paragraph{Correctness Problems.}
Large language models (LLMs) are known to exhibit biases that can be potentially harmful \citep{havaldar2023multilingual, liang2021towards, zhao2017men}.
This can occur when the data they are trained over contain biases that the LLM may pick up on.
However, typical metrics for language generation, such as the Bleu score, do not have the ability to check for them.

\vspace{-0.05in}
\paragraph{Integrity Constraints.}
We investigate how integrity constraints can be programmed using \tool{} to detect and study biases in LLMs.
We wrote a total of around 50 queries to investigate various biases, and analyze two such queries in detail.

First, we write a query B-Q1 to discover the biases in the prompts of the \Alpaca{} dataset by identifying the adjectives that are highly correlated with the occupations of farmer and engineer.
This covers the class of adjective-profession biases \cite{nangia2020crows, smith2022m, li2020unqovering, kurita2019measuring}, where certain adjectives are biased towards certain professions (e.g. brilliant scientists).
Further details and examples of this bias are provided in Section~\ref{sec: case_study_bias}.
Second, we write a query B-Q2 to quantify bias in the model response.
Here, we take adjectives associated with farmers and adjectives associated with engineers to construct prompts to determine how frequently an LLM uses the farmer adjective to describe farmers and the engineer adjective to describe engineers.
More details are included in Section \ref{sec: case_study_bias}.

\subsection{Natural Language Reasoning}

\paragraph{Overview and Setup.}
In this task, the goal is to answer reasoning questions in natural language text.
We consider two reasoning tasks: {\em date understanding}, which involves reasoning over dates and relative durations to determine the described date, and {\em arithmetic reasoning}, which involves solving word problems using basic arithmetic to compute desired quantities.
For date understanding, we use the BigBench benchmark's date understanding task (Date) \citep{bigbench} and for arithmetic reasoning we use GSM8K \citep{cobbe2021training}.
We use the top performing 7B parameter pretrained model from the Open LLM leaderboard which at the time of writing was Mistral-7B (\texttt{mistralai/Mistral-7B-v0.1}) \citep{mistral}. The prompt used for both tasks is the same prompt used by \citet{fcot}.
To decode the output of the LLM, we use a temperature of 0.4 along with the default parameters to the Huggingface generation API.

\vspace{-0.05in}
\paragraph{Correctness Problems.}
One of the major issues with LLMs is that the output generated is not guaranteed to conform to the conditions of the ground truths.
For example, LLMs should generate valid dates in response to prompts in the date understanding task, and numbers for the arithmetic reasoning task.
There have been elaborate prompting mechanisms devised to improve the reliability of LLMs such as chain-of-thought \citep{wei2022chain} and ReACT \citep{yao2022react} and systems for programming these prompting mechanisms \citep{langchain}.

\vspace{-0.05in}
\paragraph{Integrity Constraints.}
\tool{} is useful for orchestrating language model prompting due to its highly expressive user-defined functions.
Placing language model inference calls inside user-defined functions allows for queries which prompt language models and process the output.
We demonstrate the usefulness of this interface with a query to constrain the output of a language model similar to LMQL \citep{beurer2023prompting}.
We then evaluate our prompting technique and show that \tool{} can be used to constrain LLM output while maintaining accuracy as well as LMQL on two benchmark reasoning tasks.

\begin{figure}
    \centering
    \begin{subfigure}[b]{0.4\textwidth}
        \includegraphics[width=\textwidth]{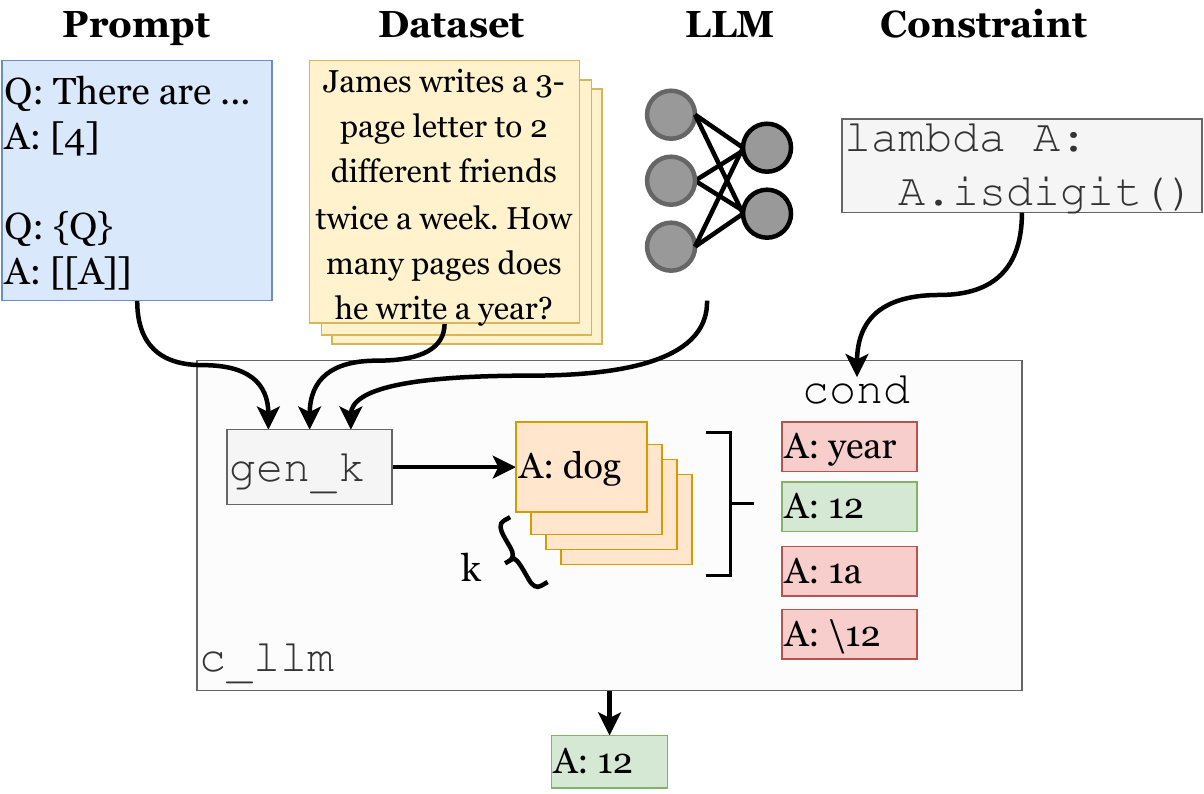}
        \caption{Constrained generation overview.}
        \label{fig:cllm-overview}
    \end{subfigure}
    \hfill
    \begin{subfigure}[b]{0.58\textwidth}
\begin{mdframed}
\begin{minted}[fontsize=\scriptsize, breaklines]{Python}
def gen_k(n, dataset, prompt, k=10):
    return Query(n, base=dataset)
        .project(parse_and_prepare_prompt)
        .project(lambda prompts, togen: (llm(prompts, k), togen), bs=10)
        .project(parse_outputs)
        .flatten()
def c_llm(n, dataset, prompt, cond, ans, k=10):
    return gen_k(n, dataset, prompt, ans, k)
        .filter(lambda prompt, gen: cond(gen))
        .group_by(lambda prompt, gen: prompt)
        .project(lambda prompt, gens: ans(gens))
\end{minted}
\end{mdframed}
    \caption{Constrained generation with \tool{}.}
    \label{fig:cllm-impl}
    \end{subfigure}
    \vspace{-0.05in}
    \caption{Using \tool{} to orchestrate prompting of language models with output constraints.}
    \label{fig:cllm}
\end{figure}

In order to achieve this, we build a helper query, shown in Figure~\ref{fig:cllm-impl}, whose operation is shown in Figure~\ref{fig:cllm-overview}.
This query samples multiple responses from an LLM and filters out the responses which violate the given constraints.
The query uses four \tool{} operations and many of the UDFs are single-lined lambda functions defined within the query specification, except for the implementation of the prompting semantics which include some use of regular expressions.

We use this helper query to write around 20 queries to enforce various constraints over the Mistral-7B LLM.
We analyze two of them in this paper.
First, we write the query L-Q1, which performs Chain-of-Thought (CoT) \citep{wei2022chain} prompting with constraints for the arithmetic reasoning task.
The second query L-Q2 similarly attempts to constrain the LLM for the date understanding task.
Both queries are explored in more detail in Section~\ref{sec:case_study_lmql}.

\section{Case Studies}
\label{sec:case_studies}

In this section, we evaluate the expressivity of \tool{} through a series of case studies for each of the above tasks. In each case study, we discuss the integrity constraints and queries from Table \ref{tab:tasks} and provide a detailed analysis of the issues discovered with those queries.

\subsection{Case Study: Prediction Error Patterns in Self-Driving Object Detectors}
\label{sec:case_study_self_driving}

\paragraph{Object Consistency Constraint} 
\begin{figure}
    \centering
    \begin{subfigure}[b]{0.32\textwidth}
        \includegraphics[width=\linewidth]{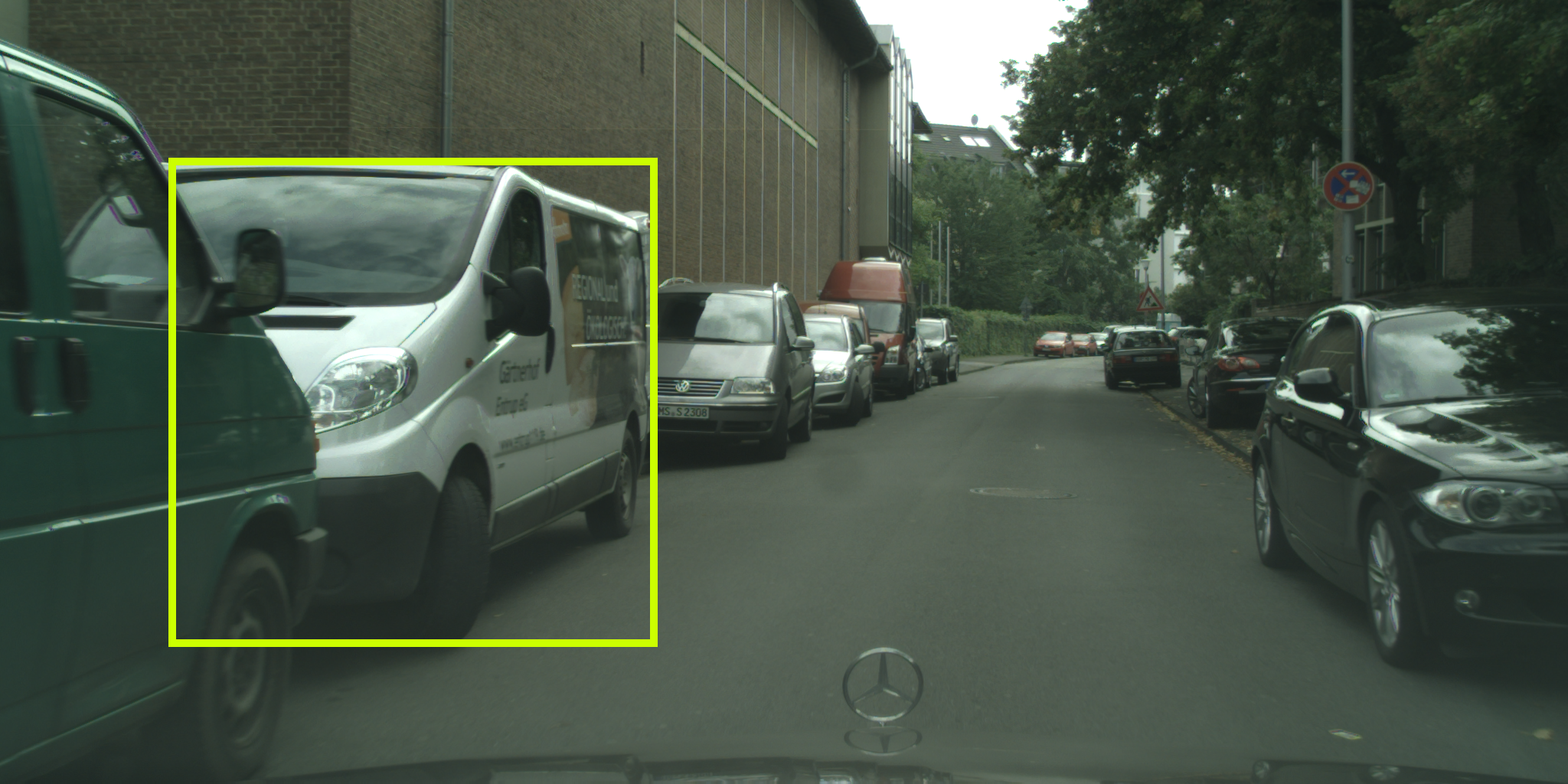}
        \caption{}
        \label{fig:consistency_f1}
    \end{subfigure}
    \hfill
    \begin{subfigure}[b]{0.32\textwidth}
        \includegraphics[width=\linewidth]{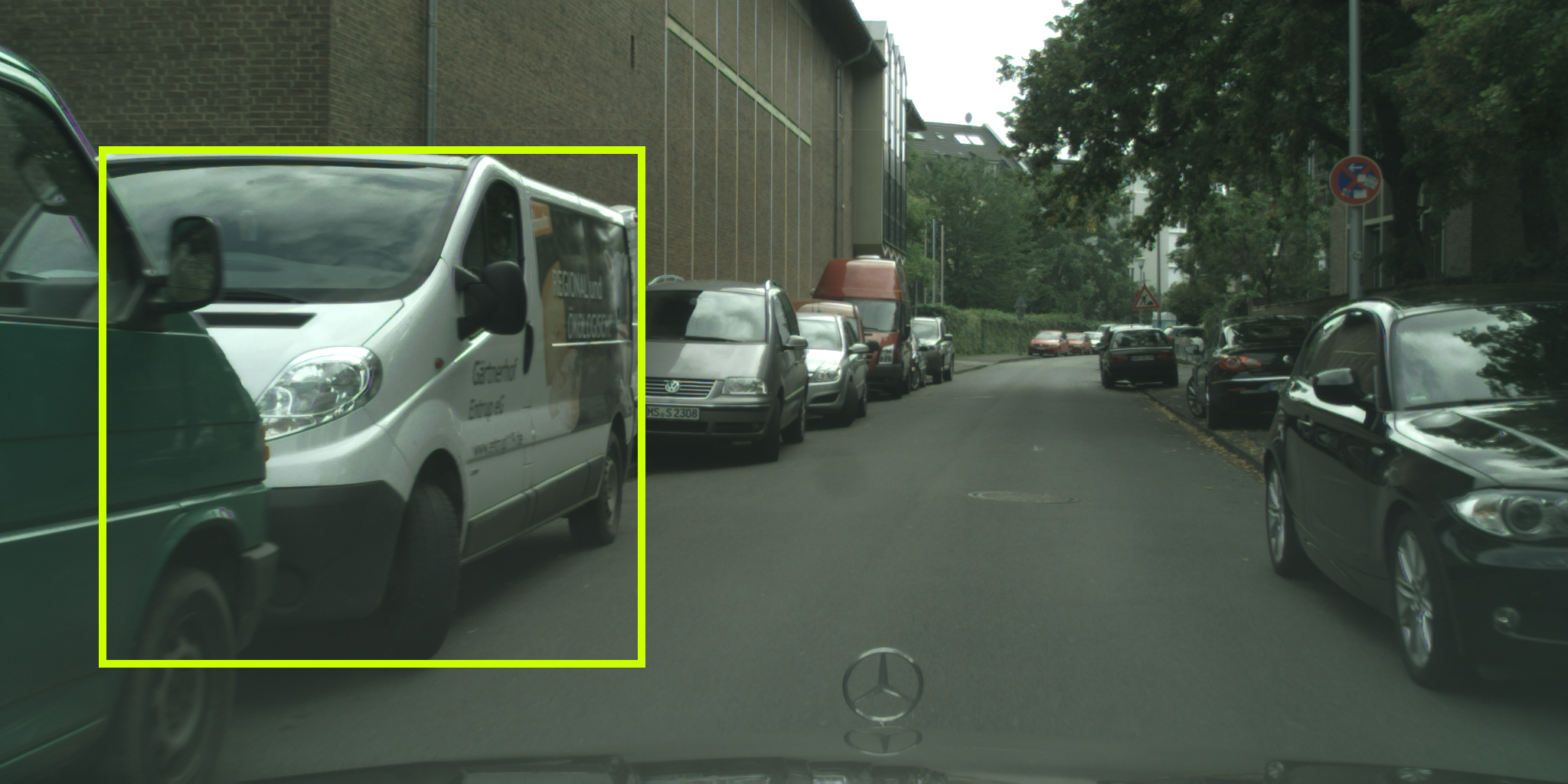}
        \caption{}
        \label{fig:consistency_f2}
    \end{subfigure}
    \hfill
    \begin{subfigure}[b]{0.32\textwidth}
        \includegraphics[width=\linewidth]{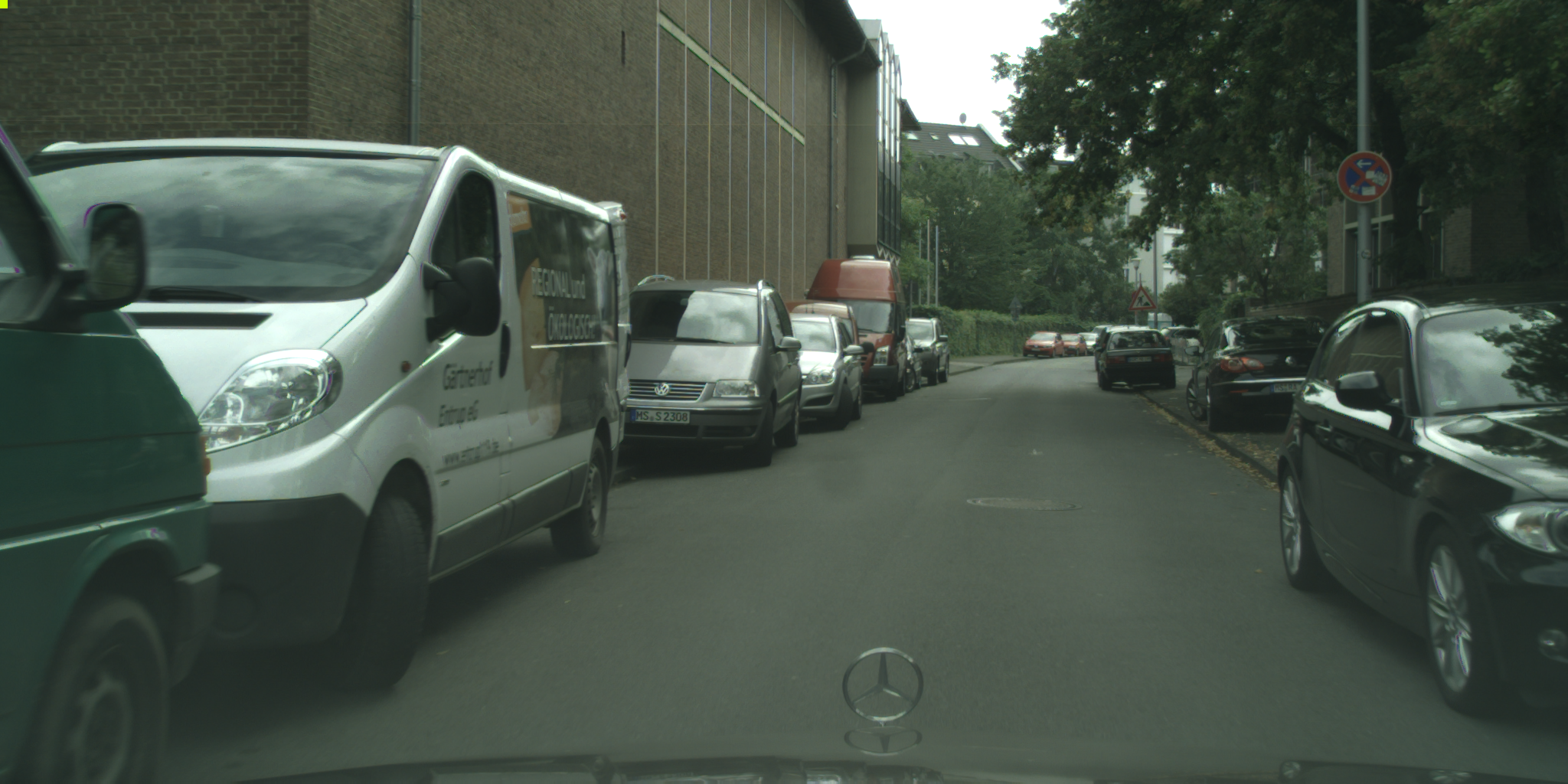}
        \caption{}
        \label{fig:consistency_f3}
    \end{subfigure}
    \vspace{-10pt}
    \caption{A violation of the object consistency constraint. Figures \ref{fig:consistency_f1} to \ref{fig:consistency_f3} are three continuous frames from the \cityscapes{} dataset. A large truck is detected in the first two frames but not in the third.}\label{fig: object_consistency}
\end{figure}
Section~\ref{sec:overview} describes a version of the object consistency constraint (S-Q1) specialized over pedestrians.
S-Q1 finds 6,264 objects that are detected in two consecutive frames and disappear in the third.
To evaluate the effectiveness of this query, we look at how well it finds model mispredictions by focusing on the 500 frames with ground truth bounding box labels provided by the Cityscapes dataset.
Of the tuples of three frames (from S-Q1), where an object disappears in the third frame, there are 113 frames with labeled ground truth bounding boxes. In 71 out of these 113 frames, the query successfully detects an object that was missed by the model. Thus, this query has 62.83\% precision. In total, there are 1388 mispredicted objects, so the query has a recall of 5.12\%.

Overall, S-Q1 has good precision but low recall since the violations of just this integrity constraint alone cannot capture a large percentage of all the model failures. On the other hand, this constraint does capture real safety-critical issues.
For example, as Figure \ref{fig: object_consistency} shows, as the camera vehicle (i.e., the vehicle from which the video is recorded) gradually approaches the white van parked on the left, the object detector suddenly fails to detect it (see Frame 3) even though it is predicted in the previous two frames. Considering the short distance between the two vehicles, such behavior by an object detector deployed in an autonomous vehicle could potentially result in an accident.

\paragraph{High-velocity constraint} 
\begin{figure}
    \centering
    \begin{subfigure}[b]{0.3\textwidth}
        \includegraphics[width=\linewidth]{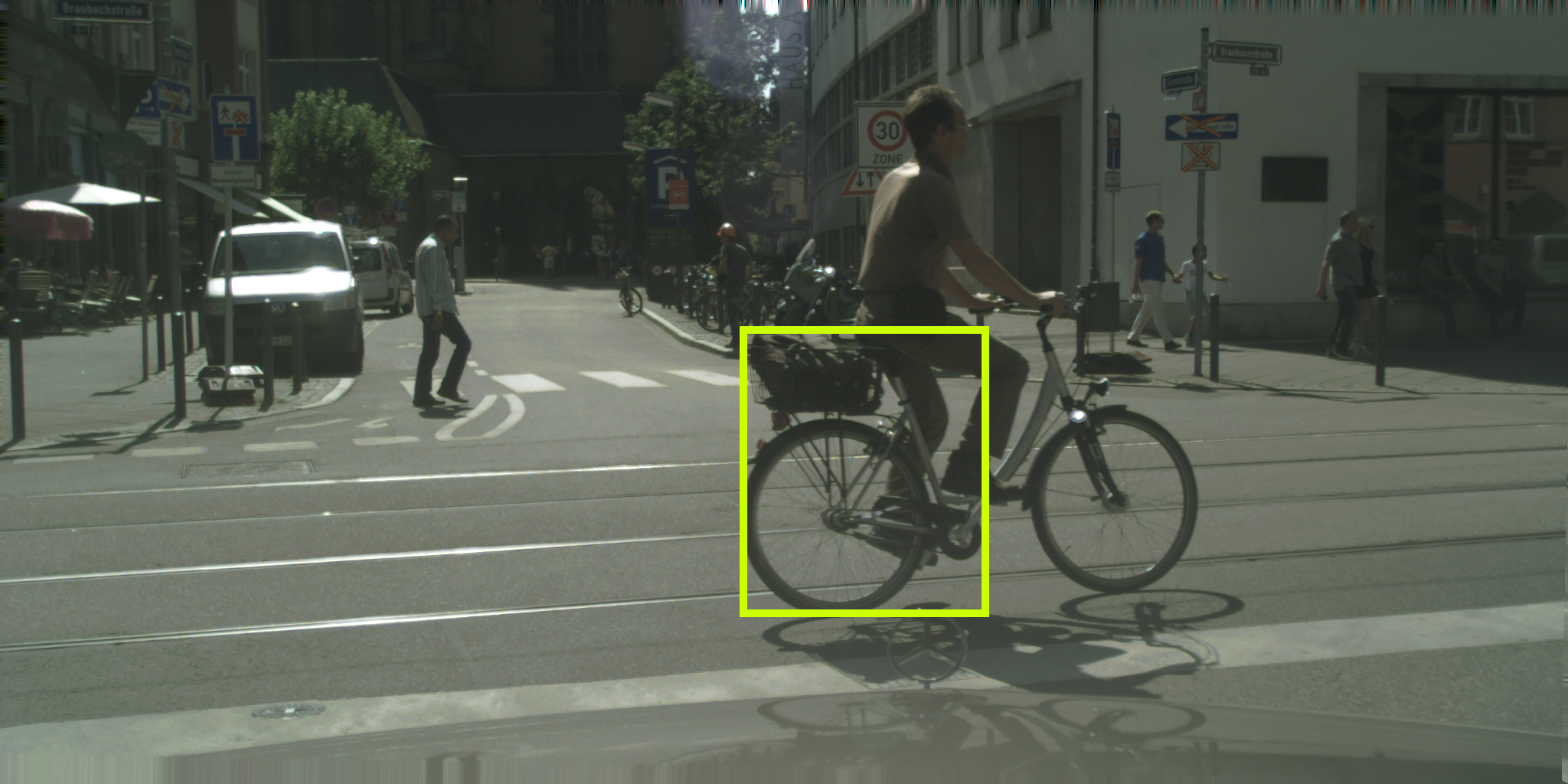}
        \caption{}
        \label{fig:high_speed_f1}
    \end{subfigure}
    \begin{subfigure}[b]{0.3\textwidth}
        \includegraphics[width=\linewidth]{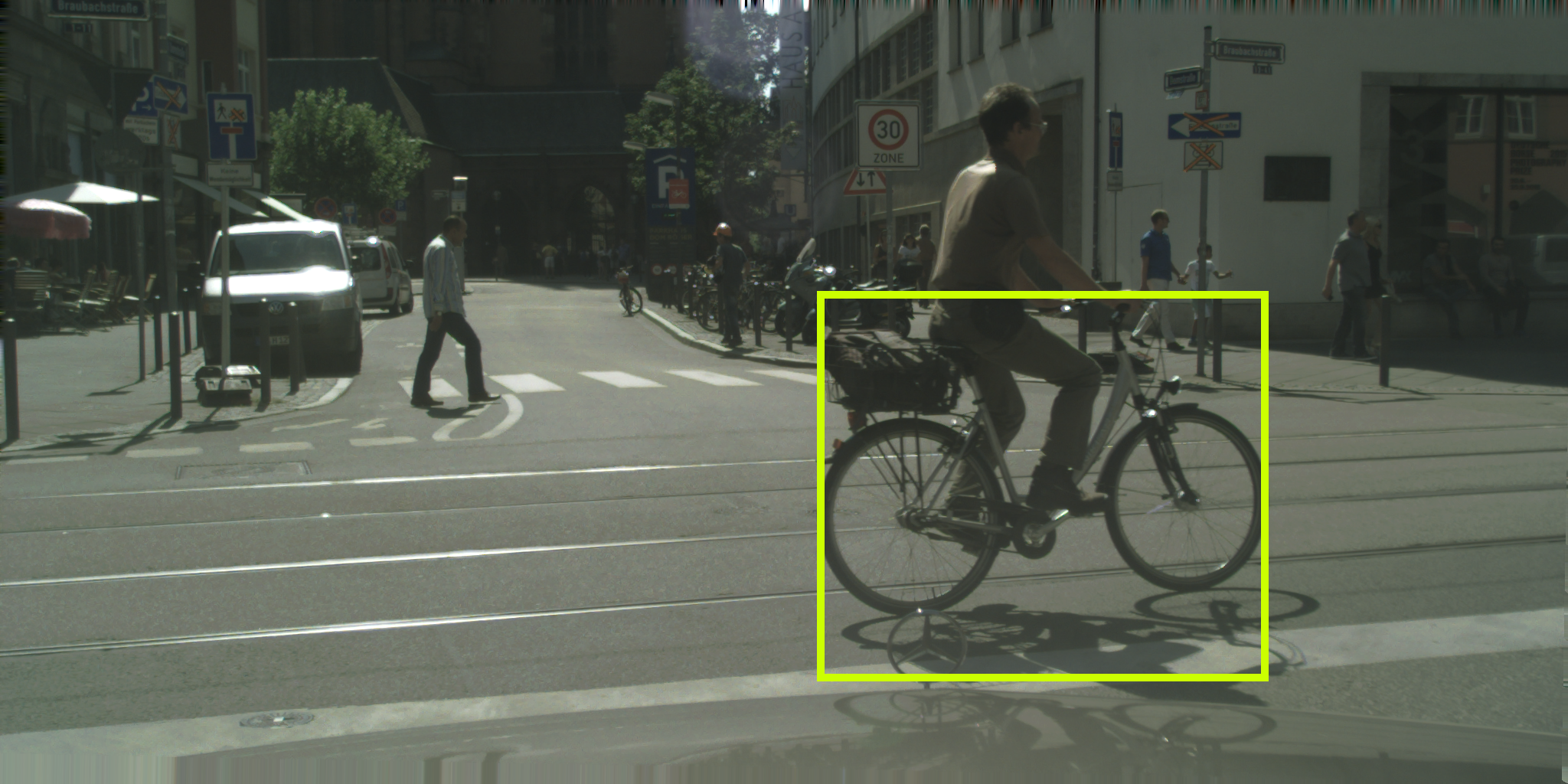}
        \caption{}
        \label{fig:high_speed_f2}
    \end{subfigure}
    \vspace{-10pt}
    \caption{A violation of the high-velocity constraint where Figures \ref{fig:high_speed_f1} and \ref{fig:high_speed_f2} are two continuous frames from the \cityscapes{} dataset. The predicted bounding boxes in these two frames are inconsistent in their positions and sizes, thus exaggerating the speed of the object.
    }\label{fig:high-velocity example}
\end{figure}
Building on top of S-Q1, we construct S-Q2 for computing the speed of objects detected in three consecutive frames and filtering them.
The speed of an object is calculated by estimating the distance the object moves across consecutive frames. 
If an object has an unusually high speed, it may indicate an inaccurate object detector prediction. For example, the bicycle shown in Figure \ref{fig:high-velocity example} is the object with the highest speed computed by S-Q2. This is primarily caused by abrupt variations in the bounding box size in the two consecutive frames. As Figure \ref{fig:high-velocity example} shows, the bounding box in frame 1 only covers a fraction of the bicycle.
On the other hand, the bounding box in frame 2 correctly identifies the entirety of the bicycle.
This results in the bicycle appearing to move an abnormally large distance within the span of two frames.
The following shows the full \tool{} query to represent this integrity constraint.
\begin{center}
\begin{minipage}{0.8\textwidth}
\begin{mdframed}
\begin{minted}[fontsize=\scriptsize]{Python}
Query('fast_vehicles', base = 'three_adj_matches')
    .filter(get_vehicles)
    .project(get_velocities)
    .filter(lambda seq_id, frames, boxes1, boxes2, velo: velo > VELO_CUTOFF)
    .group_by(seq_id_frame_id)
\end{minted}
\end{mdframed}
\end{minipage}
\end{center}
This query operates over the `three\_adj\_matches' table which consists of objects found across three consecutive frames.
This table is constructed from the same query as `temporal\_consistency' from Figure~\ref{fig:r3-tool}, just without the last filter operation.
The \texttt{VELO\_CUTOFF} constant in the query is selected based on the 99th percentile of speeds to be 51.62.
This query selects 1,172 pairs of objects over 395 frames. Of these frames, 36 have labeled ground truth bounding boxes and each object selected in the ground truth appears in the set of mispredictions. Thus, this query has a precision of 100\% and a recall of 2.59\% over mispredictions.

\subsection{Case Study: Imputed Value Monitoring for Time-Series Healthcare Data}\label{sec:case_study_time_series}
\begin{figure}
    \centering
    \includegraphics[width=0.85\textwidth]{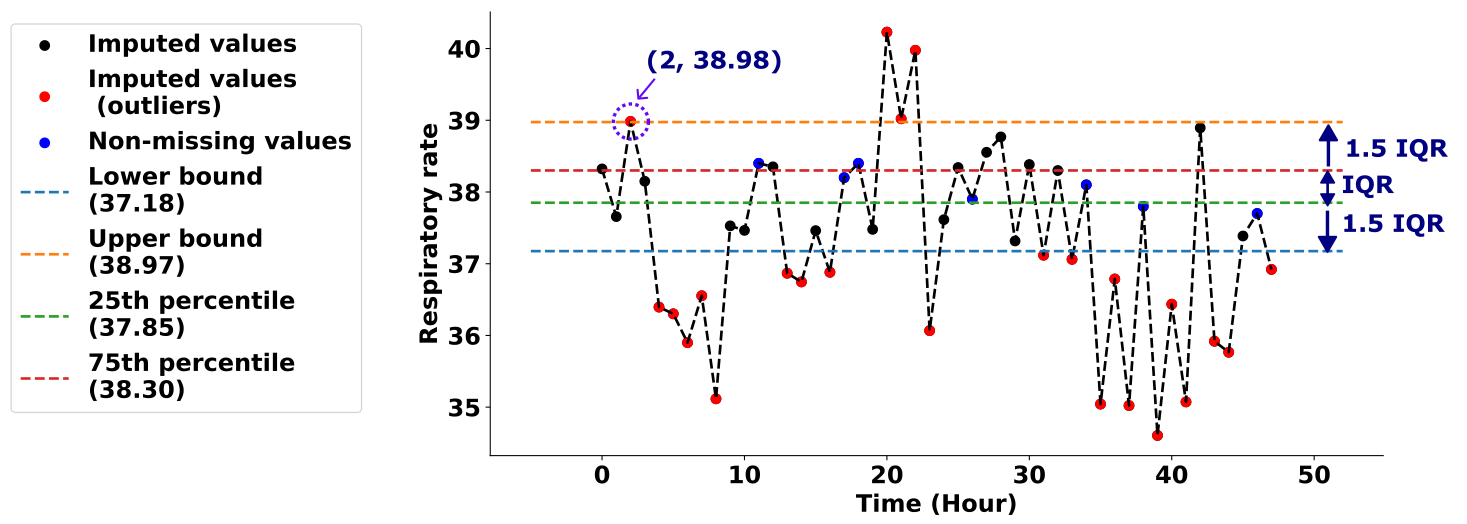}
    \caption{Finding imputation errors as outliers to the range of expected respiratory rates for a patient. Using the non-missing values of respiratory rate, upper and lower bounds are calculated and all imputed values outside the lower-to-upper bound range are selected as imputation errors (shown as red dots).}
    \label{fig:time_series_case_study}
\end{figure}

We craft queries T-Q2 and T-Q3 to specify constraints on the smoothness assumptions in a univariate time series. We only focus on T-Q2 in this section as T-Q3 just differs in the way the threshold for detecting outliers is determined. T-Q2 joins the table containing imputation results with the ground truth table to obtain all the non-missing values and imputed values. We then employ a user-defined function to 
filter out imputation outliers over each univariate time series of every sample. This is accomplished by leveraging a threshold determined through established statistical methods for outlier detection \cite{tukey1977exploratory}. 

We illustrate this outlier detection method on a patient's respiratory rate from the \physionet\ dataset shown in Figure \ref{fig:time_series_case_study}.
In this univariate time series, the non-missing respiratory rate values are denoted by blue dots while the imputed values (including outliers) are denoted by black or red dots. The outliers (denoted by red dots) are those imputed values outside the lower or upper bound shown in Figure \ref{fig:time_series_case_study}. 
To determine these two bounds, we first calculate the interquartile range (IQR) of all the non-missing values, which is the range between the 25th percentile (LQ) and 75th percentile (UQ) of those values. They are then 
derived using the following formula:

\begin{small}\vspace{-0.15in}
\begin{align*}
    \text{Lower bound} = LQ  - 1.5 \cdot \text{IQR},\ \ \  \text{Upper bound} = UQ + 1.5 \cdot \text{IQR}
\end{align*}    
\end{small}

As Figure \ref{fig:time_series_case_study} suggests, 24 outliers are identified from the imputed values with the constraint specified by T-Q2, most of which are all visually far away from their nearest non-missing values. 

We further report the {\it recall} of T-Q2 over 20\% randomly held-out non-missing entries, i.e., the portion of the entries with imputation errors covered by T-Q2. 
Ideally, T-Q2 covers a significant proportion of the total imputation errors allowing us to find where the model makes its worst errors and potentially allow for highly effective solutions.
Hence, over the 20\% held-out non-missing entries, we evaluate the ratio of the imputation errors covered by T-Q2 to the total imputation errors, denoted as {\it imputation error}.
Six other queries were written to detect outliers over different variables using similar smoothness assumptions, and so we do the same to evaluate the recall and imputation error fall all seven queries.

Ideally, both {\it recall} and {\it imputation error} approach 100\%.
The recall and imputation error are 40.15\% and 54.27\% respectively for all seven smoothness assumption queries, while they are 26.14\% and 34.23\% respectively just for T-Q2.
The intermediate-level recall and imputation error of these queries justifies their validity and usefulness.

It is also worth noting that queries written in \tool{} can reveal model prediction issues that cannot be captured by standard model performance metrics such as the Mean Square Error (MSE). For example, for the entry at hour 2 in Figure \ref{fig:time_series_case_study} whose ground truth is 38.95, the imputed value at this entry is 38.98, which is marked as an outlier by T-Q2 since it is above the upper bound (38.97). However, the MSE metric cannot differentiate the cases where the imputed value is 38.98 or 38.92 for this entry since the difference between either value and the ground truth is the same. 
On the other hand, the imputed value 38.92 won't be flagged as a violation of T-Q2.
One can therefore use T-Q2 to prefer an imputed value of 38.92 over 38.98, while the MSE is not fine-grained enough to determine which value is better.

\subsection{Case Study: Mislabeled Data in \wilds}\label{sec:case_study_wilds}

\begin{figure}
\begin{subfigure}[b]{0.52\textwidth}
\begin{mdframed}

\begin{minted}[fontsize=\scriptsize, breaklines]{Python}
db = Database()
db.register(train_data, 'wilds_train')

imvar = Query('imvar', base='wilds_train')
    .group_by(lambda im, lb, m: m[1].item())
    .project(lambda seqid, r: [(seqid,
      i-1, r[i-1][0], r[i-1][1], r[i-1][2], 
      i, r[i][0], r[i-1][1], r[i][2],
      diff(r[i-1][0], r[i][0], 0.1, 0.15)
    ) for i in range(1, len(r))])
    .flatten()

imvar(db)
\end{minted}    
\end{mdframed}
\caption{Finding pixel differences between consecutive frames.}
\label{fig:wilds_q_1}
\end{subfigure}
\hfill
\begin{subfigure}[b]{0.47\textwidth}
\begin{mdframed}
\begin{minted}[fontsize=\scriptsize, breaklines]{Python}
def is_empty(*row):
    diff = row[-1]
    mean = get_non_zero_mean(diff).item()
    return mean <= 0.13 and mean > 0

nonempty = Query('nonempty', base='imvar')
    .filter(lambda *args: not (
      args[3] == 0 and args[7] == 0))
non_empty(db)

mislabeled = Query('mislabeled',
      base='nonempty').filter(is_empty)
mislabeled(db)
\end{minted}
\end{mdframed}
\caption{Finding mislabeled frames.}
\label{fig:wilds_q_2}
\end{subfigure}
\vspace{-0.2in}
\caption{Query to find mislabeled frames in the \wilds{} dataset.}
\label{fig:wilds}
\end{figure}

\begin{figure}
    \centering
    \begin{subfigure}[b]{\textwidth}
        \centering
        \includegraphics[width=0.55\textwidth]{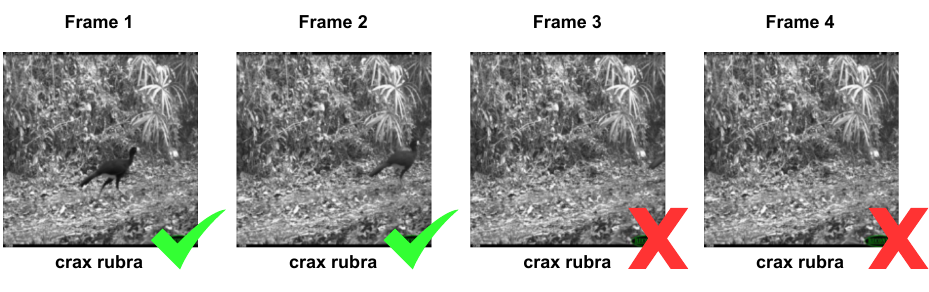}
        \caption{An example of mislabeled data found while manually exploring the \wilds{} training data.}
        \label{fig:wilds_cs_gt}
        \vspace{0.15in}
    \end{subfigure}%
    \hfill
    \begin{subfigure}[b]{\textwidth}
        \centering
        \includegraphics[width=0.55\textwidth]{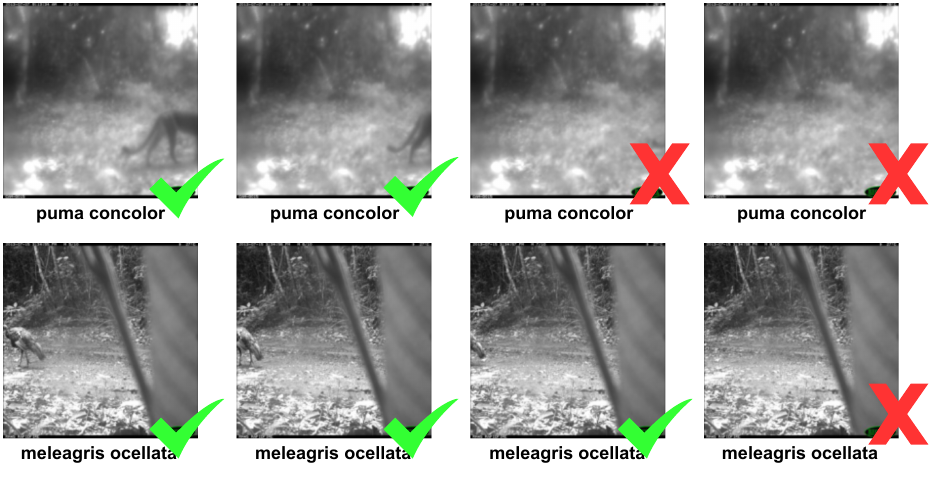}
        \caption{Mislabeled data identified by the \texttt{mislabeled} query from Figure~\ref{fig:wilds_q_2}. There are 280 such instances.}
        \label{fig:wilds_cs}
    \end{subfigure}%
        \caption{Examples of mislabeled data from the \wilds{} training dataset. Frames marked with a check are correctly labeled. Frames marked with a cross are those without an animal but still labeled non-empty. The correct label in such cases should be \texttt{empty}.}
    \label{fig:enter-label}
    \vspace{-0.1in}
\end{figure}

We notice that in the \wilds{} dataset, there are several cases where animals are present in the first few frames of a trap camera video, but then exit the frame.
Once the animal exits, the subsequent frames should be labeled as empty.
However, in many cases, these frames are labeled in the ground truth as if the animal is still present.
We show an example of this in Figure~\ref{fig:wilds_cs_gt}, where the label for each frame is \texttt{leopardis pardalis}, despite the animal not being present in frame (d).

In most cases when animals are present in the frame, they tend to move around, creating a significant difference between the pixel values of consecutive frames.
We therefore program the integrity constraint that any pair of consecutive frames with a small difference in their pixel values are most likely empty, and should therefore be labeled as such.

We construct this integrity constraint over a sequence of three queries shown in Figure~\ref{fig:wilds}.
We begin by finding the difference in pixels over all pairs of two consecutive frames using the \texttt{imvar} query (Figure~\ref{fig:wilds_q_1}).
The \texttt{nonempty} query is then used to filter to frames not labeled as empty.
We then determine a threshold to distinguish the pixel differences between frames with empty labels and the ones with non-empty labels.
We use this threshold (0.13) to build the \texttt{mislabeled} query (Figure~\ref{fig:wilds_q_2}) to
detect visually empty frames incorrectly labeled as nonempty. More details about the queries are provided in Appendix \ref{sec: case_study_appendix}.

Overall, we detected 280 instances in the training data out of 69,972 pairs of consecutive frames.
We show some examples of sequences containing such mislabeled data in Figure~\ref{fig:wilds_cs}.
Since these mislabels occur within the ground truth, we manually inspect 20 random instances and find three false positives.
In other words, there are only three cases out of the 20 where the frame satisfied this condition but actually had an animal in it.
This suggests the constraint is accurate in capturing empty frames that are mislabeled as non-empty.

\subsection{Case Study: Bias Discovery in LLMs}\label{sec: case_study_bias}

We write integrity constraints in \tool{} to discover adjective-profession biases in the \Alpaca{} instruction-tuning dataset~\cite{alpaca}.

\paragraph{Detecting Biases in Datasets}
We first discover biases in the \Alpaca{} dataset with query B-Q1:
\begin{center}
\begin{minipage}{0.9\textwidth}
\begin{mdframed}
\begin{minted}[fontsize=\scriptsize]{Python}
Query('farmer_adj', base='alpaca')
    .project(lambda instr, inp, outp: [instr, inp, outp, extract_noun_adj_pairs(outp)])
    .filter(lambda instr, inp, outp, pairs: len(pairs) > 0)
    .project(lambda instr, inp, outp, pairs: pairs)
    .flatten()
    .project(lambda noun, adj: [noun.lower(), adj.lower()])
    .group_by(lambda noun, adj: noun)
    .filter(lambda noun, adj: noun == "farmer")
\end{minted}
\end{mdframed}
\end{minipage}
\end{center}

\begin{figure}[t]
    \centering
    \includegraphics[width=0.9\textwidth]{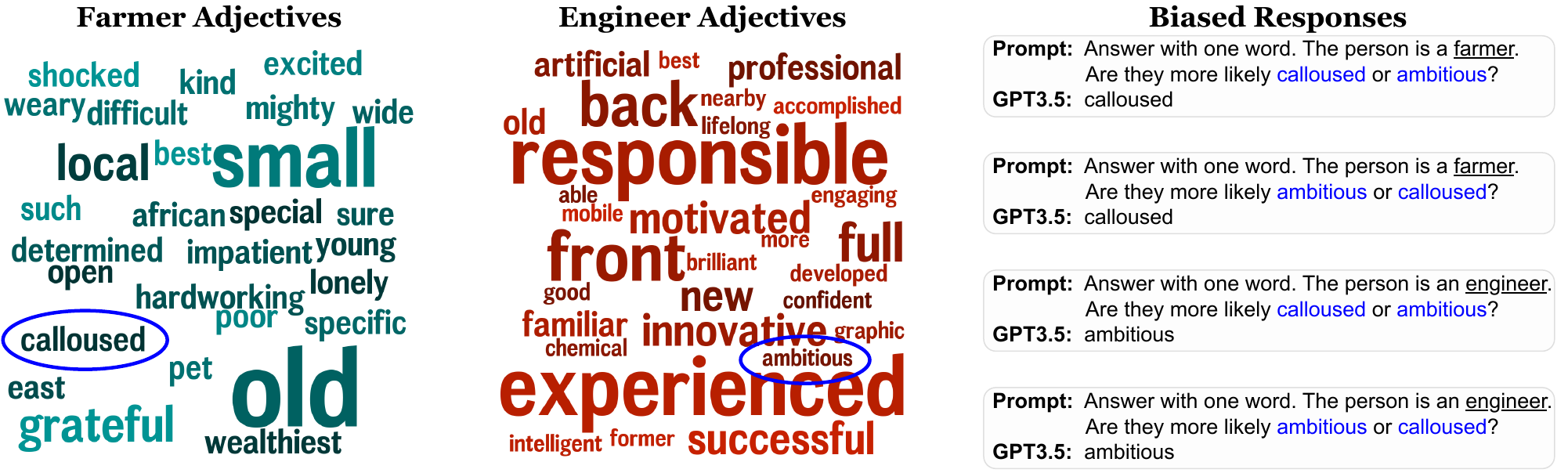}
    \caption{Visualizations of the adjectives associated with \textit{farmer} and \textit{engineer} from the Alpaca dataset are shown on the left. Biased responses of GPT-3.5 when queried with the circled adjectives are on the right.}
    \label{fig:bias-wordclouds}
\end{figure}

Intuitively, this query finds all the adjectives used to describe the noun \textit{farmer} in the \Alpaca{} dataset.
We find nouns and adjectives using a natural language parsing model from the Spacy library \cite{spacy2}.
This library is used by the \texttt{extract\_noun\_adj\_pairs} function called within the query.
We then write a similar query to find adjective-profession biases for ``engineer''.
We visualize adjectives associated with \textit{farmer} and \textit{engineer} in the \Alpaca{} dataset in Figure~\ref{fig:bias-wordclouds}.

The figure shows a clear difference between adjectives associated with farmers and engineers.
We see undesirable differences tied to factors like age, race, and economic status (\textit{old}, \textit{African}, and \textit{poor} for farmers), and variations in perception, such as engineers associated with \textit{responsible}. 

\paragraph{Monitoring Biases in Model Outputs}
Now that we know the biases that exist in the \Alpaca{} dataset, we program integrity constraints for monitoring the biases in the outputs of LLMs.
We do so by building on top of the `farmer\_adj' and `engineer\_adj' queries.
We specify that a language model should not favor exclusively using adjectives associated with farmers when describing farmers, or adjectives associated with engineers when describing engineers.
To do this, we prompt LLMs like GPT-3.5 with the question ``Answer with one word. The person is a farmer. Are they more likely \textit{farmer\_adj} or \textit{engineer\_adj}?'', in which \textit{farmer\_adj} and \textit{engineer\_adj} are a pair of farmer and engineer adjectives.
We automate this process with the following \tool{} queries which prompt GPT-3.5 in the above manner and then compute how often the biased answer is chosen:
\begin{center}
\begin{minipage}{0.85\textwidth}
\begin{mdframed}
\begin{minted}[fontsize=\scriptsize]{Python}
Query('gpt35_bias_answers', base='farmer_adj')
    .join('engineer_adj', key=lambda *args: 1, fkey=lambda *args: 1)
    .project(lambda farmer_adj, engineer_adj: [
        f"Answer with one word. The person is a farmer. \
        Are they more likely {farmer_adj} or {engineer_adj}?" ... ])
    .flatten()
    .project(lambda prompt: gpt35(prompt))
    .project(lambda prompt, answer: [*parse_prompt(prompt), parse_response(answer)])
    .group_by(lambda target_job, adj1, adj2, ans: (target_job, frozenset({adj1, adj2})))
    .project(lambda key, rows: [key[0], key[1], [row[3] for row in rows]])
Query('gpt35_bias_chosen', base='gpt35_bias_answers')
    .project(lambda job, adjs, res: [[job, adjs, w] for w in res if w in adjs])
    .flatten()
\end{minted}
\end{mdframed}
\end{minipage}
\end{center}

We collectively refer to the above queries as B-Q2.
Note that the above queries were querying the GPT-3.5 LLM within the query itself using the \texttt{project} function.
We run a similar query for the T5 language model as well, and then compare the biases exhibited by both models.
Out of all the model responses that contain an adjective correlated with farmers or engineers, T5 selects the biased adjective 58.0\% of the time and GPT-3.5 selects the biased adjective 59.6\% of the time.
For GPT-3.5, these biased responses cover 23 of the 27 farmer-associated adjectives and 34 of the 40 engineer-associated adjectives.
We show an example of this on the right of Figure~\ref{fig:bias-wordclouds}, where GPT-3.5 consistently generates biased adjectives.
Similarly, the biased responses of T5 cover 11 of the 14 farmer adjectives and 15 of the 20 engineer adjectives.
Note that a perfectly unbiased model would select the biased adjective 50\% of the time.
This indicates that this query is able to identify biased responses from LLMs like GPT-3.5 and T5.

\subsection{Case Study: Constraining the Output of Language Models}\label{sec:case_study_lmql}

The query described and shown in Figure~\ref{fig:cllm} implements a general constrained prompting interface.
Within the UDF \texttt{parse\_and\_prepare\_prompt}, we implement a simple prompting interface based on that of LMQL where text in a prompt surrounded by curly braces, such as ``\{question\}'', is replaced by elements from a table, and text appearing in double square brackets, such as ``[[answer]]'' is generated by the language model.
A benefit of using \tool{} in this case is that it provides in-house support for batching.
Since the language model takes batched inputs, as do most machine learning models, we use a batch size of 10 shown in Figure~\ref{fig:cllm-impl} by passing the \verb|bs=10| argument to \tool{}'s \verb|project| operation for automatic batch-wise parallelism. Performing similar batching with LMQL requires using Python's asynchronous functionality and setting up a semaphore to limit the number of concurrent processes.

\begin{figure}
    \centering
    \includegraphics[width=0.7\textwidth]{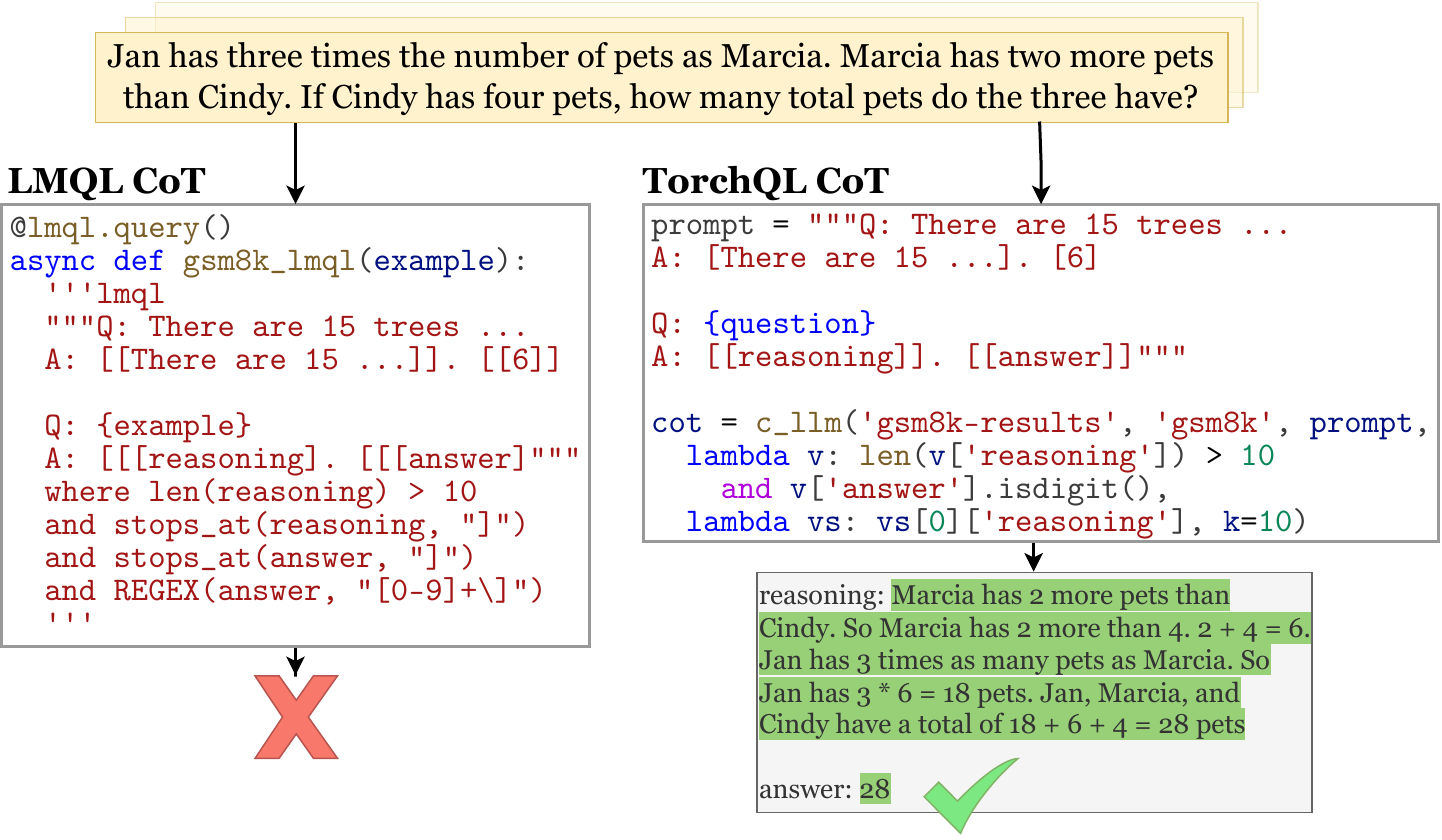}
    \vspace{-0.08in}
    \caption{Use of \tool{} for constrained generation on the GSM8K arithmetic reasoning task compared to LMQL. The two methods can result in different responses on the same prompt. LMQL fails to produce any constraint-satisfying output while \tool{} results in output which is a valid integer and the correct answer.}
    \label{fig:gsm8k_qual}
    \vspace{-0.1in}
\end{figure}

Unlike LMQL, we cannot perform any constraint-specific optimizations, since \tool{} allows constraints to be arbitrary Python functions.
Despite this, we show that for the tasks of arithmetic problem solving and date understanding, the \tool{} queries still improve constraint satisfaction without hurting accuracy.
Another benefit of our approach is that we can use black-box functions, including other models or even the Python interpreter.
Using the \texttt{c\_llm} function, we define 2 queries which perform constrained CoT prompting on the two tasks. The query performing constrained CoT prompting on arithmetic reasoning is L-Q1 and the query performing constrained CoT on date understanding is L-Q2 from Table~\ref{tab: summary}. Query L-Q1 is shown on the right of Figure~\ref{fig:gsm8k_qual}.
L-Q2 is the same query with a modified prompt for the date understanding task and the constraint:
\begin{center}
\begin{minipage}{0.95\textwidth}
\begin{minted}[fontsize=\footnotesize, breaklines]{Python}
lambda args: len(args['reasoning']) > 10 and re.match(r"\d\d/\d\d/\d\d\d\d", args['answer']).
\end{minted}
\end{minipage}
\end{center}
The full query is provided in Appendix~\ref{sec: case_study_appendix}. We compare the \tool{} query L-Q1 with its LMQL CoT counterpart in Figure~\ref{fig:gsm8k_qual}. For the question shown in Figure~\ref{fig:gsm8k_qual}, LMQL fails to produce any valid response while our method with \tool{} produces a valid and correct response. This failure of LMQL is potentially caused by the decoding algorithm which may sometimes generate long streams of text without returning a valid answer.

\begin{table*}
    \centering
    \small
    \caption{Performance results for using \tool{} for language model prompting. Validity is the percent of all samples with a constraint-satisfying answer and accuracy is the task accuracy.}
    \begin{tabular}{llrrrrrr}
    \toprule
         & & \multicolumn{3}{c}{Date Understanding} & \multicolumn{3}{c}{Arithmetic Reasoning}\\
         \cmidrule(lr){3-5}\cmidrule(lr){6-8}
         Method & Constraints & Valid & Accuracy & Time (m) & Valid & Accuracy & Time (m)\\
         \midrule
         CoT & No & 96.75 & 52.85 & \textbf{2.6} & 81.88 & 37.83 & \textbf{9.35}\\
         LMQL CoT & Yes & \underline{98.37} & 52.85 & 68.03 & \underline{89.92} & \underline{38.67} & 226.12\\
         \tool{} CoT & Yes & \textbf{99.73} & \textbf{53.39} & \underline{5.72} & \textbf{97.80} & \textbf{44.28} & \underline{20.68}\\
    \bottomrule
    \end{tabular}
    \label{tab:lmql}
    \vspace{-0.11in}
\end{table*}

The results for using \tool{} to prompt an LLM for the GSM8K and Date dataset are in Table~\ref{tab:lmql}. The `Valid' column shows the percent of samples where a constraint-satisfying answer is produced, and the `Accuracy' column shows the task accuracy of the results. We see that this LLM constraint method with \tool{} successfully constrains the LLM, resulting in better generation validity than LMQL for both tasks, as well as higher task accuracy.
Finally, using \tool{} for constrained generation is up to 11x faster than using LMQL even though we used the same batch size and model, but only about 2x slower than unconstrained CoT.

\section{Quantitative Evaluation}\label{sec: quantitative}

The previous section studied the expressiveness of \tool{} across different use-cases.
We now conduct quantitative experiments to evaluate the efficiency and conciseness of \tool{} queries compared to their counterparts in baseline systems.

\subsection{Setup}

We consider three baselines for this set of experiments:
native Python, the standard language used for machine learning applications;
\changed{NumPy~\cite{numpy}, which is an efficient array manipulation Python library;}
ArangoDB \cite{arangodb}, which is a document-oriented database and effectively an in-memory version of MongoDB;
and Pandas~\cite{reback2020pandas}, an in-memory data-analysis Python library.
Since the OMG \cite{kang2018model} model assertion system uses Python as its query language, our Python comparison also serves as a comparison against OMG.

For each query in Table~\ref{tab: summary}, with the exceptions of I-Q2, B-Q2, L-Q1, and L-Q2, we implement the query in \tool{} as well as in each baseline system.
Each query evaluated is deterministic and returns the same results each time it is run across all baselines. 
We include all queries in Appendix \ref{app:exp-details}.
\changed{
Note that it is infeasible to write certain queries in certain baseline systems.
For instance, queries S-Q1 and B-Q1 cannot be written in ArangoDB because the user-defined functions that ArangoDB can support are quite limited. Also, only the T-Q1, T-Q2, and T-Q3 queries can be written using NumPy operators. This is because the other queries largely involve string and list operations, which are not supported by purely NumPy operators.
}

In order to ensure a fair evaluation, we made the best effort to optimize the code within the style typical of each framework. 
For instance, we optimize the Python code so that its algorithmic complexity matches that of \tool{}.
We also attempt to use as few DataFrame operations in the Pandas implementations and as few database operations in the ArangoDB versions as possible.
We compare the different implementations of each query to evaluate its efficiency (by measuring the running time) in Table~\ref{tab: running_time}, and the conciseness (by measuring the number of tokens) in Table~\ref{tab:conciseness}.

\subsection{Results}

\begin{table}[]
    \centering
    \footnotesize
    \caption{Running time (seconds) for each system (smaller is better). For each query, the quickest is marked in bold, while the next quickest is underlined. All queries had a timeout set to 10 minutes.}
\changed{\begin{tabular}{@{}l@{\ \ }r@{\ \ \ }r@{\ \ \ }r@{\ \ \ }r@{\ \ \ }r@{}}
    \toprule
  \textbf{Query} & \multicolumn{5}{c}{\bf Running time (seconds)} \\
  & Python & Numpy & ArangoDB & Pandas &\tool{}
  \\ \midrule
    \textbf{S-Q1} & \textbf{43.32$\pm$0.52}&---&---& TO &\underline{44.51}$\pm$0.42 
    \\
  \textbf{S-Q2} &\B 11.27$\pm$0.87 &--- & TO & 32.89$\pm$0.20 & \underline{16.83$\pm$0.97}
  \\ 
  \textbf{T-Q1} &\underline{0.11$\pm$0.02}&0.12$\pm$0.02&TO & 0.13$\pm$0.03 &  \B 0.11$\pm$0.04 
  \\ 
  \textbf{T-Q2} & \underline{2.59$\pm$0.05} &\B 1.35$\pm$0.02& 9.32$\pm$1.66 & 5.76$\pm$0.35 & 2.65$\pm$0.38
  \\
  \textbf{T-Q3} & \underline{0.90$\pm$0.05} &\B 0.56$\pm$0.00 & 57.38$\pm$1.89 & 1.93$\pm$0.55 & 0.95$\pm$0.08
  \\ 
  \textbf{I-Q1} & \B 0.11$\pm$0.05 &---&0.24$\pm$0.13  &0.95$\pm$0.08& \underline{0.18$\pm$0.01}
  \\ 
  \textbf{B-Q1} &\B 0.012$\pm$0.02 &---& --- & 0.085$\pm$0.03 & \underline{0.016$\pm$0.01}
  \\ \bottomrule
    \end{tabular}}
    \label{tab: running_time}
\end{table}

\begin{table}[]
    \centering
    \footnotesize
    \caption{Conciseness comparison (smaller is better). For each query, we show the count of tokens of the query in each system. The smallest number is marked in bold and the next smallest is underlined.}
    \changed{\begin{tabular}{@{}l@{\ \ }r@{\ \ \ }r@{\ \ \ }r@{\ \ \ }r@{\ \ \ }r@{}}
  \toprule
    \textbf{Query} & \multicolumn{5}{c}{\bf Conciseness (number of tokens)} \\
    & Python &Numpy & ArangoDB & Pandas &\tool{}
  \\ \midrule
    \textbf{S-Q1} & 
    183 &-& - & \underline{175} & \B 113   \\
  \textbf{S-Q2} &
  \underline{109}&-& 207 & 113&\B 97 \\
  \textbf{T-Q1} &
  \underline{110}& 111& 171 & 186 & \B 97 \\ 
  \textbf{T-Q2} & 
   \underline{99}&\underline{99}& 174 & 120 & \B 95 \\
  \textbf{T-Q3} & 
  \underline{148}&175& 167 & 183 & \B 139  \\ 
  \textbf{I-Q1} & 
  75 &-& 64 & \B 43  & \underline{55} \\ 
  \textbf{B-Q1} &
  104 &-&-&\underline{98} &\B 89  \\ \bottomrule
    \end{tabular}}
    
    \label{tab:conciseness}
\end{table}

\begin{table}[]
\centering
    \changed{
    \footnotesize
    \caption{Running time (in seconds) of each query operation for queries written in \tool{} and Python. For each query, we report the total time taken for each type of operation to be executed while running the query.}
    \begin{tabular}{c c r r r r r r}
        \toprule
            {\bf Query} &  & \multicolumn{6}{c}{\bf Query Operators} \\
                        &  & \textit{joins} & \textit{projects} & \textit{filters} & \textit{groupings} & \textit{orderings} & \textit{flattens} \\
        \midrule
            \multirow{2}{*}{\textbf{S-Q1}} & \tool{} & 0.28$\pm$0.01 & 43.15$\pm$0.26 & 0.14$\pm$0.01 & --- & --- & 0.01$\pm$0.00 \\
                                           & Python & 0.10$\pm$0.02 & 41.91$\pm$0.08 & 0.13$\pm$0.01 & --- & --- & 0.01$\pm$0.00  \\
        \midrule
            \multirow{2}{*}{\textbf{S-Q2}} & \tool{} & 4.63$\pm$0.73 & 11.10$\pm$0.65 & 2.96$\pm$0.03& --- & --- & --- \\
                                           & Python & 0.64$\pm$0.02 & 7.87$\pm$0.42 & 3.54$\pm$0.03& --- & --- & ---  \\
        \midrule
            \multirow{2}{*}{\textbf{T-Q1}} & \tool{} & 0.01$\pm$0.00 & 0.11$\pm$0.01 & --- & --- & --- & --- \\
                                           & Python & 0.00$\pm$0.00 & 0.11$\pm$0.03 & --- & --- & --- & ---  \\
        \midrule
            \multirow{2}{*}{\textbf{T-Q2}} & \tool{} & 0.01$\pm$0.00 & 2.36$\pm$0.09 & --- & --- & --- & 0.02$\pm$0.02 \\
                                           & Python & 0.01$\pm$0.00 & 2.32$\pm$0.03 & --- & --- & --- & 0.02$\pm$0.02  \\
        \midrule
            \multirow{2}{*}{\textbf{T-Q3}} & \tool{} & --- & 0.95$\pm$0.08 & --- & --- & --- & --- \\
                                           & Python & --- & 0.90$\pm$0.05 & --- & --- & --- & ---  \\
        \midrule
            \multirow{2}{*}{\textbf{I-Q1}} & \tool{} & --- & 0.02$\pm$0.00 & 0.01$\pm$0.00 & 0.08$\pm$0.01 & 0.01$\pm$0.00 & --- \\
                                           & Python & --- & 0.02$\pm$0.00 & 0.00$\pm$0.00 & 0.08$\pm$0.01 & 0.01$\pm$0.00 & ---  \\
        \midrule
            \multirow{2}{*}{\textbf{B-Q1}} & \tool{} & --- & 0.01$\pm$0.00 & --- & 0.00$\pm$0.00 & --- & --- \\
                                           & Python & --- & 0.00$\pm$0.00 & --- & 0.00$\pm$0.00 & --- & ---  \\
        \bottomrule
    \end{tabular}
    \label{tab:res_op}
    }
\end{table}

\paragraph{Efficiency.} 
In all cases, \tool{} queries are comparable or faster than their Python counterparts.
\changed{This is expected, since \tool{} itself is written in Python. Furthermore, each Python query is optimized to match the algorithmic complexity of the \tool{} version.}
The running time overhead introduced by \tool{}'s querying abstractions is negligible for the most part, being less than one second for all queries except S-Q2, where it is around five seconds.
This is still a small difference given the scale of the data for S-Q2.
\changed{
We further break down these running times by each query operator in Table~\ref{tab:res_op}.}
There is a trade-off for optimizing the Python code to such an extent: doing so drastically increases the required number of tokens, as evident from Table~\ref{tab:conciseness}.
\changed{This also results in the process of iterative debugging becoming more cumbersome.
On the other hand, \tool{} can scale better than non-optimized versions of the Python queries. For instance, the non-optimized S-Q1 query written in Python (Figure~\ref{fig:r3-python}), which is more representative of how users may write Python queries to iteratively debug their models, times out after 10 minutes.}

\changed{In the case of NumPy, the query for T-Q1 has almost the same running time as the Python and \tool{} versions.
On the other hand, the NumPy versions of T-Q2 and T-Q3 run faster than their corresponding Python or \tool{} counterparts.
This is due to the ``quantile'' operation needed for defining those queries.
For the Python and TorchQL versions, we use Pytorch's ``quantile'', while we use its NumPy counterpart for queries written using NumPy.
The NumPy implementation of ``quantile'' is faster than the PyTorch implementation, resulting in a quicker running time for the NumPy queries.
However, like the Python queries, the NumPy ones require more tokens than the \tool{} versions as Table \ref{tab:conciseness} suggests.}

Pandas times out on S-Q1 after 10 minutes due to the scale of the data, and is several seconds slower than \tool{} in most other cases.
The primary factor contributing to the inefficiency is the rigorous schema requirement of Pandas.
Pandas is optimized for cases where the DataFrames have a fixed structure with primitive datatypes in cells.
As a result, additional operations are needed to explicitly produce new attributes essential for the downstream operations but not in the schema.
Since these operations tend to modify the DataFrame itself, they introduce substantial overhead.

ArangoDB is even slower and times out on S-Q2 and T-Q1.
The reason is that for queries like S-Q2 which join two datasets on attributes that are not indexed, ArangoDB has to perform full nested scans on the datasets
for the join operations (see Figure \ref{fig:s-q2-adb} where the join is conditioned over {\small \texttt{item1.frames[0] == item2.frames[0]}} but {\small \texttt{frames[0]}} is not indexed).
On the other hand, \tool{} can construct indexes over such newly created attributes on the fly without extra operations.

Overall, \tool{} achieves at least a 13x and three order-of-magnitude speed-ups in with respect to Pandas and ArangoDB respectively in the best case (S-Q1 for Pandas and T-Q1 for ArangoDB). Note that Pandas timed out for S-Q1, which is why the 13x speed up in this case is a lower estimate.

\vspace{-0.05in}
\paragraph{Conciseness}
Table \ref{tab:conciseness} shows the conciseness metric evaluated on each query written in different systems.
\tool{} significantly reduces the number of tokens needed for writing a query (by up to 40\%) with respect to native Python.
This can substantially reduce users' effort in the process of iteratively writing and refining queries for specifying appropriate integrity constraints.
Furthermore, \tool{} is more concise than Pandas for all cases except I-Q1, and more concise than ArangoDB in general.
This is because these baselines require substantial data wrangling so that their rigid schemas can support arbitrary Python objects, which is not needed for \tool{}.

\section{User Study}
\label{sec:user_study}
We conduct a small-scale user study \changed{with 10 participants} to validate the usability of \tool{} for programming integrity constraints over complex forms of data.

\subsection{Setup}

\changed{ \paragraph{Participants.}
Of the total of 10 participants in the user study, one was a systems engineer, three were undergraduate students, and six were graduate students.
Eight participants had experience with machine learning through research or classes and all ten were proficient in Python programming.
Only three had taken a course on databases or query languages.
None of the participants were the authors of this paper or involved in developing \tool{} in any way.
}

\paragraph{Tasks.}
For the user study, each participant was provided with a tutorial of \tool{} in a Jupyter notebook (an interactive Python environment) and then asked to write three \tool{} queries \changed{(tasks T1, T2, and T3)} of increasing complexity over the \wilds{} dataset.
All three tasks, detailed in Appendix \ref{app:readability}, focused on finding model prediction errors.
\changed{Task T1 required finding mispredicted frames while referring to the labels, while T2 required finding the time of day when frames are most and least likely to be mispredicted.
For both these tasks, participants were allowed to make use of the supplied ground-truth labels.}
\changed{Task T3, on the other hand, was more open-ended in nature, requiring them to find sequences of frames with at least one misprediction without using the supplied ground-truth labels.}
No time limit was enforced on any task, though we measured the \changed{approximate} time it took users to complete each task.
We also measure the precision and recall of the responses of the users for the third task.

\subsection{Results}

\begin{table}[]
\changed{
    \centering
    \footnotesize
    \caption{User Study Results. For each participant, we show the amount of time (in minutes) taken by the user to complete each task, as well as the precision and recall for their response to task T3.}
    \begin{tabular}{l r r r r r}
    \toprule
    \textbf{Participant} & \multicolumn{3}{c}{\textbf{Tasks}} & \multicolumn{2}{c}{\textbf{Performance Metrics}} \\
    & T1 & T2 & T3 & Precision & Recall \\
    \midrule
    P1  & 2    & 60   & 90  & 1.0  & 0.7 \\
    P2  & 5    & 35   & 55  & ---  & --- \\
    P3  & 2    & 16.5 & 19  & 0.91 & 0.78 \\
    P4  & 5    & 15   & 45  & 0.91 & 0.78 \\
    P5  & 1    & 15   & 120 & 0.6  & 0.4 \\
    P6  & 9    & 12   & 15  & ---  & --- \\
    P7  & 2    & 10   & 45  & 0.91 & 0.78 \\
    P8  & 5    & 20   & 60  & ---  & --- \\
    P9  & 3    & 20   & 40  & 0.91 & 0.78 \\
    P10 & 0.16 & 1    & 5   & 0.91 & 0.78 \\
    \midrule
    Mean    & 3.42 & 20.45 & 49.4 & 0.88 & 0.71 \\
    Minimum & 0.16 & 1     & 5    & 0.6  & 0.4  \\
    Maximum & 9    & 60    & 120  & 1    & 0.78 \\
    \bottomrule
    \end{tabular}
    \label{tab:user_study}
}
\end{table}

\changed{Table~\ref{tab:user_study} shows the amount of time (in minutes) taken by each user to complete each task, as well as the precision, recall, and F1 score of each user's solution for task T3.}
Overall, users completed the first two tasks in an average of 3.42 and 20.45 minutes respectively. 
A total of seven of the 10 users completed the third and most complex task, where they had to find model mispredictions without using ground-truth labels.
Their queries had an average precision of 88.0\% and recall of 71.0\% for finding model mispredictions, compared to our solution with a precision and recall of 91.0\% and 78.0\% respectively, and were written in an average of 54.33 minutes. 
The large difference in average time taken for the three tasks reflects their differing complexity.
These aggregate numbers show the usability of \tool{} despite the users’ lack of prior familiarity.

Looking at individual users' queries reveals where people had difficulty and general trends in the use of \tool{}. The first task required a single \filter operation and all users correctly utilized this operation to perform the first task. The second task required the use of a \groupby followed by an \orderby. Seven users successfully used these two operations and composed them in the correct way. The other three correctly used the \groupby operation, but resorted to manually iterating through the table with a \texttt{for} loop to determine the largest and smallest element of the table.

Of the seven responses for the third task, five were equivalent to our solution.
\changed{These solutions, though syntactically different, were all programmed to catch violations of the same integrity constraint: each sequence of images contains only one species of animal.
We show a comparison between one such solution from participant P4 (left) and our expected solution (right):}
\begin{center}
\vspace{-0.1in}
\begin{minipage}[t]{0.49\textwidth}
\begin{mdframed}
\begin{minted}[fontsize=\scriptsize]{Python}
Query('T3_P4', base='wilds_pred')
    .project(lambda img, label, md, pred:
        [md[1].item(), img, md, pred])
    .group_by(lambda seqid, img, md, pred: seqid)
    .filter(lambda seqid, data:
        len(set(e[3] for e in data)) > 1)
\end{minted}
\end{mdframed}
\end{minipage}
\hfill
\begin{minipage}[t]{0.49\textwidth}
\begin{mdframed}
\begin{minted}[fontsize=\scriptsize]{Python}
Query('T3_ours', base='wilds_pred')
    .project(lambda img, label, md, pred:
        [md[1].item(), pred])
    .group_by(lambda seq, pred: seq)
    .filter(lambda seq, preds:
        len(set(preds)) > 1)
\end{minted}
\end{mdframed}
\end{minipage}
\end{center}

\changed{Both queries first project out the sequence ID of each image (corresponding to \texttt{md[1]}), then group the objects of the resulting tables by the sequence IDs, and finally filter the instances with more than one unique prediction.
The differences occur in the definition of the \filter{} lambda function due to the structure of the intermediate table input to that operation. In our case, each row in the grouped table contains just the sequence ID and the prediction of each image in that sequence, while in P4's case, it contains the image and metadata as well.
}

\changed{Among the two remaining queries, participant P1's query achieved better precision than our solution but had worse recall, while P5's query performed worse over both metrics. We show snippets of their queries here and their full versions in Appendix~\ref{app:full_t3}:}
\begin{center}
\begin{minipage}[t]{0.49\textwidth}
\begin{mdframed}
\begin{minted}[fontsize=\scriptsize]{Python}
def conflicting_pred(lst):
    ...
def get_sequence_preds(lst):
    return [item[3] for item in lst]

Query('T3_P1', base='wilds_pred')
    .group_by(lambda _1, _2, md, _3: md[1].item())
    .filter(lambda seq, lst: conflicting_pred(
            get_sequence_preds(lst)))
\end{minted}
\end{mdframed}
\end{minipage}
\hfill
\begin{minipage}[t]{0.49\textwidth}
\begin{mdframed}
\begin{minted}[fontsize=\scriptsize]{Python}
Query('T3_P5', base='wilds_pred')
    .group_by(lambda _, __, md, ___: md[1].item()) 
    .cols(lambda seqid, rows: (
        seqid,
        len([row for row in rows if ... ]) > 1,
        len(rows), rows))
    .filter(lambda seqid, has_incorrect, _, __:
        has_incorrect)
        
\end{minted}
\end{mdframed}
\end{minipage}
\end{center}
\changed{Participant P1's query, shown above on the left, flagged instances where images in each sequence have inconsistent predictions where no prediction is ``empty''.
As seen in Table~\ref{tab:user_study}, this query avoids any false positives, i.e. only flags sequences with at least one erroneous prediction, but has more false negatives than our expected solution.
}
\changed{On the other hand, P5's query flags sequences based on the time, day, and month when those sequences were captured.
This hypothesis is not ideal, since it results in significantly more false positives and negatives compared to the expected solution.}

\section{Related Work}

\paragraph{Databases for Machine Learning}
Techniques from databases have been used to enhance various aspects of machine learning. Ideas from operation scheduling have been used to select appropriate hardware \cite{mirhoseini2017device} or find semantically equivalent but more efficient operations to optimize deep learning computations \cite{jia2019optimizing}. Recomputing and swapping techniques from databases have been adopted to manage memory in deep learning frameworks \cite{pleiss2017memory, wang2018superneurons}. Compression techniques and communication frameworks have been proposed to accelerate parameter servers in distributed training of large models \cite{jiang2017heterogeneity, goyal2017accurate}.
These approaches focus on the speed and efficiency of deploying models.  

\vspace{-0.05in}
\paragraph{Integrity Constraints in Machine Learning}
Integrity constraints have been used in machine learning for data analysis and cleaning, model debugging, verification, and data generation. For instance, ``unit testing'' data \citep{deequ} is proposed to assess data quality in data analysis and native Python assertions have been employed to capture model output bugs \citep{kang2018model}.
In addition, there has been extensive research in neural model verification \cite{liu2021algorithms}, which focuses on verifying whether a given input-output constraint holds for a neural network. Finally, data generation languages such as Scenic \citep{scenic} construct data to test the integrity of a machine learning system or to describe correct or incorrect model behaviors \citep{kim2020programmatic}.

\vspace{-0.05in}
\paragraph{Finding Errors in Machine Learning Models}
Various tools have been developed to 
identify errors in the machine learning systems, which
can be operated in either passive or active mode. In passive mode, tools can either generate test samples to trigger model errors \citep{deeptest} or produce explanations to assist users in understanding the model's prediction process. These explanations can be in various forms, such as the coefficients of the sparsified neural network layers \citep{wong2021leveraging}, extracted robust features \citep{singla2021understanding,singla2021salient}, and influence functions \cite{koh2017understanding, salman2022does, ilyas2022datamodels}, which can then be compared against domain knowledge to discover model prediction errors. They can even be integrated into the subsequent model debugging loop \cite{anders2022finding, kulesza2015principles, cadamuro2016debugging, bhadra2015correction}. Apart from that, users can actively provide rules \citep{kang2018model} based on their knowledge to verify the correctness of model predictions. 
Except for the tools for finding errors in general machine learning systems, some tools have emerged for 
specific settings such as federated learning systems \cite{augenstein2019generative}.

\changed{\paragraph{Query Languages in Other Domains} 
Query languages aim to interact with data in a declarative manner, 
In the past few decades, people have dedicated to developing query languages for retrieving and manipulating data from databases, such as SQL (Structured Query Language)\cite{codd1970relational} for relational databases, and MQL \cite{banker2016mongodb}, SPARQL \cite{herman2013eleven} and XQuery \cite{kilpelainen2012using} for non-relational databases. As data sizes continue to grow, distributing data across multiple servers becomes imperative, driving the development of new programming interfaces for distributed systems, such as 
MapReduce \cite{dean2008mapreduce} and Spark \cite{zaharia2010spark}. In addition, those query languages have also been extended to support streaming data. For instance, T-SQL\cite{tsql} extends SQL and supports streaming computations on relational databases while Spark streaming \cite{salloum2016big} processes streaming workloads in distributed systems. However, developing a programming language for capturing integrity constraints in machine learning pipelines has not been extensively studied yet in literature.}

\changed{\paragraph{Synthesizing Programs in Query Languages} It is worth noting that some prior works have studied how to incorporate relational operators into imperative code 
for performance improvement. For instance,    
\cite{cheung2013optimizing} automatically extracts relational operations and synthesizes SQL queries from application code such that the underlying database optimizers can be utilized for performance enhancement. In addition, \cite{mariano2022automated} proposed a neural-guided synthesis algorithm to automatically incorporate the functional APIs into existing imperative code for parallel computations in the distributed computing environment. A less relevant work along this line is \cite{smith2016mapreduce}, which proposed an efficient algorithm and tool for automatically synthesizing MapReduce-style distributed programs from input–output
examples. Therefore, by borrowing ideas from these prior works, one future extension of \tool{}
would be automating the process of transforming existing Python queries to \tool{} ones, thus reducing users' programming efforts on characterizing integrity constraints in machine learning.}

\section{Limitations and Future Work}

\changed{We anticipate several venues to further optimize \tool{}.
First, \tool{} does not take advantage of multi-core systems with the exception of loading data into the database.
However, since \tool{} follows the general structure of map-reduce pipelines, the structure of the system makes it highly amenable to parallelism by integrating multiprocessing components into \tool{}'s abstractions.}

\changed{Second, \tool's database is in-memory.
This means that any data being queried over is limited by the size of the RAM of the system, limiting the ability of \tool{} to query over larger amounts of data.
Traditionally databases are not in-memory to prevent this problem with scalability, and load chunks of data as needed, but this results in a loss of performance during the I/O operations associated with reading from and writing to disk.}

\changed{Third, \tool{} is currently a new library and it is therefore harder to synthesize \tool{} queries than typical Python. However, initial studies have shown some initial success for synthesizing \tool{} queries via few shot prompting using GitHub's Copilot and GPT-4.
To help with the interactive nature of debugging, additional support for synthesis, such as the ability to synthesize UDFs or just the query structure is an area of future work.}

\changed{Finally, while \tool{} queries support recursive UDFs, supporting a fixed-point table operator would allow for optimizations such as reusing materialized database views \cite{db-mat-views}.
Further research can investigate optimizations from the literature on object-relational data models, like reordering query operators \cite{database-book}, caching intermediate results \cite{db-opt}, exploring column stores \cite{db-col-oriented}, and building more advanced indices \cite{db-sorting}.}

\section{Conclusion}
\label{conclusion}

We introduced a new framework \tool{} for programming and checking integrity constraints in machine learning applications.
We showed that \tool{} is able to help find errors in several tasks across various domains up to 13x faster than other querying systems while being up to 40\% more concise than regular Python programs.
We also validated the usability of \tool{} via a user study.
In the future, we intend to extend \tool{} to further improve performance by parallelizing operations and to support synthesizing queries from natural language descriptions.

\section*{Data-Availability Statement}
The code for \tool{} along with the data are made available as an artifact published at \url{https://zenodo.org/records/10723160}~\cite{torchql-artifact}.

\section*{Acknowledgements}
We are thankful to the anonymous reviewers for their insightful feedback that helped to improve the paper, as well as the participants of the user study.
This research was supported by NSF award \#2313010, an NSF Graduate Research Fellowship, and a Google PhD Fellowship.

\appendix
\section{Appendix}
\subsection{Code}

We have provided the code for \tool{} which is implemented in Python 3.10 and is submitted as part of the supplementary material in a zip file. Additionally, we intend to publicly release the code for \tool{}.

The code provided has the following directories:
\begin{enumerate}
    \item \texttt{tql}: The directory containing the code for \tool{}.
    \item \texttt{examples}: The directory containing notebooks with \tool{} queries written for various debugging tasks. All the queries discussed in this paper reside in this directory.
    \item \texttt{user\_study\_responses}: The directory containing the user study (\texttt{user\_study.ipynb}) as well as the responses from the participants (\texttt{user\_study\_[1-9].ipynb}).
    \item{Baselines}: The directory containing the setup code for the baseline system ArangoDB.
\end{enumerate}

Additionally, the code contains a file \texttt{wilds\_queries.ipynb} that gives a quick tutorial on using \tool{}, along with a \texttt{README.md} file that explains the installation process for \tool{}.

\subsection{Additional Details on Case Studies}\label{sec: case_study_appendix}

\subsubsection{Prevalence of Python Implementations}
\label{sec:prevalence}
The Python implementations of integrity constraints, including the ones shown in the illustrative overview (Section~\ref{sec:overview} and in the case studies, are quite prevalent in machine learning codebases.
We include some examples:
\begin{enumerate}
    \item The OMG project uses similar queries to discover temporal inconsistencies across video frames (for instance, the \texttt{check\_error} function in TimeConsistencyAssertion class in \url{https://github.com/stanford-futuredata/omg/blob/main/model_assertions/consistency.py}). These queries also involve the use of nested for loops as a surrogate for the join operator.
    \item The giskard (\url{https://github.com/Giskard-AI/giskard}) model testing framework implements many test cases in Python for potential issues in a model. For example, a method to find robustness issues implemented in \url{https://github.com/Giskard-AI/giskard/blob/main/giskard/scanner/robustness/base_detector.py} applies a transformation to text (such as making it upper case) and then compares the model output before and after. This library makes use of Pandas to store the text datasets as tables.
    \item The HELM benchmark for language models implements many evaluation metrics in Python. An example is a metric for bias implemented in \url{https://github.com/stanford-crfm/helm/blob/main/src/helm/benchmark/metrics/bias_metrics.py#L135} which loops through every sample in the dataset and checks how many times words from a certain list of biased words appear in the sample.
\end{enumerate}

\subsubsection{Object Detection}
We construct the integrity constraint mentioned in Section \ref{sec:case_study_wilds} with a few queries. We begin by finding the difference in pixels over consecutive frames:

\begin{center}
\begin{minipage}{0.8\textwidth}
\begin{mdframed}
\begin{minted}[fontsize=\footnotesize, breaklines]{Python}
db = Database()
db.register_dataset(wilds_train_data, 'wilds_train')

image_var = Query('image_var', base='wilds_train')
        .group_by(lambda img, label, md: md[1].item())
        .project(lambda seqid, rows: [(seqid, i-1, rows[i-1][0],
                rows[i-1][1], rows[i-1][2], i, rows[i][0], rows[i-1][1],
                rows[i][2], img_diff(rows[i-1][0], rows[i][0], 0.1, 0.15)
            ) for i in range(1, len(rows))])
        .flatten()
image_var(db)
\end{minted}    
\end{mdframed}
\end{minipage}
\end{center}

We write the query over the \texttt{wilds\_train} table, where each entry contains a frame, its label, and any associated metadata.
We first group each entry by the sequence ID of the frame, and then project out the relevant fields of consecutive frames along with their pixel difference calculated by the \texttt{img\_diff} function.

We then plot a histogram of these differences for frames labeled empty and those labeled non-empty.
This histogram, along with some trial-and-error, lets us determine an appropriate threshold (0.13) and write the following query to detect empty, but mislabeled, frames:

\begin{center}
\begin{minipage}{0.7\textwidth}
\begin{mdframed}
\begin{minted}[fontsize=\footnotesize, breaklines]{Python}
def mean_in_threshold(*row):
    diff = row[-1]
    mean = get_non_zero_mean(diff).item()
    
    return mean <= 0.13 and mean > 0

non_empty = Query('image_var_nonempty', base='image_var')
        .filter(lambda *args: not (args[3] == 0 and args[7] == 0))
non_empty(db)

mislabeled = Query('mislabeled', base='image_var_nonempty')
        .filter(mean_in_threshold)
mislabeled(db)
\end{minted}
\end{mdframed}
\end{minipage}
\end{center}

We look for frames where the mean of pixel differences is greater than 0 (to avoid instances where the consecutive frames are simply duplicates) and not more than 0.13.

We provide all the \tool{} queries (including the above query) used for the Object Detection task in a python notebook \url{"examples/self_driving_examples/time_series_all_queries.ipynb"} in the submitted code. Due to the substantial size and licensing constraints of the \cityscapes{} dataset, we do not include it along with the submitted code. But you can download it from \url{https://www.cityscapes-dataset.com/}.

\subsubsection{Data Imputation}
We provide all the \tool{} queries (including the intermediate ones throughout the explorations) used for the Data imputation task in a python notebook \url{"examples/time\_series\_explorations/time\_series\_all\_queries.ipynb"} in the submitted code. We uploaded the data needed for this notebook so that this notebook is runnable.

\subsubsection{Image Classification}
We provide all the \tool{} queries (including the intermediate ones throughout the explorations) used for the Image Classification task in two python notebooks: \url{"examples/wilds_examples/wilds_baseline_comparison.ipynb"} (for I-Q1) and \url{"examples/wilds_examples/wilds_empty_frames.ipynb"} (for I-Q2).
These two notebooks can automatically download the data that is needed for the experiments.

\subsubsection{Text Generation}
We provide all the \tool{} queries (including the intermediate ones throughout the explorations) used for the Text Generation task in a python notebook \url{"examples/bias_examples/bias_queries.ipynb"} in the submitted code. This notebook can automatically load the input data from \url{"alpaca/"} folder for experiments.

\subsubsection{Natural Language Reasoning}
We provide all the \tool{} queries (including the intermediate ones throughout the explorations) used for the Text Generation task in a Python notebook \url{"examples/lmql_examples/llm.ipynb"} in the submitted code. This notebook can automatically download the model and data for experiments.

The L-Q2 query is also provided below, but it is also found in uncompressed form in the above Python notebook.
\begin{center}
\begin{minipage}{0.7\textwidth}
\begin{mdframed}
\begin{minted}[fontsize=\footnotesize, breaklines]{Python}
prompt = """Q: 2015 is coming in 36 hours. ...
A: [If 2015 is ...]. So the answer is [01/05/2015].

{question} [[reasoning]]. So the answer is [[answer]]"""

cot = c_llm('date-lmql', 'date', prompt,
    lambda args: len(args['reasoning']) > 10 and re.match(r"\d\d/\d\d/\d\d\d\d", args['answer']),
    lambda gens: gens[0],
    k=10)
\end{minted}
\end{mdframed}
\end{minipage}
\end{center}

\subsection{Additional Details for Quantitative Evaluations}\label{app:exp-details}
\subsubsection{Compute Resources}
For each of the following experiments, we use a local server with four Nvidia 2080 Ti GPUs and 80 Intel Xeon Gold 6248 CPUs. We only use the GPUs for evaluating the object detection model used in the experiments in Appendix~\ref{appendix: obj_detection} and for evaluating the T5 model for the experiments in Appendix~\ref{app:bias-details}.

\subsubsection{Object Detection}\label{appendix: obj_detection}
As mentioned in Section \ref{sec:benchmarks}, we use the Cityscapes training set, which is composed of 15K dash-cam frames in total. We use a subset of 33 video sequences consisting of 1000 frames for the experiments. In total, we wrote 57 queries for debugging the object detection model. These queries ranged from queries to find dangerous situations to queries that processed the data for use by other queries. All 57 queries are available in \verb|examples/self_driving_explore.ipynb| in the supplementary materials. We comprehensively evaluate two of these queries, S-Q1 and S-Q2, using \tool{} and other systems (Pandas, ArangoDB, and Python). The detailed implementations of S-Q1 and S-Q2 are included in Figure \ref{fig:s-q1-full} and Figure \ref{fig:s-q2-full} respectively. In what follows, we briefly explain how these two queries are constructed with the operators introduced in Section \ref{sec:operator_semantics}.

\paragraph{S-Q1} As Figure \ref{fig:s-q1-mdb} shows, this query takes the output of the OneFormer model \citep{jain2022oneformer}, represented by a named table ``preds'', as the input to this query. Note that we use one additional ``register'' operator to set up this table.  Each row of the named table ``preds'' carries four fields, including the video sequence ID, the frame ID within each video sequence, the semantic segmentation result, and the object detection result. At the beginning of this query, we perform two ``join'' operations over three copies of ``out'' with join keys specified by three functions, ``fid\_plus1'', ``fid'', and ``fid\_plus2'', which results in a table where each row consists of three continuous frames from the same video sequence. Then we apply a function to each row in this table with the ``project'' operator and invoke two user-defined functions ``match\_three\_bboxes'' and ``zip\_adj\_preds'', which ends up with a list of matched bounding boxes corresponding to the same object across three contiguous frames. Each row in the result so far contains a list of matched bounding boxes which is not convenient for further analysis. We therefore perform a ``flatten'' operation to split each row into multiple rows where each one corresponds to one object with matched bounding boxes across three frames. In the end, the ``filter'' operator is applied to select the objects with the same predicted labels in the first two frames and no detection in the third frame. The final result is stored in a table named ``temporal\_consistency''.

\paragraph{S-Q2} As Figure \ref{fig:s-q2-mdb} shows, this query joins the output of S-Q1 with itself using the ``join'' operator with the sequence ID and frame ID (denoted by ``row[0]'' and ``row[1][0]'' resp.) as the join key to obtain all pairs of objects within the same frame. From the resulting pairs of objects, we identify those that are both predicted as vehicles in two contiguous frames by invoking one user-defined function ``is\_vehicle'' in a filter operator. In this filter operation, ``row[2][0]'' and ``row[2][1]'' represent the predicted labels of the first object in two contiguous frames while ``row[6][0]'' and ``row[6][1]'' represent the predicted labels of the second object in the same two frames. In the subsequent ``project'' operation, we estimate the speed of those pairs of vehicles by leveraging the differences of their bounding box positions by using a user-defined function ``'get\_velo', in which ``row[3][0]'', ``row[3][1]'' represents the bounding boxes of the first object in two contiguous frames while ``row[7][0]'', ``row[7][1]'' represents the bounding boxes of the first object in the same two frames.

\begin{figure}
\begin{subfigure}[b]{\textwidth}
\begin{mdframed}
\begin{minted}[fontsize=\fontsize{5}{6}\selectfont, breaklines]{Python}
Query('temp_consistency', base='preds')
    .join('preds', key=fid_plus1, fkey=fid)
    .join('preds', key=fid_plus2, fkey=fid) # join three continuous frames
    .project(match_three_bboxes) # applying "match_three_bboxes"
    .project(zip_adj_preds) # applying "zip_adj_preds"
    .flatten() # flatten
    .filter(lambda sid, frames, lb, bx: lb[0] == lb[1] and lb[2] == "No Match" and in_center(bx[0]))
\end{minted}    
\end{mdframed}
\caption{S-Q1 in \tool{}}
\label{fig:s-q1-mdb}
\end{subfigure}
\hfill
\begin{subfigure}[b]{\textwidth}
\begin{mdframed}
\begin{minted}[fontsize=\fontsize{5}{6}\selectfont, breaklines]{Python}
output_preds_df = pd.DataFrame(db.tables["preds"], columns=['seq_id', 'frame_id', 'pred_labels', 'pred_bboxes'])
# join three continuous frames
output_preds_df['frame_id_p1'] = fid_plus1(output_preds_df['frame_id'])
output_preds_df['frame_id_p2'] = fid_plus2(output_preds_df['frame_id'])
output_preds_df = output_preds_df.merge(output_preds_df, left_on=['seq_id', 'frame_id_p1'], right_on=['seq_id', 'frame_id'], how='left', suffixes=('', '_2')) \
.merge(output_preds_df, left_on=['seq_id', 'frame_id_p2'], right_on=['seq_id', 'frame_id'], how='left', suffixes=('', '_3')).dropna()
# applying "match_three_bboxes"
output_preds_df[['labels', 'boxes']] = output_preds_df.apply(lambda row: match_three_adj_bboxes(row['pred_bboxes'], row['pred_bboxes_2']
row['pred_bboxes_3'], row['pred_labels'], row['pred_labels_2'], row['pred_labels_3']), axis=1,result_type='expand')
# applying "zip_adj_preds"
output_preds_df['frames'] = output_preds_df.apply(lambda row: [row['frame_id'], row['frame_id_2'], row['frame_id_3']], axis=1)
df_l = output_preds_df.apply(lambda row: list(zip(*zip_adj_preds(row['seq_id'], row['frames'], row['labels'], row['boxes'])[0])), axis=1
result_type='expand')
# flatten
output_preds_df['frames'] = output_preds_df['frames'].explode([0, 1, 2, 3])
# filtering
df_l.rename(columns={0: 'seq_id', 1: 'frames', 2: 'labels', 3: 'boxes'}, inplace=True)
df_c = pd.DataFrame(df_l['labels'].to_list())
df_b = pd.DataFrame(df_l['boxes'].to_list())
df_l = df_l.reset_index()
df_inconsistencies = df_l[(df_c[0] == df_c[1]) & (df_c[2] == "No Match") & (df_b[0].apply(is_center))]
\end{minted}
\end{mdframed}
\caption{S-Q1 in Pandas}
\label{fig:s-q1-pandas}
\end{subfigure}
\hfill
\begin{subfigure}[b]{\textwidth}
\begin{mdframed}
\begin{minted}[fontsize=\fontsize{5}{6}\selectfont, breaklines]{Python}
key1_mappings, key2_mappings, key3_mappings = dict(), dict(), dict()
# build join index
for row in list(db.tables["preds"]):
    key1, key2 = fid(*row), fid_plus1(*row)
    if not key1 in key1_mappings:
        key1_mappings[key1] = []
    key1_mappings[key1].append(row)
    if not key2 in key2_mappings:
        key2_mappings[key2] = []
    key2_mappings[key2].append(row)
# join first two frames
for key1 in key1_mappings:
    if key1 in key2_mappings:
        for row1 in key1_mappings[key1]:
            for row2 in key2_mappings[key1]:
                seq_id1, frame_id1, labels1, bboxes1 = row1
                seq_id2, frame_id2, labels2, bboxes2 = row2
                row = [seq_id1, [frame_id1, frame_id2], [labels1, labels2], [bboxes1, bboxes2]]
                key3 = fid_plus2(*row)
                if not key3 in key3_mappings:
                    key3_mappings[key3] = []
                    key3_mappings[key3].append(row)
res_ls = []
# join with the third frames
for key3 in key3_mappings:
    if key3 in key2_mappings:
        for row1 in key3_mappings[key3]:
            for row2 in key2_mappings[key3]:
                seq_id1, frames, labels, bboxes = row1
                seq_id3, frame_id3, labels3, bboxes3 = row2
                frames = frames + [frame_id3]
                labels = labels + [labels3]
                bboxes = bboxes + [bboxes3]
                # apply match_three_adj_bboxes and zip_adj_preds
                curr_res_ls=zip_adj_preds(seq_id1, frames, *match_three_adj_bboxes(*bboxes, *labels))
                for sub_ls in curr_res_ls:
                    for item in sub_ls:
                        seq_id, frames, labels, boxes = item
                        # filtering
                        if labels[0] == labels[1] and labels[2] == "No Match" and is_center(boxes[0]):
                            res_ls.append(item)
\end{minted}    
\end{mdframed}
\caption{S-Q1 in Python}
\label{fig:s-q1-python}
\end{subfigure}
\caption{Implementation of S-Q1 in different query systems}\label{fig:s-q1-full}
\vspace{-0.15in}
\end{figure}

\begin{figure}
\begin{subfigure}[b]{\textwidth}
\begin{mdframed}
\begin{minted}[fontsize=\fontsize{5}{6}\selectfont, breaklines]{Python}
Query('paired_velocity', base='three_adj_matches')
    .join('three_adj_matches', key=lambda idx, *row: (row[0], row[1][0]), fkey=lambda idx, *row: (row[0], row[1][0])) # join two copies of flat_matches
    .filter(lambda row: (is_vehicle(row[2][0]) and is_vehicle(row[2][1]) and is_vehicle(row[6][0]) and is_vehicle(row[6][1])
    and tuple(row[3][0]) != tuple(row[7][0]))) # filtering
    .project(lambda row: [seq_id, [row[1][0], row[1][1]], row[3][0], row[3][1], row[7][0], row[7][1], get_velo(row[3][0], row[3][1]), get_velo(row[7][0], row[7][1])]) # projection
\end{minted}    
\end{mdframed}
\caption{S-Q2 in \tool{}}
\label{fig:s-q2-mdb}
\end{subfigure}
\hfill
\begin{subfigure}[b]{\textwidth}
\begin{mdframed}
\begin{minted}[fontsize=\fontsize{5}{6}\selectfont, breaklines]{Python}
flat_matches_df = pd.DataFrame(list(db.tables["three_adj_matches"]), columns=["seq_id", "frames", "labels", "boxes"])
#join two copies of flat_matches
flat_matches_df["meta_0"] = flat_matches_df["frames"].apply(lambda item: item[0])
join_flat_matches_df = flat_matches_df.merge(flat_matches_df, left_on=["seq_id", "meta_0"], right_on=["seq_id", "meta_0"], suffixes=
("_1", "_2"))
#filtering
selected_boolean = list(join_flat_matches_df.apply(lambda row: (is_vehicle(row["labels_1"][0]) and is_vehicle(row["labels_1"][1]) and
is_vehicle(row["labels_2"][0]) and is_vehicle(row["labels_2"][1])
and tuple(row["boxes_1"][0]) != tuple(row["boxes_2"][0])),axis=1))
#projection
selected_join_flat_matches_df = join_flat_matches_df[selected_boolean]
selected_join_flat_matches_df["res"] = selected_join_flat_matches_df.apply(lambda row:(row["seq_id"], [row["frames_1"][0],
row["frames_1"][1]], row["boxes_1"][0], row["boxes_1"][1], row["boxes_2"][0], row["boxes_2"][1], get_velo(row["boxes_1"][0],
row["boxes_1"][1]), get_velo(row["boxes_2"][0], row["boxes_2"][1])), axis = 1)
\end{minted}
\end{mdframed}
\caption{S-Q2 in Pandas}
\label{fig:s-q2-pandas}
\end{subfigure}
\hfill
\begin{subfigure}[b]{\textwidth}
\begin{mdframed}
\begin{minted}[fontsize=\fontsize{5}{6}\selectfont, breaklines]{Python}
res_ls = []
# build join index
for idx in range(len(db.tables["three_adj_matches"])):
    seq_id, frames, labels, boxes = flat_matches_ls[idx]
    key = (seq_id, frames[0])
    if key not in indexed_flat_matches:
        indexed_flat_matches[key] = []
        indexed_flat_matches[key].append(flat_matches_ls[idx])
# join starts
for key in indexed_flat_matches:
    val_ls = indexed_flat_matches[key]
    for item1 in val_ls:
        for item2 in val_ls:
            seq_id, frames1, labels1, boxes1 = item1
            _, frames2, labels2, boxes2 = item2
            if is_vehicle(labels1[0]) and is_vehicle(labels1[1]) and is_vehicle(labels2[0]) and is_vehicle(labels2[1]) and tuple(boxes1[0]) != tuple(boxes2[0]):
                res_ls.append([seq_id, [frames1[0], frames1[1]], boxes1[0], boxes1[1], boxes2[0], boxes2[1], get_velo(boxes1[0], boxes1[1], get_velo(boxes2[0], boxes2[1])])
# join ends
\end{minted}    
\end{mdframed}
\caption{S-Q2 in Python}
\label{fig:s-q2-python}
\end{subfigure}
\begin{subfigure}[b]{\textwidth}
\begin{mdframed}
\begin{minted}[fontsize=\fontsize{5}{6}\selectfont, breaklines]{Python}
adb = Arango_db()
adb.init()
adb.register_dataset_dir(db.tables["three_adj_matched"], "three_adj_matched", ["idx", "seq_id", "frames", "labels", "boxes"])
query = (" For item1 in three_adj_matched "
+ " For item2 in three_adj_matched "
+ " Filter item1.seq_id == item2.seq_id and item1.frames[0] == item2.frames[0]" # join two copies of flat_matches
+ " and item1.labels[0] IN ['car','truck','bus', 'motorcycle', 'bicycle']"
+ " and item2.labels[0] IN ['car','truck','bus', 'motorcycle', 'bicycle']"
+ " and item1.idx != item2.idx" # filtering
# projection
+ " return {'seq_id':item1.seq_id, 'frames':[item1.frames[0], item1.frames[1]], 'boxes':[item1.boxes[0], item1.boxes[1], item2.boxes[0], item2.boxes[
+ " 'velocity1': [(item1.boxes[0][0] + item1.boxes[0][2])/2 - (item1.boxes[1][0] + item1.boxes[1][2])/2, (item1.boxes[0][1] + item1.boxes[0][3])/2 -
(item1.boxes[1][1] + item1.boxes[1][3])/2],"
+ " 'velocity2': [(item2.boxes[0][0] + item2.boxes[0][2])/2 - (item2.boxes[1][0] + item2.boxes[1][2])/2, (item2.boxes[0][1] + item2.boxes[0][3])/2 -
(item2.boxes[1][1] + item2.boxes[1][3])/2] }")
query_res = adb.execute_query(query)
\end{minted}    
\end{mdframed}
\caption{S-Q2 in ArangoDB}
\label{fig:s-q2-adb}
\end{subfigure}
\caption{Implementation of S-Q2 in different query systems}\label{fig:s-q2-full}
\vspace{-0.15in}
\end{figure}

\subsubsection{Data Imputation}
As mentioned in Section \ref{sec:benchmarks}, the \physionet\ dataset \cite{silva2012predicting} is used, which is composed of de-identified patients' records in Intensive Care Units (ICU). By reusing the pre-process scripts from \cite{tashiro2021csdi}, we obtain 4000 time-series samples, each of which contains 35 features collected from 48 time stamps. 

However, note that those records are collected sporadically, which ends up with highly sparse medical time series sequences with a substantial number of missing entries. Various neural networks have been proposed to impute those missing entries to construct seemingly complete time series for the down-stream tasks, which could violate domain knowledge or intuitions, like the smoothness assumption on typical medical time series data \cite{you2018time}. We therefore proposed T-Q1, T-Q2 and T-Q3 to capture such bugs in the imputation results. The details of the implementations of those three queries in different query systems are included in Figure \ref{fig:T-q1-full}-Figure \ref{fig:T-q3-full}. Similar to Appendix \ref{appendix: obj_detection}, we also describe how to construct these three queries as follows.

\begin{figure}
\begin{subfigure}[b]{\textwidth}
\begin{mdframed}
\begin{minted}[fontsize=\fontsize{5}{6}\selectfont, breaklines]{Python}
db=Database()
db.register_dataset(torch.permute(train_valid_samples, (2,1,0)),
"train_valid_samples")
db.register_dataset(torch.permute(mask, (2,1,0)), "mask")
(Query("consistency_bound_mask", base="train_valid_samples") # register
    .join("mask") # join
    .project(compute_gap_res0) # compute threshold
    .project(aggregate_rows0)).run(db)
\end{minted}    
\end{mdframed}
\caption{T-Q1 in \tool{}}
\label{fig:T-q1-mdb}
\end{subfigure}
\hfill
\begin{subfigure}[b]{\textwidth}
\begin{mdframed}
\begin{minted}[fontsize=\fontsize{5}{6}\selectfont, breaklines]{Python}
train_valid_data = train_valid_samples.reshape(-1, train_valid_samples.shape[2])
train_valid_masks = mask.reshape(-1, mask.shape[2])
midx = pd.MultiIndex.from_product([np.arange(train_valid_samples.shape[0]),
np.arange(train_valid_samples.shape[1])])
train_valid_data_pd = pd.DataFrame(data = train_valid_data.numpy(), index=midx, columns=["feat_" + str(idx) for idx in range(train_valid_samples.shape[2])])
train_valid_mask_pd = pd.DataFrame(data = mask.numpy(), index=midx, columns=["feat_m_" + str(idx) for idx in range(train_valid_samples.shape[2])])
gap_res_all_feat = [] # register
join_res = train_valid_data_pd.merge(train_valid_mask_pd, left_index=True, right_index=True) #join
for feat_id in range(train_valid_samples.shape[2]):
    data = join_res["feat_"+ str(feat_id)]
    mask = join_res["feat_m_"+ str(feat_id)]
    curr_selected_res_gaps = compute_gap_res0_numpy(data, mask) # compute threshold
    gap_res_for_curr_feat = aggregate_rows0(curr_selected_res_gaps) # compute statistics of thresholds
    gap_res_all_feat.append(gap_res_for_curr_feat)
\end{minted}
\end{mdframed}
\caption{T-Q1 in Pandas}
\label{fig:T-q1-pandas}
\end{subfigure}
\hfill
\begin{subfigure}[b]{\textwidth}
\begin{mdframed}
\begin{minted}[fontsize=\fontsize{5}{6}\selectfont, breaklines]{Python}
ratio_range_max_by_attr = [] # register
for feat_id in range(train_valid_samples.shape[-1]): # iterate over each feature
    curr_data, curr_mask = torch.t(train_valid_samples[:, :, feat_id]), torch.t(mask[:,:,feat_id])
    gap_value = compute_gap_res0(curr_data, curr_mask) # compute threshold
    ratio_range_max_by_attr.append(aggregate_rows0(gap_value)) # compute statistics of threshold
\end{minted}    
\end{mdframed}
\caption{T-Q1 in Python}
\label{fig:T-q1-python}
\end{subfigure}
\hfill
\begin{subfigure}[b]{\textwidth}
\begin{mdframed}
\begin{minted}[fontsize=\fontsize{5}{6}\selectfont, breaklines]{Python}
ratio_range_max_by_attr = [] # register
for feat_id in range(train_valid_values.shape[-1]): # iterate over each feature
    curr_data, curr_mask = np.transpose(train_valid_values[:, :, feat_id]), np.transpose(train_valid_masks[:,:, feat_id]) 
    gap_value = compute_gap_res0_numpy(curr_data, curr_mask)  # compute threshold
    ratio_range_max_by_attr.append(aggregate_rows0(gap_value))  # compute statistics of threshold
\end{minted}    
\end{mdframed}
\caption{T-Q1 in Numpy}
\label{fig:T-q1-numpy}
\end{subfigure}
\begin{subfigure}[b]{\textwidth}
\begin{mdframed}
\begin{minted}[fontsize=\fontsize{5}{6}\selectfont, breaklines]{Python}
db = Arango_db()
db.init()
db.register_dataset_3D(train_valid_samples.permute((2,0,1)), "train_valid_samples")
reshaped_mask = mask.permute(2,0,1)
reshaped_mask_ids = [[reshaped_mask[idx][sub_idx].nonzero().view(-1) for sub_idx in
range(len(reshaped_mask[idx]))] for idx in range(len(reshaped_mask))]
db.register_dataset_3D(reshaped_mask_ids, "mask")
query = ("Let mask_ids = (For item in mask"
+ " For sub_item in item.data"
+ " For idx in sub_item.sub_idx"
+ " Let sub_sub_item = slice(sub_item.data, idx, 3)"
+ " FILTER sub_sub_item[0] == sub_sub_item[1] - 1 and sub_sub_item[1] == sub_sub_item[2] - 1"
# equivalent to "mask = torch.logical_and(torch.logical_and(mask, mask_prev), mask_next)"
+ " return {'id': item.id, 'second_id': sub_item.id, 'sub_idx': idx, 'data': sub_sub_item})"
+ " Let gap_res = (For sub_item in mask_ids"
+ " For sample in train_valid_samples"
+ " For sub_sample in sample.data"
+ " FILTER sub_item.id == sample.id and sub_sample.id == sub_item.second_id" #join
# compute threshold
+ " Let gap = MAX([ABS(sub_sample.data[sub_item.data[0]] - sub_sample.data[sub_item.data[1]]) -
ABS(sub_sample.data[sub_item.data[0]] - sub_sample.data[sub_item.data[2]]),
ABS(sub_sample.data[sub_item.data[2]] - sub_sample.data[sub_item.data[1]]) -
ABS(sub_sample.data[sub_item.data[0]] - sub_sample.data[sub_item.data[2]])]) " \
+ " return {'id': sub_item.id, 'sub_id': sub_item.second_id, 'data': gap})"
# compute statistics of thresholds
+ " For item in gap_res collect id = item.id into groups=item.data return {'id': id, 'sub_id':'sub_id',
'group': groups}")
query_result = db.execute_query(query)
\end{minted}    
\end{mdframed}
\caption{T-Q1 in ArangoDB}
\label{fig:T-q1-adb}
\end{subfigure}
\caption{Implementation of T-Q1 in different query systems}\label{fig:T-q1-full}
\vspace{-0.15in}
\end{figure}

\begin{figure}
\begin{subfigure}[b]{\textwidth}
\begin{mdframed}
\begin{minted}[fontsize=\fontsize{5}{6}\selectfont, breaklines]{Python}
db = Database()
db.register_dataset(test_imputed_x, "test_imputed_x")
db.register_dataset(test_dataset, "test_dataset")
# Query starts from here
res = (Query("test_imputation", base="test_dataset")
.join("test_imputed_x") # join
.project(lambda observed_values, observed_mask, _, gt_mask, test_imputed_x: [list(zip(*[torch.t(test_imputed_x), torch.t(gt_mask)]))])
.flatten() # flatten operation so that we can iterate over each feature
.project(check_sequence_feat_range_on_time_series0_single_feat)).run(db) # compute threshold and determine violations
\end{minted}    
\end{mdframed}
\caption{T-Q2 in \tool{}}
\label{fig:T-q2-mdb}
\end{subfigure}
\hfill
\begin{subfigure}[b]{\textwidth}
\begin{mdframed}
\begin{minted}[fontsize=\fontsize{5}{6}\selectfont, breaklines]{Python}
test_data = test_imputed_x.reshape(-1, test_imputed_x.shape[2])
test_gt_masks = test_gt_masks.reshape(-1, test_gt_masks.shape[2])
midx = pd.MultiIndex.from_product([np.arange(test_imputed_x.shape[0]), np.arange(test_imputed_x.shape[1])])
flatten_test_imputed_x_pd = pd.DataFrame(data = test_data.numpy(), index=midx, columns=["feat_" + str(idx) for idx in range(test_data.shape[1])])
flatten_test_mask_pd = pd.DataFrame(data = test_gt_masks.numpy(), index=midx, columns=["feat_m_" + str(idx) for idx in range(test_data.shape[1])])
# Query starts from here
final_res = []
flatten_test_imputed_x_pd = flatten_test_imputed_x_pd.merge(flatten_test_mask_pd, left_index=True, right_index=True) # join
# One additional step to collect values along time dimensions
grouped_test_imputed_x_pd = flatten_test_imputed_x_pd.reset_index().groupby("level_0").aggregate(col_agg_mappings)
# Iterate over each feature
for sample_idx in range(len(grouped_test_imputed_x_pd)):
    curr_res_ls = []
    for feat_idx in range(test_imputed_x.shape[2]):
        # compute threshold and determine violations
        curr_res = check_sequence_feat_range_on_time_series0_single_feat(torch.tensor(grouped_test_imputed_x_pd.loc[sample_idx, "feat_" + str(feat_idx)]), torch.tensor(grouped_test_imputed_x_pd.loc[sample_idx, "feat_m_" + str(feat_idx)]))
        curr_res_ls.append(curr_res)
    final_res.append(torch.stack(curr_res_ls, dim=1))
final_res = torch.stack(list(final_res))
\end{minted}
\end{mdframed}
\caption{T-Q2 in Pandas}
\label{fig:T-q2-pandas}
\end{subfigure}
\hfill
\begin{subfigure}[b]{\textwidth}
\begin{mdframed}
\begin{minted}[fontsize=\fontsize{5}{6}\selectfont, breaklines]{Python}
full_error_mask = []
for idx in range(len(test_dataset)): # Iterate over each sample
    observed_values, observed_mask, _, gt_mask = test_dataset[idx]
    curr_imputed_x = test_imputed_x[idx] # Join test_dataset and imputed_x at each sample
    curr_error_mask_ls = []
    for feat_id in range(observed_values.shape[-1]): # Iterate over each feature
        curr_error_mask = check_sequence_feat_range_on_time_series0_single_feat(curr_imputed_x[:,feat_id], gt_mask[:,feat_id]) # compute threshold and determine violations
        curr_error_mask_ls.append(curr_error_mask)
    full_error_mask.append(torch.stack(curr_error_mask_ls, dim=1))
full_error_mask = torch.stack(full_error_mask)
\end{minted}    
\end{mdframed}
\caption{T-Q2 in Python}
\label{fig:T-q2-python}
\end{subfigure}
\hfill
\begin{subfigure}[b]{\textwidth}
\begin{mdframed}
\begin{minted}[fontsize=\fontsize{5}{6}\selectfont, breaklines]{Python}
full_error_mask = [] # register
for idx in range(len(test_dataset)): # Iterate over each sample
    observed_values, observed_mask, _, gt_mask = test_dataset[idx]
    curr_imputed_x = test_imputed_x[idx]  # Join test_dataset and imputed_x at each sample
    curr_error_mask_ls = []
    for feat_id in range(observed_values.shape[-1]): # Iterate over each feature
        curr_error_mask = check_sequence_feat_range_on_time_series0_single_feat_numpy(curr_imputed_x[:,feat_id], gt_mask[:,feat_id]) # compute threshold and determine violation
        curr_error_mask_ls.append(curr_error_mask)
    full_error_mask.append(np.stack(curr_error_mask_ls, axis=1))
full_error_mask = np.stack(full_error_mask)
\end{minted}    
\end{mdframed}
\caption{T-Q2 in Numpy}
\label{fig:T-q2-numpy}
\end{subfigure}
\begin{subfigure}[b]{\textwidth}
\begin{mdframed}
\begin{minted}[fontsize=\fontsize{5}{6}\selectfont, breaklines]{Python}
def convert_query_res_to_tensors(query_res, data):
    res_mask = torch.zeros_like(data)
    for idx in range(len(list(query_res))):
        id1 = list(query_res)[idx]["id"]
        id2 = list(query_res)[idx]["sub_id"]
        grouped_data = list(query_res)[idx]["group_data"]
        res_mask[id1, id2] = check_sequence_feat_range_on_time_series0_single_feat_adb(data[id1,id2], grouped_data)
    return res_mask.permute(0,2,1)
db = Arango_db()
db.init()
db.register_dataset_3D(test_imputed_x.permute(0,2,1), "test_imputed_x")
reshaped_mask = test_gt_masks.permute(0,2,1)
reshaped_mask_ids = [[reshaped_mask[idx][sub_idx].nonzero().view(-1) for sub_idx in range(len(reshaped_mask[idx]))] for idx in
range(len(reshaped_mask))]
db.register_dataset_3D(reshaped_mask_ids, "test_gt_masks")
# Query starts from here
query= ("Let gt_val_array = (For val in test_imputed_x"
+ " For sub_val in val.data For mask in test_gt_masks For sub_mask in mask.data Filter val.id == mask.id and sub_val.id == sub_mask.id" # join
+ " For mask_id in sub_mask.data return {'id': val.id, 'sub_id': sub_val.id, 'data': sub_val.data[mask_id]})" # find out all the existing ground-truths
+ " For item in gt_val_array collect id = item.id, sub_id = item.sub_id into groups=item.data return {'id': id, 'sub_id': sub_id, 'group_data': groups}")
query_result= db.execute_query(query)
final_res = convert_query_res_to_tensors(query_result, test_imputed_x.permute(0,2,1)) # additional python script for computing thresholds a
determine violations
\end{minted}    
\end{mdframed}
\caption{T-Q2 in ArangoDB}
\label{fig:T-q2-adb}
\end{subfigure}
\caption{Implementation of T-Q2 in different query systems}\label{fig:T-q2-full}
\vspace{-0.15in}
\end{figure}

\begin{figure}
\begin{subfigure}[b]{\textwidth}
\begin{mdframed}
\begin{minted}[fontsize=\fontsize{5}{6}\selectfont, breaklines]{Python}
res = (Query("sliding_window_error_mask", base='test_imputed_x') # register
# unfold each sample to get subsequences of a fixed length, compute threshold and determine violations
.project(lambda x: compare_local_sliding_window_bound0(x.unfold(0, width, 1), test_imputed_x.shape[1],
test_imputed_x.shape[2]))).run(db)
\end{minted}    
\end{mdframed}
\caption{T-Q3 in \tool{}}
\label{fig:T-q3-mdb}
\end{subfigure}
\hfill
\begin{subfigure}[b]{\textwidth}
\begin{mdframed}
\begin{minted}[fontsize=\fontsize{5}{6}\selectfont, breaklines]{Python}
test_data = test_imputed_x.reshape(-1, test_imputed_x.shape[2])
test_gt_masks = test_gt_masks.reshape(-1, test_gt_masks.shape[2])
midx = pd.MultiIndex.from_product([np.arange(test_imputed_x.shape[0]), np.arange(test_imputed_x.shape[1])])
flatten_test_imputed_x_pd = pd.DataFrame(data = test_data.numpy(), index=midx, columns=["feat_" + str(idx) for idx in range(test_data.shape[1])])
flatten_test_mask_pd = pd.DataFrame(data = test_gt_masks.numpy(), index=midx, columns=["feat_m_" + str(idx) for idx in range(test_data.shape[1])])
# Query starts from here
final_res = []
flatten_test_imputed_x_pd = flatten_test_imputed_x_pd.merge(flatten_test_mask_pd, left_index=True, right_index=True) # join
# One additional step to collect values along time dimensions
grouped_test_imputed_x_pd = flatten_test_imputed_x_pd.reset_index().groupby("level_0").aggregate(col_agg_mappings)
# Iterate over each feature
for sample_idx in range(len(grouped_test_imputed_x_pd)):
    curr_res_ls = []
    for feat_idx in range(test_imputed_x.shape[2]):
        # compute threshold and determine violations
        curr_res = check_sequence_feat_range_on_time_series0_single_feat(torch.tensor(grouped_test_imputed_x_pd.loc[sample_idx, "feat_" + str(feat_idx)]), torch.tensor(grouped_test_imputed_x_pd.loc[sample_idx, "feat_m_" + str(feat_idx)]))
        curr_res_ls.append(curr_res)
    final_res.append(torch.stack(curr_res_ls, dim=1))
final_res = torch.stack(list(final_res))
\end{minted}
\end{mdframed}
\caption{T-Q3 in Pandas}
\label{fig:T-q3-pandas}
\end{subfigure}
\hfill
\begin{subfigure}[b]{\textwidth}
\begin{mdframed}
\begin{minted}[fontsize=\fontsize{5}{6}\selectfont, breaklines]{Python}
full_error_mask= [] # register
for idx in range(len(test_imputed_x)): # Iterate over each sample
    # unfold each sample to get subsequences of a fixed length
    unfolded_imputed_x = test_imputed_x[idx].unfold(0, width, 1)
    # compute threshold and determine violations
    full_error_mask.append(compare_local_sliding_window_bound0(unfolded_imputed_x, test_imputed_x.shape[1], test_imputed_x.shape[2]))
full_error_mask = torch.stack(list(full_error_mask)).reshape(test_imputed_x.shape[0], -1, test_imputed_x.shape[1]).permute((0,2,1))
\end{minted}    
\end{mdframed}
\caption{T-Q3 in Python}
\label{fig:T-q3-python}
\end{subfigure}
\begin{subfigure}[b]{\textwidth}
\begin{mdframed}
\begin{minted}[fontsize=\fontsize{5}{6}\selectfont, breaklines]{Python}
full_error_mask = []
for idx in range(len(test_imputed_x)): # Iterate over each sample
    # unfold each sample to get subsequences of a fixed length
    # unfolded_imputed_x = test_imputed_x[idx].unfold(0, width, 1)
    unfolded_imputed_x = np.lib.stride_tricks.as_strided(test_imputed_x[idx], shape=(test_imputed_x[idx].shape[0] - width + 1, width, test_imputed_x[idx].shape[1]), strides=(test_imputed_x[idx].strides[0],) + test_imputed_x[idx].strides)
    unfolded_imputed_x = np.transpose(unfolded_imputed_x, (0, 2,1))
    # compute threshold and determine violations
    full_error_mask.append(compare_local_sliding_window_bound0_numpy(unfolded_imputed_x, test_imputed_x.shape[1], test_imputed_x.shape[2]))
full_error_mask = np.transpose(np.stack(list(full_error_mask)).reshape(test_imputed_x.shape[0], -1, test_imputed_x.shape[1]), (0,2,1))

\end{minted}    
\end{mdframed}
\caption{T-Q3 in Numpy}
\label{fig:T-q3-numpy}
\end{subfigure}
\begin{subfigure}[b]{\textwidth}
\begin{mdframed}
\begin{minted}[fontsize=\fontsize{5}{6}\selectfont, breaklines]{Python}
def convert_query_res_to_tensors(query_res, data):
    res_mask = torch.zeros_like(data)
    for idx in range(len(list(query_res))):
        id1 = list(query_res)[idx]["id"]
        id2 = list(query_res)[idx]["sub_id"]
        grouped_data = list(query_res)[idx]["group_data"]
        res_mask[id1, id2] = check_sequence_feat_range_on_time_series0_single_feat_adb(data[id1,id2], grouped_data)
    return res_mask.permute(0,2,1)
db = Arango_db()
db.init()
db.register_dataset_3D(test_imputed_x.permute(0,2,1), "test_imputed_x")
reshaped_mask = test_gt_masks.permute(0,2,1)
reshaped_mask_ids = [[reshaped_mask[idx][sub_idx].nonzero().view(-1) for sub_idx in range(len(reshaped_mask[idx]))] for idx in range(len(reshaped_mask))]
db.register_dataset_3D(reshaped_mask_ids, "test_gt_masks")
# Query starts from here
query= ("Let gt_val_array = (For val in test_imputed_x"
+ " For sub_val in val.data"
+ " For mask in test_gt_masks"
+ " For sub_mask in mask.data"
+ " Filter val.id == mask.id and sub_val.id == sub_mask.id" # join
+ " For mask_id in sub_mask.data"
+ " return {'id': val.id, 'sub_id': sub_val.id, 'data': sub_val.data[mask_id]})" # find out all the existing ground-truths
+ " For item in gt_val_array "
+ " collect id = item.id, sub_id = item.sub_id into groups=item.data return {'id': id, 'sub_id': sub_id, 'group_data': groups}")
query_result= db.execute_query(query)
final_res = convert_query_res_to_tensors(query_result, test_imputed_x.permute(0,2,1)) # additional python script for computing thresholds a
determine violations
\end{minted}    
\end{mdframed}
\caption{T-Q3 in ArangoDB}
\label{fig:T-q3-adb}
\end{subfigure}
\caption{Implementation of T-Q3 in different query systems}\label{fig:T-q3-full}
\vspace{-0.15in}
\end{figure}

\paragraph{T-Q1} 
There are two input tables for the query T-Q2. One is the ``test\_dataset'' table which consists of two fields, the ``observed\_values'' variable and ``mask'' variable, which represent all the feature values (the missing values are replaced with 0) and the mask indicating the positions of the missing values respectively. The second input is the ``test\_imputed\_x'' table which stores the imputation results as a 2D matrix. This query starts with joining these two tables with a join operator and projects on two fields, ``test\_imputed\_x'' and ``mask'' on each row of the join result with a ``project'' operator. Note that the fields ``test\_imputed\_x'' and ``mask'' are from one time series sample, which are thus both 2D matrices. We then apply a transpose operation (``torch.t'') and a zip operator to concatenate each column of these two 2D matrices and then use one flatten operator to split the resulting 2D matrices into multiple single-feature time series along with their mask information.

The input of this query includes two named tables, i.e., ``train\_valid\_samples'' and ``mask'' which are both 3D tensors of the same shape. The named table ``train\_valid\_samples'' stores the medical records of each patient, which are composed of multiple measurements (such as blood pressure) as features across multiple time steps. Therefore, the three dimensions of this named table are sample dimension, temporal dimension and feature dimension respectively. The named table ``mask'' is a 0-1 3D tensor for recording which entry in ``train\_valid\_samples'' is missing or not. T-Q1 starts by joining those two named tables with one join operation, which is followed by invoking two user-defined functions ``compute\_gap\_res0'' and ``aggregate\_rows0'' in two continuous ``project'' operations. The function ``compute\_gap\_res0'' iterates each feature to compute the gap between one entry and its neighbors along temporal dimension while the function ``aggregate\_rows0'' compute the statistics of those gap values for each feature. The final results are stored in a named table ``consistency\_bound\_mask''.

\paragraph{T-Q2} The input of this query includes two named tables, ``test\_dataset'' and ``test\_imputed\_x''. Each row of ``test\_dataset'' stores each patient's record, which is composed of four fields, i.e., ``observed\_values'', ``observed\_mask'', ``time\_steps'' and ``gt\_mask''. The variable ``observed\_values'' stores temporal changes of multiple features, which is thus a 2D tensor. Due to the existence of missing values,  we use a 0-1 mask tensor, ``observed\_mask'' to indicate which entry in ``observed\_values'' is visible and ``gt\_mask'' records the subset of the visible entries (other visible entries are reserved for performance evaluations). For the named table ``test\_imputed\_x'', each row of this table is also a 2D tensor storing all the imputed values by using the time series imputation model. So T-Q2 first joins ``test\_dataset'' and ``test\_imputed\_x'' so that we can obtain both the mask tensors and the imputation result for each sample. Then for each sample, we apply ``project'' operation and perform transpose operation to facilitate the following ``flatten'' operations along each single-feature temporal sequence. For each such sequence, we apply a user-defined function ``check\_sequence\_feat\_range\_on\_time\_series0\_single\_feat'' to compute the inter-quartile range by using all the non-missing entries in ``test\_imputed\_x'' and capture the outliers based on that range. The final results are stored in a named table ``test\_imputation''.

\paragraph{T-Q3} This query is similar to T-Q2 except that 1) the outliers are determined by only using the named table ``test\_imputed\_x''; 2) only sub-sequences (rather than the entire sequence) are considered, which are obtained by using the function ``unfold''.

\begin{figure}
\begin{subfigure}[b]{\textwidth}
\begin{mdframed}
\begin{minted}[fontsize=\fontsize{5}{6}\selectfont, breaklines]{Python}
(Query('multilabel_seq_train', base='wilds_train2') # register
    .group_by(lambda idx, *args: args[1][1]) # group by sequence id
    # Iterate each sequence, compute number of unique labels and collect all the labels
    .project(lambda seqid, rows: (seqid, len(set([ row[0] for row in rows ])), [ row[0] for row in rows ]))
    # Select those sequences with more than one unique labels
    .filter(lambda seqid, numuniq, uniqlabels: numuniq > 1)
    # Sort the sequences and return their label list
    .order_by(lambda seqid, numuniq, uniqlabels: numuniq, reverse=True)).run(db)
\end{minted}    
\end{mdframed}
\caption{I-Q1 in \tool{}}
\label{fig:I-q1-mdb}
\end{subfigure}
\hfill
\begin{subfigure}[b]{\textwidth}
\begin{mdframed}
\begin{minted}[fontsize=\fontsize{5}{6}\selectfont, breaklines]{Python}
wilds_train_pd = pd.DataFrame(db.tables["wilds_train2"], columns=wilds_train_cln_ls)
# group by sequence id and compute number of unique labels and collect all the labels
grouped_join_train_pd= wilds_train_pd.groupby("meta_1").agg({"label": list,
"label":pd.Series.nunique}).rename(columns={0:"label_group", 1: "label_group_lenth"})
# Select those sequences with more than one unique labels
grouped_join_train_pd = grouped_join_train_pd[grouped_join_train_pd["label_group_lenth"] > 1]
# Sort the sequences and return their label list
grouped_join_train_pd = grouped_join_train_pd.sort_values("label_group_lenth")
\end{minted}
\end{mdframed}
\caption{I-Q1 in Pandas}
\label{fig:I-q1-pandas}
\end{subfigure}
\hfill
\begin{subfigure}[b]{\textwidth}
\begin{mdframed}
\begin{minted}[fontsize=\fontsize{5}{6}\selectfont, breaklines]{Python}
group_mappings = dict()
# group by sequence id
for item in db.tables["wilds_train2"]:
    label, meta = item
    if not meta[1] in group_mappings:
        group_mappings[meta[1]] = []
        group_mappings[meta[1]].append((label, meta))
seq_id_ls = []
seq_group_info_ls = []
seq_length_ls = []
# Iterate each sequence, and collect all the labels
for seq_id in group_mappings:
    # compute number of unique labels
    numuniq = len(set([item[0] for item in group_mappings[seq_id]]))
    # Select those sequences with more than one unique labels
    if numuniq > 1:
        seq_id_ls.append(seq_id)
        # collect all the labels
        seq_group_info_ls.append(group_mappings[seq_id])
        seq_length_ls.append(numuniq)
# Sort the sequences and return their label list
sorted_idx = np.argsort(np.array(seq_length_ls))
sorted_seq_id_ls = [seq_id_ls[idx] for idx in sorted_idx]
sorted_seq_group_info_ls = [seq_group_info_ls[idx] for idx in sorted_idx]
\end{minted}    
\end{mdframed}
\caption{I-Q1 in Python}
\label{fig:I-q1-python}
\end{subfigure}
\begin{subfigure}[b]{\textwidth}
\begin{mdframed}
\begin{minted}[fontsize=\fontsize{5}{6}\selectfont, breaklines]{Python}
adb.register_dataset_dir(db.tables["wilds_train2"], "wilds_train", ["label", "meta"])
query = ("Let grouped_join_pred = ( For item in wilds_train "
+ " collect pred = item.meta[1] INTO group=item.label" # group by sequence id
# compute number of unique labels and collect all the labels
+ " return {'pred': pred, 'label_group': group, 'label_group_lenth': COUNT_UNIQUE(group) } )"
+ " For item in grouped_join_pred "
+ " sort item.label_group_lenth DESC" # Sort the sequences and return their label list
+ " Filter item.label_group_lenth > 1" # Select those sequences with more than one unique labels
+ " return { 'pred': item.pred, 'label_group': item.label_group,
'label_group_lenth':item.label_group_lenth}")
query_res= adb.execute_query(query)
\end{minted}    
\end{mdframed}
\caption{I-Q1 in ArangoDB}
\label{fig:I-q1-adb}
\end{subfigure}
\caption{Implementation of I-Q1 in different query systems}\label{fig:I-q1-full}
\vspace{-0.15in}
\end{figure}

\begin{figure}[ht!]
\begin{subfigure}[b]{\textwidth}
\begin{mdframed}
\begin{minted}[fontsize=\fontsize{5}{6}\selectfont, breaklines]{Python}
(Query('t5_bias_all', base='t5_bias_answers') # register
# parse prompts and response
    .project(lambda prompt, answer: [*parse_prompt(prompt), parse_response(answer)])
    # group by the noun-adjective pairs
    .group_by(lambda idx, is_farmer, word1, word2, res: (is_farmer, frozenset({word1, word2})))
    # For each noun-adjective pairs, return bias
    .project(lambda key, rows: ['farmer' if key[0] else 'engineer', key[1], [row[3] for row in rows]])).run(db)
\end{minted}    
\end{mdframed}
\caption{B-Q1 in \tool{}}
\label{fig:B-q1-mdb}
\end{subfigure}
\hfill
\begin{subfigure}[b]{\textwidth}
\begin{mdframed}
\begin{minted}[fontsize=\fontsize{5}{6}\selectfont, breaklines]{Python}
t5_bias_answers_df = pd.DataFrame(list(db.tables["t5_bias_answers"]), columns=["prompt", "answer"])
# parse prompts and response
parse_res = t5_bias_answers_df.apply(lambda row: [*parse_prompt(row["prompt"]),
parse_response(row["answer"])], axis=1)
parse_res_df = pd.DataFrame(list(parse_res), columns=["is_farmer", "word1", "word2", "res"])
# one additional step for using one column to contain group key
parse_res_df["group_key"]= parse_res_df.apply(lambda row: (row["is_farmer"], frozenset({row["word1"],
row["word2"]})), axis=1)
# group by the noun-adjective pairs
grouped_res = parse_res_df.groupby("group_key")["res"].apply(list).reset_index()
# one additional step to have another column to include the flag whether it is farmer or engineer
grouped_res["noun"] = grouped_res["group_key"].apply(lambda item: "farmer" if item[0] else "engineer")
\end{minted}
\end{mdframed}
\caption{B-Q1 in Pandas}
\label{fig:B-q1-pandas}
\end{subfigure}
\hfill
\begin{subfigure}[b]{\textwidth}
\begin{mdframed}
\begin{minted}[fontsize=\fontsize{5}{6}\selectfont, breaklines]{Python}
key_row_mappings = dict()
t5_bias_answers_ls = list(db.tables["t5_bias_answers"])
# start group by
for idx in range(len(t5_bias_answers_ls)):
    prompt, answer = t5_bias_answers_ls[idx]
    # parse prompts and response
    is_farmer, word1, word2, res = [*parse_prompt(prompt), parse_response(answer)]
    key = (is_farmer, frozenset({word1, word2}))
    if not key in key_row_mappings:
        key_row_mappings[key] = []
        if is_farmer:
            noun = "farmer"
    else:
        noun = "engineer"
    # For each noun-adjective pairs, return bias
    key_row_mappings[key].append([noun, key[1], res])
# unzip the results
for key in key_row_mappings:
    key_row_mappings[key] = zip(*key_row_mappings[key])
\end{minted}    
\end{mdframed}
\caption{B-Q1 in Python}
\label{fig:B-q1-python}
\end{subfigure}
\begin{subfigure}[b]{\textwidth}
\begin{mdframed}
\begin{minted}[fontsize=\fontsize{5}{6}\selectfont, breaklines]{Python}
adb.register_dataset_dir(db.tables["wilds_train2"], "wilds_train", ["label", "meta"])
query = ("Let grouped_join_pred = ( For item in wilds_train "
+ " collect pred = item.meta[1] INTO group=item.label" # group by sequence id
# compute number of unique labels and collect all the labels
+ " return {'pred': pred, 'label_group': group, 'label_group_lenth': COUNT_UNIQUE(group) } )"
+ " For item in grouped_join_pred "
+ " sort item.label_group_lenth DESC" # Sort the sequences and return their label list
+ " Filter item.label_group_lenth > 1" # Select those sequences with more than one unique labels
+ " return { 'pred': item.pred, 'label_group': item.label_group,
'label_group_lenth':item.label_group_lenth}")
query_res= adb.execute_query(query)
\end{minted}    
\end{mdframed}
\caption{B-Q1 in ArangoDB}
\label{fig:B-q1-adb}
\end{subfigure}
\vspace{-0.1in}
\caption{Implementation of B-Q1 in different query systems}\label{fig:B-q1-full}
\vspace{-0.15in}
\end{figure}

\subsubsection{Image Classification} For this task, we examine the \wilds{} dataset and the predictions of the baseline ERM algorithm \citep{beery2020iwildcam}. \wilds{} dataset is composed of 129809 images in the training set and 14961 images in the validation set. As mentioned in Section \ref{sec:benchmarks}, since the predictions by the state-of-the-art models are made over each individual frame, the consistency of predictions across continuous frames is not considered. We therefore consider I-Q1 to query subsequences of frames to identify whether the same prediction label exists by counting how many different predictions those sequences contain. The detailed implementation of I-Q1 is included in Figure \ref{fig:I-q1-full} and how to construct I-Q1 with the operators defined in Section \ref{sec:operator_semantics} is described as follows.

As Figure \ref{fig:I-q1-mdb} shows, I-Q1 takes a named table ``wilds\_train2'' as input. Each row of this dataset corresponds to one frame in the \wilds{} dataset, which is composed of two fields, i.e., the ground-truth label and an array of metadata. So this query first groups by the second field of the metadata, i.e., ``args[1][1]'', which represents the sequence ID of this frame (recall that \wilds{} is composed of multiple video sequences). Then the number of unique ground-truth labels as well as all of these labels are collected within each sequence with the ``project'' operation. This can then facilitate the following ``filter'' operation to only retain those video sequences with at least two unique labels. The final result returns all these sequences which are sorted by the number of their unique labels.

\subsubsection{Text Generation}\label{app:bias-details}
For this task, we discover the bias in Alpaca instruction fine-tuning dataset \cite{alpaca} (\Alpaca\ for short), which consists of 52K instruction-response pairs generated from GPT-3.5 where the instructions were generated using the Self-Instruct technique \cite{selfinstruct}. We then evaluate T5 with all 52K instructions to create another set of 52K instruction-answer pairs for T5. As described in Section \ref{sec:benchmarks}, we start with the high-level hypothesis that there may be biases in the associations of occupations and the adjectives to describe different occupations such as farmer and engineer. After finding adjectives which are more associated with certain occupations than others, we confirm these biases through a counterfactual experiment by prompting the model and analyzing the results. Both promptings of the model and analyzing the results are done within \tool{}, where we describe how to prompt the model below and then discuss query B-Q1 which is used for the analysis of the counterfactual experiment. 
The detailed implementations of B-Q1 in different query systems are presented in Figure \ref{fig:B-q1-full}.

As shown in Figure \ref{fig:B-q1-mdb}, B-Q1 takes the output from the above query, ``t5\_bias\_answers,'' as input. Each row of this table stores a prompt-response pair, in which the prompt is from the \Alpaca\ dataset and the response is generated by the language model. Query B-Q1 is then used to analyze the responses. The query iterates over each prompt-response pair and parses the prompts and answers with the function ``parse\_prompt'' and ``parse\_response'' respectively in the ``project'' operation. For the ``parse\_prompt'' function, it returns a flag variable to indicate whether the prompt contains the word ``farmer'' and two additional words as the possible occupations that the prompt mentions (denoted by ``word1'' and ``word2''). For the function ``parse\_response'', it simply parses the response into a list of tokens, which is denoted by a variable ``res''. In the subsequent step, the query groups by the flag variable ``is\_farmer'' and the combination of ``word1'' and ``word2'', and then collects the parsing results of each response sentence, i.e., ``res'' from the group\_by operation, within each group.

\section{Additional Details of the User Study}

\subsection{Full Queries for Task T3}
\label{app:full_t3}

We show the full query for task T3 written by participant P1:
\begin{center}
\begin{minipage}{0.85\textwidth}
\begin{mdframed}
\begin{minted}[fontsize=\scriptsize]{Python}
def first_nonzero(lst):
    for i in range(len(lst)):
        if lst[i] != 0:
            return lst[i]
    return 0

def contains_conflicting_pred(lst):
    n = first_nonzero(lst)
    for i in range(len(lst)):
        if lst[i] != n and lst[i] != 0:
            return True
    return False

def get_sequence_preds(lst):
    return [item[3] for item in lst]

Query('T3_P1', base='wilds_pred')
    .group_by(lambda img, label, md, pred: md[1].item())
    .filter(lambda seq, lst: contains_conflicting_pred(get_sequence_preds(lst)))
\end{minted}
\end{mdframed}
\end{minipage}
\end{center}

Below is the full query for participant P5:
\begin{center}
\begin{minipage}{0.85\textwidth}
\begin{mdframed}
\begin{minted}[fontsize=\scriptsize]{Python}
Query('T3_P5', base='wilds_pred')
    .group_by(lambda img, _, metadata, pred: metadata[1].item())
    .cols(lambda seqid, rows: (seqid, len([row for row in rows
            if row[2][5] in [21, 20, 3, 2, 5, 19] or row[2][4] in [29, 17, 19] or
            row[2][3] in [9, 4] or row[2][6] in [31, 32, 37, 35]]) > 1,
        len(rows), rows))
    .filter(lambda seqid, has_incorrect, _, __: has_incorrect)
    .cols(lambda seqid, _, __, ___: seqid)
\end{minted}
\end{mdframed}
\end{minipage}
\end{center}

\subsection{Additional Readability Task in the User Study}\label{app:readability}
We further expand the user study to evaluate the readability of the queries written in different systems. Specifically, users are presented with two queries written in \tool{}, native Python and Pandas, and asked to score the readability of these two queries on a scale of 1 to 10. These queries are the SQ-1 and IQ-1 queries.

We collected preliminary responses from five respondents. 
The average readability scores among them are 8.8, 3.4 and 4.6 for \tool{}, Python and Pandas respectively. Hence, 
in general, \tool{} was said to be more readable than Pandas, which was slightly more readable than Python.

\bibliography{references}

\end{document}